\pdfoutput=1
\documentclass[apj, iop, numberedappendix, appendixfloats]{emulateapj}
\usepackage{graphicx}
\usepackage{enumerate}
\usepackage[english]{babel}
\usepackage[T1]{fontenc}
\usepackage{hyperref}
\usepackage{amsmath}
\usepackage[utf8]{inputenc} 
\usepackage{apjfonts}
\usepackage{natbib}
\usepackage{longtable}
\usepackage{threeparttablex}
\usepackage{multirow}

\makeatletter
\let\oldlt\longtable
\let\endoldlt\endlongtable
\def\longtable{\@ifnextchar[\longtable@i \longtable@ii}
\def\longtable@i[#1]{\begin{figure}[t]
\onecolumn
\begin{minipage}{0.5\textwidth}
\oldlt[#1]
}
\def\longtable@ii{\begin{figure}[t]
\onecolumn
\begin{minipage}{0.5\textwidth}
\oldlt
}
\def\endlongtable{\endoldlt
\end{minipage}
\twocolumn
\end{figure}}
\makeatother

\newcommand{\myemail}{\textit{matteo.cerruti@lpnhe.in2p3.fr}}
\newcommand{\caitlinemail}{\textit{caajohns@ucsc.edu}}
\newcommand{\wystanmail}{\textit{wystan.benbow@cfa.harvard.edu}}
\newcommand{\veritas}{VERITAS}
\newcommand{\magic}{MAGIC}
\newcommand{\hess}{H.E.S.S.}
\newcommand{\fermilat}{\textit{Fermi}-LAT}

\begin{document}
\title{Upper limits from five years of blazar observations\\ with the \veritas\ Cherenkov telescopes}
\shorttitle{Upper limits from five years of \veritas\ blazar observations}

\author{
S.~Archambault\altaffilmark{1},
A.~Archer\altaffilmark{2},
W.~Benbow\altaffilmark{3},
R.~Bird\altaffilmark{4},
J.~Biteau\altaffilmark{5},
M.~Buchovecky\altaffilmark{6},
J.~H.~Buckley\altaffilmark{2},
V.~Bugaev\altaffilmark{2},
K.~Byrum\altaffilmark{7},
M.~Cerruti\altaffilmark{3,$\star$},
X.~Chen\altaffilmark{8,9},
L.~Ciupik\altaffilmark{10},
M.~P.~Connolly\altaffilmark{11},
W.~Cui\altaffilmark{12},
J.~D.~Eisch\altaffilmark{13},
M.~Errando\altaffilmark{14},
A.~Falcone\altaffilmark{15},
Q.~Feng\altaffilmark{12},
J.~P.~Finley\altaffilmark{12},
H.~Fleischhack\altaffilmark{9},
P.~Fortin\altaffilmark{3},
L.~Fortson\altaffilmark{16},
A.~Furniss\altaffilmark{17},
G.~H.~Gillanders\altaffilmark{11},
S.~Griffin\altaffilmark{1},
J.~Grube\altaffilmark{10},
G.~Gyuk\altaffilmark{10},
M.~H\"{u}tten\altaffilmark{9},
N.~H{\aa}kansson\altaffilmark{8},
D.~Hanna\altaffilmark{1},
J.~Holder\altaffilmark{18},
T.~B.~Humensky\altaffilmark{19},
C.~A.~Johnson\altaffilmark{5},
P.~Kaaret\altaffilmark{20},
P.~Kar\altaffilmark{21},
N.~Kelley-Hoskins\altaffilmark{9},
M.~Kertzman\altaffilmark{22},
D.~Kieda\altaffilmark{21},
M.~Krause\altaffilmark{9},
F.~Krennrich\altaffilmark{13},
S.~Kumar\altaffilmark{18},
M.~J.~Lang\altaffilmark{11},
G.~Maier\altaffilmark{9},
S.~McArthur\altaffilmark{12},
A.~McCann\altaffilmark{1},
K.~Meagher\altaffilmark{23},
P.~Moriarty\altaffilmark{11},
R.~Mukherjee\altaffilmark{14},
T.~Nguyen\altaffilmark{23},
D.~Nieto\altaffilmark{19},
A.~O'Faol\'{a}in de Bhr\'{o}ithe\altaffilmark{9},
R.~A.~Ong\altaffilmark{6},
A.~N.~Otte\altaffilmark{23},
N.~Park\altaffilmark{24},
J.~S.~Perkins\altaffilmark{25},
A.~Pichel\altaffilmark{26},
M.~Pohl\altaffilmark{8,9},
A.~Popkow\altaffilmark{6},
E.~Pueschel\altaffilmark{4},
J.~Quinn\altaffilmark{4},
K.~Ragan\altaffilmark{1},
P.~T.~Reynolds\altaffilmark{27},
G.~T.~Richards\altaffilmark{23},
E.~Roache\altaffilmark{3},
A.~C.~Rovero\altaffilmark{26},
M.~Santander\altaffilmark{14},
G.~H.~Sembroski\altaffilmark{12},
K.~Shahinyan\altaffilmark{16},
A.~W.~Smith\altaffilmark{28},
D.~Staszak\altaffilmark{1},
I.~Telezhinsky\altaffilmark{8,9},
J.~V.~Tucci\altaffilmark{12},
J.~Tyler\altaffilmark{1},
S.~Vincent\altaffilmark{9},
S.~P.~Wakely\altaffilmark{24},
O.~M.~Weiner\altaffilmark{19},
A.~Weinstein\altaffilmark{13},
D.~A.~Williams\altaffilmark{5},
B.~Zitzer\altaffilmark{7}\\ (the VERITAS collaboration), 
and M.~Fumagalli\altaffilmark{29},
J.~X.~Prochaska\altaffilmark{30}
}

\altaffiltext{1}{Physics Department, McGill University, Montreal, QC H3A 2T8, Canada}
\altaffiltext{2}{Department of Physics, Washington University, St. Louis, MO 63130, USA}
\altaffiltext{3}{Fred Lawrence Whipple Observatory, Harvard-Smithsonian Center for Astrophysics, Amado, AZ 85645, USA\\ \wystanmail}
\altaffiltext{4}{School of Physics, University College Dublin, Belfield, Dublin 4, Ireland}
\altaffiltext{5}{Santa Cruz Institute for Particle Physics and Department of Physics, University of California, Santa Cruz, CA 95064, USA\\ \caitlinemail}
\altaffiltext{6}{Department of Physics and Astronomy, University of California, Los Angeles, CA 90095, USA}
\altaffiltext{7}{Argonne National Laboratory, 9700 S. Cass Avenue, Argonne, IL 60439, USA}
\altaffiltext{8}{Institute of Physics and Astronomy, University of Potsdam, 14476 Potsdam-Golm, Germany}
\altaffiltext{9}{DESY, Platanenallee 6, 15738 Zeuthen, Germany}
\altaffiltext{10}{Astronomy Department, Adler Planetarium and Astronomy Museum, Chicago, IL 60605, USA}
\altaffiltext{11}{School of Physics, National University of Ireland Galway, University Road, Galway, Ireland}
\altaffiltext{12}{Department of Physics and Astronomy, Purdue University, West Lafayette, IN 47907, USA}
\altaffiltext{13}{Department of Physics and Astronomy, Iowa State University, Ames, IA 50011, USA}
\altaffiltext{14}{Department of Physics and Astronomy, Barnard College, Columbia University, NY 10027, USA}
\altaffiltext{15}{Department of Astronomy and Astrophysics, 525 Davey Lab, Pennsylvania State University, University Park, PA 16802, USA}
\altaffiltext{16}{School of Physics and Astronomy, University of Minnesota, Minneapolis, MN 55455, USA}
\altaffiltext{17}{Department of Physics, California State University - East Bay, Hayward, CA 94542, USA}
\altaffiltext{18}{Department of Physics and Astronomy and the Bartol Research Institute, University of Delaware, Newark, DE 19716, USA}
\altaffiltext{19}{Physics Department, Columbia University, New York, NY 10027, USA}
\altaffiltext{20}{Department of Physics and Astronomy, University of Iowa, Van Allen Hall, Iowa City, IA 52242, USA}
\altaffiltext{21}{Department of Physics and Astronomy, University of Utah, Salt Lake City, UT 84112, USA}
\altaffiltext{22}{Department of Physics and Astronomy, DePauw University, Greencastle, IN 46135-0037, USA}
\altaffiltext{23}{School of Physics and Center for Relativistic Astrophysics, Georgia Institute of Technology, 837 State Street NW, Atlanta, GA 30332-0430}
\altaffiltext{24}{Enrico Fermi Institute, University of Chicago, Chicago, IL 60637, USA}
\altaffiltext{25}{N.A.S.A./Goddard Space-Flight Center, Code 661, Greenbelt, MD 20771, USA}
\altaffiltext{26}{Instituto de Astronomia y Fisica del Espacio, Casilla de Correo 67 - Sucursal 28, (C1428ZAA) Ciudad Aut\'{o}noma de Buenos Aires, Argentina}
\altaffiltext{27}{Department of Physical Sciences, Cork Institute of Technology, Bishopstown, Cork, Ireland}
\altaffiltext{28}{University of Maryland, College Park / NASA GSFC, College Park, MD 20742, USA}
\altaffiltext{29}{Institute for Computational Cosmology and Centre for Extragalactic Astronomy, Department of Physics, Durham University, South Road, Durham, DH1 3LE, UK}
\altaffiltext{30}{Department of Astronomy and Astrophysics, University of California, Santa Cruz, CA 95064, USA}
\altaffiltext{$\star$}{Now at Sorbonne Universit\'{e}s, UPMC Universit\'{e} Paris 06, Universit\'{e} Paris Diderot, Sorbonne Paris Cit\'{e}, CNRS, Laboratoire de Physique Nucl\'{e}aire et de Hautes Energies (LPNHE), 4 place Jussieu, F-75252, Paris Cedex 5, France\\ \myemail}

\begin{abstract}
Between the beginning of its full-scale scientific operations in 2007 and 2012, the \veritas\ Cherenkov telescope array observed more than 130 blazars; of these, 26 were detected as very-high-energy (VHE; E$>$100 GeV) $\gamma \textrm{-ray}$ sources. In this work, we present the analysis results of  a sample of 114 undetected objects. The observations constitute a total live-time of $\sim$ 570 hours. The sample includes several unidentified \textit{Fermi}-Large Area Telescope (LAT) sources (located at high Galactic latitude) as well as all the sources from the second \fermilat\ catalog which are contained within the field of view of the \veritas\ observations. We have also performed optical spectroscopy measurements in order to estimate the redshift of some of these blazars that do not have a spectroscopic distance estimate. We present new optical spectra from the Kast instrument on the Shane telescope at the Lick observatory for 18 blazars included in this work, which allowed for the successful measurement or constraint on the redshift of four of them. For each of the blazars included in our sample we provide the flux upper limit in the \veritas\ energy band. We also study the properties of the significance distributions and we present the result of a stacked analysis of the data-set, which shows a 4 $\sigma$ excess. \vspace{0.3cm}
\end{abstract}


\section{Introduction}
\label{section1}
\setcounter{footnote}{0}

The current generation of imaging atmospheric Cherenkov telescopes (IACTs), sensitive to very-high-energy (VHE; E$>$100 GeV) $\gamma$-ray photons, has significantly increased our knowledge of blazars. They represent by far the class of objects which dominates the VHE extragalactic sky. Among the 66 extragalactic VHE sources currently detected,\footnote{For a recent review, see e.g. \citet{Gunes13}; for an updated list of known VHE sources see \url{http://tevcat.uchicago.edu}} about 90$\%$ of them are blazars.\\

In the framework of the unified model of active galactic nuclei (AGN), blazars are radio-loud AGN, characterized by a pair of relativistic jets of plasma emitted along the polar axis of the super-massive black-hole powering the system,  aligned along the line of sight of the observer \citep[see][]{Urry95}. The observational properties of blazars include a spectral energy distribution (SED) characterized by a non-thermal continuum from radio to $\gamma$-rays, extreme temporal variability and strong polarization. These properties can be explained by considering that the emission from the jet, enhanced by relativistic effects, dominates the SED \citep[][]{Angel80}. Spectroscopic measurements in optical and UV reveal that two distinct sub-classes of blazars exist: BL Lac objects, characterized by a featureless optical/UV spectrum, and flat-spectrum radio-quasars (FSRQs), which show instead broad emission lines. The traditional division between the two classes of objects is an equivalent width of the emission lines equal to 5$ \ensuremath{\mathring{\text{A}}}$ \citep[see][]{Stickel91}. The two sub-classes are also characterized by different luminosity and redshift distributions. The FSRQs are on average brighter and located at higher redshifts \citep[see e.g.][]{Padovani92, Massaro09}.  In the unified AGN model, this dichotomy is associated with a similar dichotomy seen in radio-galaxies. FSRQs are considered the blazar version of FR II radio-galaxies \citep{FR74}, while BL Lac objects correspond to FR I \citep{Urry95}.\\ 

 Two broad non-thermal components characterize blazar SEDs. The first, peaking between millimeter and X-rays, is attributed to synchrotron emission by a population of electrons/positrons in the blazar jet. The second, peaking in $\gamma$-rays, is associated, in leptonic models, with inverse Compton scattering between the same leptons and their own synchrotron emission \citep[synchtron-self-Compton model, SSC,][]{Konigl81}, or an external photon field, such as the emission from the dusty torus, the accretion disc or the broad-line-region \citep[see][]{Sikora94}. Alternatively, in hadronic scenarios, the second SED component is attributed to synchrotron emission by protons, or by secondary particles produced in p-$\gamma$ interactions \citep{Mucke01}.
 The position of the first peak is used to further classify BL Lac objects into low and high-frequency-peaked BL Lac objects (LBL/HBL), depending on whether the peak frequency is located in infrared/optical or UV/X-rays, respectively. The transition between LBL and HBL is smooth, and a population of intermediate-frequency-peaked BL Lac objects exists as well \citep[IBL, with $10^{14}\ \text{Hz}\ $<$\ \nu_{syn}\ $<$\ 10^{15}$ Hz, see][]{Laurent98}.  While BL Lac objects present a variety of synchrotron peak frequencies, FSRQs are almost all characterized by a low-frequency peak. This classification can be seen as a more recent version of the older classification \citep[see e.g.][]{Padovani95} of blazars into radio-selected objects (RBLs, which are more likely FSRQs and LBLs) and X-ray-selected objects (XBLs, which are more likely HBLs)\footnote{The classification of BL Lac objects as LBL/IBL/HBL is sometimes replaced by LSP/ISP/HSP \citep[low/intermediate/high-synchrotron-peaked blazars, see][]{Abdo10}, making explicit reference to the synchrotron origin of the first component of the SED. For BL Lac objects, the two triplets of acronyms can be considered as synonyms. }. \\
 
The measurement of blazar spectral properties at VHE is important not only to characterize the blazar emission itself, but also to indirectly study the extragalactic background-light  (EBL) in the infrared and visible bands due to the absorption that it causes on VHE photons via $e^\pm$ pair-production, \citep[see][]{Salamon98}. VHE observations can also be used to put limits on the strength of the intergalactic magnetic field according to the non-detection of the emission from the cascade triggered by the interaction of the  pairs with the cosmic-microwave-background \citep[see e.g.][]{Taylor11}.\\

Even though the detection and the measurement of the VHE spectrum of a blazar is of paramount importance for the comprehension of the physics of relativistic jets in AGN and for cosmological studies, a non-detection in the VHE regime can also be extremely useful. It can constrain the source emission model, especially when the flux upper-limit is significantly lower than the extrapolation of the \fermilat\ \citep{FermiLAT}  measurement in high-energy $\gamma$-rays (HE; 100 MeV$<$E$<$100 GeV) up to VHE, implying the presence of a spectral cut-off. It can also constrain the variability properties of the source at VHE if the blazar has been previously detected or is detected at a later time during a higher flux state. Additionally, for the upcoming Cherenkov Telescope Array \citep[CTA,][]{CTA}, it will be useful to have the information from all past observations performed by the current generation of IACTs, in order to make predictions for expected outcomes \citep[see e.g.][]{Sol13}.\\

All three major IACTs (\hess, \magic\ and \veritas) have published upper-limits on blazar VHE emission, including more than 70 sources in their lists \citep{HESSUL2005, HESSUL2008, MAGICUL, Fermi6UL, HESSUL2014}. Past IACTs, such as CAT, HEGRA and Whipple, have also presented the results from undetected blazars \citep{Piron00, HegraUL, Falcone04, WhippleUL}, though their limits have in general been superseded by the current instruments. At higher energies, upper limits on blazars have also been estimated using the air-shower technique with Milagro \citep{milagroul}.\\

The study of blazars is complicated by the uncertainty in their redshifts. In fact, the almost featureless spectra of BL Lac objects imply that a redshift estimate can be obtained only via absorption features from the host galaxy or clouds in the intergalactic medium, or via molecular emission lines \citep[see][]{Fuma12}, or via a less precise photometric estimate (comparing the luminosity of the host galaxy to samples of giant elliptical galaxies). If the blazar belongs to a group of galaxies, its redshift can also be estimated by studying the non-active companions \citep[see][]{Muriel15}. For VHE studies, the knowledge of the redshift of the blazar is very important, because the absorption by the EBL increases with the distance of the object. Currently, the farthest detected VHE blazar is the gravitationally-lensed quasar S3 0218+357 \citep{S30218} at $z=0.944$, closely followed by the FSRQ PKS 1441+25  \citep{1441magic, 1441veritas} at $z = 0.939$,  while the farthest, persistent (i.e. detected not only during episodic flaring activity) VHE blazar is PKS~1424+240 \citep[$z>0.61$, see][]{PKS1424Veritas, Furniss13, PKS1424Veritas2}. To improve the constraints on the distance of some VHE candidates, new redshift estimates obtained with the Kast spectrograph at Lick Observatory (see Section \ref{section2} and Appendix \ref{appendixredshift}) are presented together with the VHE upper limits from \veritas.\\

In this paper we present the results of the analysis of most of the non-detected blazars observed by \veritas\ from 2007 (the beginning of full-scale scientific operations) to August 2012 \citep[before the upgrade of the VERITAS array, see][]{Kieda13}. \veritas\ upper limits on six VHE candidates were presented by \citet{Fermi6UL}. The sample also includes several unidentified \fermilat\ objects \citep[located at high Galactic latitude, and most probably associated with unidentified AGN, see][]{Ackermann12, Mirabal12, DoertErrando}.\\

The paper is organized as follows: in Section \ref{section2} we provide the details of the properties of the sample and the source selection; the details of the \veritas\ data analysis and results are provided in Section \ref{section3}; a stacked analysis of the data is presented in Section \ref{section6}, studying the full data-set as well as sub-data-sets defined by redshifts and blazar classes; the conclusions are in Section \ref{section7}.\\

\section{The sample and new redshift estimates}
\label{section2}

Blazars targeted by \veritas\ as VHE source candidates were selected according to a variety of criteria. Early source selections were based on blazar X-ray or radio catalogs, while more recent candidates have also been selected on the basis of their  \fermilat\ spectral characteristics \citep{Fermi0FGL, 1FGL, 2FGL} or on their association with clusters of HE $\gamma$-rays \citep[see][]{verj0521}. The target list includes:\\
\begin{itemize} 
\item all the nearby ($z < 0.3$) HBL/IBL recommended as potential VHE emitters by \citet{SDS, Perlman00} and \citet{CG}   
\item the X-ray brightest nearby ($z < 0.3$) HBL in the Sedentary \citep{Sedentary} and ROXA \citep{ROXA} surveys
\item four distant ($z > 0.3$) BL Lac objects recommended by \citet{CG} and \citet{Costamante07}
\item all nearby ($z < 0.3$) blazars detected by \textit{EGRET} \citep{Egret}
\item several FSRQs recommended as potential VHE emitters by \citet{Perlman00} and \citet{Padovani02}
\item two high-frequency-peaked FSRQs \citep[B2 0321+33 and Mrk 1218, see][]{Perlman00, Falcone04}, which have also been classified as Seyfert-1 galaxies \citep[see][]{Osterbrock83, FermiNLS1}
\item the brightest \fermilat\ sources after extrapolation into the \veritas\ energy band \citep{Fermi0FGL, 1FGL, 2FGL}
\item sources associated with clusters of HE $\gamma$-rays, but not included in \fermilat\ catalogs, similar to \textit{VER~J0521+211} \citep{verj0521}. \\
\end{itemize}

In addition, several targets have been observed by \veritas\ only as targets of opportunity (ToO), following flare alerts by multi-wavelength partners \citep[see][]{Errando11}.\\ 

The sources included in our sample are listed in Table \ref{Sourcelist} (for sources without VHE detection, 82 targets) and Table \ref{knownVHElist} (for known VHE emitters, detected by other instruments, or by \veritas\ after 2012 and during flaring activity, 11 targets). Sources are listed in order of increasing Right Ascension (R.A.). For every target, we indicate the name, the coordinates (R.A. and Dec, in J2000), the catalog redshift $z$, the blazar class, the \veritas\ dead-time corrected exposure (see Section \ref{section3}), the average zenith angle of \veritas\ observations, the dates of \veritas\ observations (in MJD) and on which basis the source was selected as a VHE candidate. Names and coordinates are taken from the \textit{SIMBAD} database.\footnote{\url{http://simbad.u-strasbg.fr/simbad/}} The references for the redshift estimates (and their uncertainties) and the blazar class are provided in the table notes. For every source, the archival SED from the \textit{ASDC SED Builder} tool\footnote{\url{http://tools.asdc.asi.it/SED/}} has been visually inspected. It is used to classify all BL Lac objects marked as HBLs.
The total number of hours of \veritas\ data analyzed is about 570, which represents about 60\% of a single \veritas\ yearly observing season, i.e. about one tenth of all good quality \veritas\ data taken from 2007 to 2012. \\

The field-of-view (FoV) of the \veritas\ telescope array is about $3.5^\circ$ and for every observation there is a chance that, in addition to the targeted blazar, other $\gamma$-ray sources are contained within the FoV. For every target included in our sample we checked if other known $\gamma$-ray sources \citep[included in the 2-year \fermilat\ catalog, 2FGL, see][]{2FGL} were present in the FoV. Twenty-one 2FGL sources were indirectly observed by \veritas\ through proximity to the blazar of interest, and are listed in Table  \ref{2FGLlist}. We indicate as well the counterpart name (from the 2FGL catalog), the coordinates (R.A. and Dec) of the counterpart, the redshift and the blazar sub-class (if known). The majority of these additional 2FGL sources are AGN without classification.\\

\subsection{New redshift measurements}

In order to measure the distance of some $\gamma$-ray blazars, we observed 18 of the \veritas\ targets using the dual-arm Kast spectrograph at the Cassegrain focus of the Shane telescope at Lick Observatory. For all the observations presented here, the instrument was configured with the 600/4310 grism on the blue arm, and the 600/7500 grating on the red arm, the D55 dichroic, and a $2''$ slit. The dichroic crossover creates an instrumental gap located at $\sim$5500 $ \ensuremath{\mathring{\text{A}}}$ and affects approximately  200~$ \ensuremath{\mathring{\text{A}}}$ of the spectrum. In the spectroscopic figures \ref{fig:one} - \ref{fig:six} (Appendix A), this gap is shown on each spectrum. While the absolute fluxes are shown in the plots, the flux calibration is the least certain aspect of the spectra. In several spectra there is a gap in flux across the dichroic; this is purely the result of calibrations and is not intrinsic to the AGN. The wavelength coverage is from $\sim$3450 $ \ensuremath{\mathring{\text{A}}}$ to $\sim$8000 $ \ensuremath{\mathring{\text{A}}}$, but tellurics and fringing mask features above 6850 $\ensuremath{\mathring{\text{A}}}$. We do not show this contaminated portion in the spectra. The targets, observation dates and exposures are summarized in Table \ref{newredshift}. The corresponding standard star and the signal-to-noise ratio ($S/N$) are also included in this table.  The data were reduced following standard techniques with the Low-Redux pipeline\footnote{\url{http://www.ucolick.org/\~xavier/LowRedux/}}. Each spectrum was inspected visually for absorption or emission features. Features are noted in the spectral plots, Figures \ref{fig:one} - \ref{fig:six}, and in Table \ref{newredshift}. For 14 of the sources observed, there are no spectral features that allow redshift measurements or redshift limits. By definition, BL Lac objects have no or weak spectral lines, so this high rate of non-detections is somewhat expected. Below are the sources for which features were found that allow for redshift determinations (see Figure \ref{fig:one}):\\
\begin{itemize}
 \item RGB J0250+172
 
Observations of the source were obtained on August 15, 2010 (UT) and resulted in the detection of galactic features at $z=0.243$. We detected Ca II (H, equivalent width (EW) $=870 \pm 160 \ \textrm{m} \mathring{\text{A}}$; K, $EW=1340 \pm 190 \ \textrm{m} \mathring{\text{A}}$), G band ($EW=650 \pm 180 \ \textrm{m} \mathring{\text{A}}$) and Mg I ($EW=530 \pm 120 \ \textrm{m} \mathring{\text{A}}$) absorption.
 
In the literature, \citet{Bauer00} quotes a redshift of z=1.10 for RGB J0250+172; however, there is no information on spectroscopic lines provided within the reference. \citet{Nilsson03} presents optical images of BL Lac objects, including RGB J0250+172. They find that the object is clearly resolved and state that $z=1.10$ is too high because it results in a host galaxy that is exceedingly bright ($M_{R} < -29.0$). Based on fits to the observed light profile, a redshift estimate of $z=0.25$ is provided, which is similar to the value measured within this work.\\

  \item 1ES 1118+424
  
  Observations of the source were obtained on February 14, 2013 (UT) and resulted in the detection of galactic features at $z=0.230$. We detected Ca II (H, $EW= 11500 \pm 1500 \ \textrm{m} \mathring{\text{A}}$; K, $EW=13000 \pm 1700 \ \textrm{m} \mathring{\text{A}}$), G band ($EW=3790 \pm 862 \ \textrm{m} \mathring{\text{A}}$), Ca I ($EW= 1208 \pm 199 \ \textrm{m} \mathring{\text{A}}$) and Mg I ($EW= 2896 \pm 254 \ \textrm{m} \mathring{\text{A}}$) absorption lines. 
  
   In the literature, the redshift for 1ES 1118+424 is quoted as $z=0.124$ from a private communication \citep[see][]{Falomo99}. However, \citet{Falomo99} derive a lower limit of $z>0.5$ based on images taken using the Nordic Optical Telescope, where the source is unresolved. They simulate an elliptical host galaxy with $M_{R} = -23.8$ and an effective radius of 10 kpc to determine the lowest redshift at which it would not be resolved. These galactic parameters are what they find from other BL Lac objects in their study, and they note that assuming a less luminous and smaller host galaxy would result in a lower redshift estimate. \\

  \item RBS 1366 (=1E 1415+25.9) 
  
  Observations of the source obtained on May 30, 2014 (UT)  resulted in the detection of galactic features at $z=0.237$. We detected Ca II (H, $EW= 1300 \pm 130 \ \textrm{m} \mathring{\text{A}}$; K, $EW= 1570 \pm 140 \ \textrm{m} \mathring{\text{A}}$), G band ($EW= 210 \pm 160 \ \textrm{m} \mathring{\text{A}}$), Ca I ($EW= 710 \pm 64 \ \textrm{m} \mathring{\text{A}}$) and Mg I ($EW= 1900 \pm 100 \ \textrm{m} \mathring{\text{A}}$) absorption. 
    
 \citet{Halpern86} also measure a redshift of $z=0.237$ based on Ca II, G band, Fe I, Mg I and Na absorption. This source displays the significant variability associated with BL Lac objects.  The spectrum taken in 2013 has a lower overall flux than the spectrum taken in 2014 (see Table \ref{newredshift}), indicating that the source might have been in different flux states.\\
  
  \item 1ES 2321+419

Observations of the source obtained on October 28, 2014 (UT) resulted in the detection of absorption features at $z=0.267$. We detected Ca II (H, $EW=260 \pm 47 \ \textrm{m} \mathring{\text{A}}$; K, $EW= 180 \pm 52 \ \textrm{m} \mathring{\text{A}}$) and Mg II (2796 $\mathring{\text{A}}$, $EW= 740 \pm 83 \ \textrm{m} \mathring{\text{A}}$; 2803 $\mathring{\text{A}}$, $EW= 510 \pm 67 \ \textrm{m} \mathring{\text{A}}$) absorption. Because the Ca II absorption is narrow, and there are no other galactic features, only a lower limit can be set on the redshift of the source. Additionally, there is potentially Mg II absorption at a higher redshift, $z=0.346$. 

In the literature,  \cite{Falomo99} derive a lower limit for this source of $z > 0.45$ using the same technique as for 1ES 1118+424. While our value is not inconsistent, it is considerably lower than that placed based on assumptions about the host galaxy.\\
  
  \end{itemize}

\section{\veritas\ observations and data analysis}
\label{section3}
The \veritas\ (Very Energetic Radiation Imaging Telescope Array System) telescope array is composed of four IACTs of 12-m diameter each, located at the Fred Lawrence Whipple Observatory, on the slopes of Mount Hopkins, in southern Arizona (31$^\circ$ 40$^\prime$ N, 110$^\circ$ 57$^\prime$ W). Each telescope has a segmented mirror which focuses light onto a camera composed of 499 photomultipliers located at the focal plane. The instrument FoV is 3.5$^\circ$. For further details on the \veritas\ instrument see \citet{Holder06, Holder11}.\\

The telescopes measure the faint Cherenkov light induced by the electromagnetic showers triggered by the interaction of the $\gamma$-ray photons with the Earth atmosphere. Similar cascades triggered by cosmic rays are also detected by \veritas, and can be rejected by applying specific cuts on the shape of the Cherenkov image \citep{Hillas85}.\\

\newpage

\begin{ThreePartTable}
\renewcommand\TPTminimum{\textwidth}
  \begin{TableNotes}
  \vspace{0.2cm}
  \item[a] Unconstrained redshifts are indicated with a hyphen ($-$). If the redshift value is uncertain it is followed by a colon ($:$).\\ Redshift references: \tablenotemark{1} {\protect\cite{Fischer98}};\tablenotemark{2} {\protect\cite{Rau2012}}; \tablenotemark{3}{\protect\cite{Laurent98}}; \tablenotemark{4}{\protect\cite{Perlman96}}; \tablenotemark{5}{\protect\cite{Lawrence86}}; \tablenotemark{6}{\protect\cite{Shaw13}; \tablenotemark{7}{\protect\cite{Meisner10}};\tablenotemark{8}{\protect\cite{Marcha96}};\tablenotemark{9}{\protect\cite{Bohringer00}};\tablenotemark{10}{\protect\cite{Sbarufatti05}};\tablenotemark{11}{\protect\cite{Cohen87}};\tablenotemark{12}{this work};\tablenotemark{13}{\protect\cite{Piranomonte07}};\tablenotemark{14}{\protect\cite{Bauer00}};\tablenotemark{15}{\protect\cite{Lavaux11}};\tablenotemark{16}{\protect\cite{Carswell74}};\tablenotemark{17}{\protect\cite{Falomo91}};\tablenotemark{18}{\protect\cite{Osterbrock83}};\tablenotemark{19}{\protect\cite{Stickel89}};\tablenotemark{20}{\protect\cite{Plotkin10}};\tablenotemark{21}{\protect\cite{Nilsson03}};\tablenotemark{22}{\protect\cite{Stocke91}}; \tablenotemark{23}{\protect\cite{Cao99}}; \tablenotemark{24}{\protect\cite{Sbarufatti09}};\tablenotemark{25}{\protect\cite{Burbidge77}};\tablenotemark{26}{\protect\cite{White00}};\tablenotemark{27}{\protect\cite{Padovani95b}};\tablenotemark{28}{\protect\cite{Sandrinelli13}};\tablenotemark{29}{\protect\cite{Jannuzi93}};\tablenotemark{30}{\protect\cite{Henstock97}};\tablenotemark{31}{\protect\cite{Shaw09}}; \tablenotemark{32}{\protect\cite{Erac04}};\tablenotemark{33}{\protect\cite{Stickel88}}; \tablenotemark{34}{\protect\cite{Sbarufatti06}}; \tablenotemark{35}{\protect\cite{Smith76}};\tablenotemark{36}{\protect\cite{Falomo99}}}
  \vspace{0.2cm}
\item[b] Blazars of unknown type are indicated with a hyphen ($-$).\\ Blazar type references: \tablenotemark{37}{\protect\cite{LaurentIBL}};  \tablenotemark{38}{\protect\cite{Giommi12}}; \tablenotemark{39}{\protect\cite{Chandra12}}; \tablenotemark{40}{\protect\cite{0235Fermi}}; \tablenotemark{41}{\protect\cite{Nieppola06}};  \tablenotemark{42}{\protect\cite{FermiNLS1}}; \tablenotemark{43}{\protect\cite{Massaro12}}; \tablenotemark{44}{\protect\cite{Rani11}}; \tablenotemark{45}{\protect\cite{Osterbrock83}};  \tablenotemark{46}{\protect\cite{Impey88}}; \tablenotemark{47}{\protect\cite{Cornwell86}};  \tablenotemark{48}{\protect\cite{Ajello14}};  \tablenotemark{49}{\protect\cite{Massaro03}}; \tablenotemark{50}{\protect\cite{Padovani02}}; \tablenotemark{51}{\protect\cite{Komossa06}}; \tablenotemark{52}{\protect\cite{Li10}}; \tablenotemark{53}{\protect\cite{Drink97}}; \tablenotemark{54}{\protect\cite{Abdo10}}; \tablenotemark{55}{\protect\cite{Lister11}};\tablenotemark{56}{\protect\cite{Smith76}}
  \vspace{0.2cm}
\item[c] Source selection references: \textit{1FGL}, \protect\cite{1FGL}; \textit{2FGL}, \protect\cite{2FGL};  \textit{B97}, \protect\cite{Brinkmann97}; \textit{CG02}, \protect\cite{CG}; \textit{C07}, \protect\cite{Costamante07};   \textit{F04}, \protect\cite{Falcone04}; \textit{M01}, \protect\cite{Egret}; \textit{N06}, \protect\cite{Nieppola06};  \textit{ROXA}, \protect\cite{ROXA}; \textit{S96},  \protect\cite{SDS};   \textit{SHBL}, \protect\cite{Sedentary}; \textit{P00}, \protect\cite{Perlman00};   \textit{P02}, \protect\cite{Padovani02}; \textit{ToO}, Target of Opportunity 
\vspace{2cm}
  \end{TableNotes}
\begin{longtable*}{|c|c|c|c|c|c|c|c|c|}
\tablecaption{List of sources observed by \veritas \label{Sourcelist}} 
\tablehead{
 Source name & R.A. & Dec & $z^a$ & Type$^b$ & Exposure & $\hat{\theta}_{zenith}$ & MJDs & Selection$^c$  \\
 \hline
    & [hr min sec] & [deg min sec] &  & & [hr] & [deg] & [-50000] & 
    }
    \insertTableNotes
    \endlastfoot
\hline 
 \multirow{4}{*}{RBS 0042} &	 \multirow{4}{*}{00 18 27.8}&	 \multirow{4}{*}{+29 47 32}&	\multirow{4}{*}{0.100:\tablenotemark{1}}&	 \multirow{4}{*}{HBL}&	\multirow{4}{*}{7.1} & \multirow{4}{*}{15} & 4731/32/33/40 & \multirow{4}{*}{SHBL} \\	
 & & & & & & & 4741/42/46/73 & \\	
  & & & & & & & 5089/90/91 & \\
    & & & & & & & 5131/43 & \\		
    \hline
 \multirow{2}{*}{RBS 0082}	 & \multirow{2}{*}{00 35 14.7}&	 \multirow{2}{*}{+15 15 04}&	\multirow{2}{*}{1.28:\tablenotemark{2}}	&	\multirow{2}{*}{HBL}&	\multirow{2}{*}{6.3} & \multirow{2}{*}{19} & 5100/01/02& \multirow{2}{*}{SHBL}\\	
  & & & & & & & 5119/20/29/30 & \\	
  \hline
\multirow{9}{*}{1ES 0037+405}	 & \multirow{9}{*}{00 40 13.8}	 & \multirow{9}{*}{+40 50 05} &	 \multirow{9}{*}{-}	&	 \multirow{9}{*}{HBL	}& \multirow{9}{*}{36.0}  & \multirow{9}{*}{14} & 4767/68/71/72& \multirow{9}{*}{ToO}\\ 
  & & & & & & & 4773/89/91 & \\
      & & & & & & & 4800/02/22/29/46 & \\	
      & & & & & & & 5156/57/58/59 & \\
        & & & & & & & 5457/70/71/72 & \\
          & & & & & & & 5475/76/77/78/79 & \\
            & & & & & & & 5495/96/97/98 & \\
              & & & & & & & 5500/01/23/26 & \\
                & & & & & & & 5546/54/56 & \\
\hline
1RXS J0045.3+2127&	 00 45 19.2	 &+21 27 43&	 - &	 	 HBL&	1.2 & 22 & 5512 & N06, 1FGL\\
\hline			
\multirow{2}{*}{RGB J0110+418}&	\multirow{2}{*}{01 10 04.9}	& \multirow{2}{*}{+41 49 51}&	\multirow{2}{*}{0.096\tablenotemark{3}}&	 	\multirow{2}{*}{IBL\tablenotemark{38}}&	\multirow{2}{*}{4.0}	 & \multirow{2}{*}{13} & 4832/33 & \multirow{2}{*}{P00}\\	
  & & & & & & & 5866/67/68/96 & \\
  \hline	 
\multirow{4}{*}{1ES 0120+340}	& \multirow{4}{*}{01 23 08.7}&	 \multirow{4}{*}{+34 20 51}&	\multirow{4}{*}{0.272\tablenotemark{4}}&	 	 \multirow{4}{*}{HBL}&	\multirow{4}{*}{5.9} & \multirow{4}{*}{16} & 4383/84/93/94 & \multirow{4}{*}{SHBL, CG02} \\
  & & & & & & & 4414/37/5171 & \\
    & & & & & & & 5470/73/99 & \\
      & & & & & & & 5512/26 & \\	
      \hline
QSO 0133+476	& 01 36 58.6&	 +47 51 29&	0.859\tablenotemark{5}&		FSRQ\tablenotemark{39}&	0.8 & 40 & 4508 & ToO\\	
\hline
B2 0200+30&	 02 03 45.6&	 +30 41 30	& 0.761\tablenotemark{6} &- &	1.8 & 21& 5569/70/88 & ToO\\	
\hline
\multirow{2}{*}{CGRaBS J0211+1051}&	 \multirow{2}{*}{02 11 13.1}&	 \multirow{2}{*}{+10 51 35}&	\multirow{2}{*}{0.20:\tablenotemark{7}}	& 	\multirow{2}{*}{LBL\tablenotemark{40}}&	\multirow{2}{*}{4.0} & \multirow{2}{*}{31}& 5588/89/90/91 & \multirow{2}{*}{ToO}\\	
    & & & & & & & 5595/99 & \\
    \hline
\multirow{3}{*}{RGB J0214+517}	& \multirow{3}{*}{02 14 17.9}	& \multirow{3}{*}{+51 44 52}&	\multirow{3}{*}{0.049\tablenotemark{8}}&	 	\multirow{3}{*}{IBL\tablenotemark{41}}&	\multirow{3}{*}{5.1} & \multirow{3}{*}{22} & 4773/89/90/91/94& \multirow{3}{*}{P00, CG02}\\	
        & & & & & & & 4800/5156/81 & \\
              & & & & & & & 5201/04/5489 & \\
            \hline
RBS 0298	& 02 16 30.9&	 +23 15 13&	0.289\tablenotemark{9} &	 	 HBL&	3.1 & 14& 4731/32/33/73& SHBL\\	
\hline
RBS 0319	& 02 27 16.6&	 +02 02 00&	0.457\tablenotemark{10} &	 	 HBL	&0.3	 &31  & 4412 & SHBL\\	
\hline
AO 0235+16	& 02 38 38.9	 &+16 36 59&	0.94\tablenotemark{11}&	 	LBL\tablenotemark{41}&	4.3 & 21 & 4737/38/39/42/45& 1FGL, ToO\\	
\hline
\multirow{2}{*}{RGB J0250+172}&	 \multirow{2}{*}{02 50 38.0}&	\multirow{2}{*}{ +17 12 08}&	\multirow{2}{*}{0.243\tablenotemark{12}}&	 	\multirow{2}{*}{IBL\tablenotemark{38}}&	\multirow{2}{*}{5.1} & \multirow{2}{*}{22} & 5144/59/5472 & \multirow{2}{*}{1FGL}\\
        & & & & & & & 5528/42/71/98& \\	
 \hline
 \multirow{4}{*}{2FGL J0312.8+2013}&	 \multirow{4}{*}{03 12 23.0	}& \multirow{4}{*}{+20 07 50}&	 \multirow{4}{*}{-}&	 	 \multirow{4}{*}{-}&	\multirow{4}{*}{9.7} & \multirow{4}{*}{14} & 5209 & \multirow{4}{*}{2FGL}\\
     & & & & & & & 5830/33/34/40 & \\	
         & & & & & & & 5855/56/57/58 & \\
             & & & & & & & 5860/61/62 & \\
\hline 
\multirow{2}{*}{RGB J0314+247}&	 \multirow{2}{*}{03 14 02.7}&	 \multirow{2}{*}{+24 44 33}&	\multirow{2}{*}{0.056\tablenotemark{3}}&	 	\multirow{2}{*}{LBL\tablenotemark{42}}&	\multirow{2}{*}{3.1} &\multirow{2}{*}{28} & 4441 & \multirow{2}{*}{P00} \\
    & & & & & & & 4761/62/63/64/73 & \\
\hline	
RGB J0314+063&	 03 14 23.9&	 +06 19 57&	 	-& 	 HBL&	0.3 & 26 & 5868 & SHBL\\
\hline	
\multirow{3}{*}{RGB J0321+236}&	 \multirow{3}{*}{03 22 00.0}&	 \multirow{3}{*}{+23 36 11}&	 \multirow{3}{*}{-}&	 	\multirow{3}{*}{IBL\tablenotemark{38}}&	\multirow{3}{*}{9.2} & \multirow{3}{*}{12}& 5501/02/03/04& \multirow{3}{*}{1FGL}\\
    & & & & & & & 5507/08/10/11 & \\
        & & & & & & & 5512/13/14 & \\	
\hline
\multirow{3}{*}{B2 0321+33}	& \multirow{3}{*}{03 24 41.2}&	 \multirow{3}{*}{+34 10 46}&	\multirow{3}{*}{0.061\tablenotemark{8}}&	 & \multirow{3}{*}{9.1} & \multirow{3}{*}{12} & 4409/12/16/37& \multirow{3}{*}{P00, F04}\\
    & & & & FSRQ/ & & & 4438/39/40/47 &  \\
        & & & & \ \ NLS1\tablenotemark{43}& & & 4448/49/50/64 & \\	
 \hline	
1FGL J0333.7+2919&	 03 33 49.2&	 +29 16 32&	 -&		IBL\tablenotemark{44}&	0.6 & 3 & 5571 & 1FGL\\	
\hline
\multirow{4}{*}{1RXS J044127.8+150455}&	 \multirow{4}{*}{04 41 27.4}&	 \multirow{4}{*}{+15 04 56}&	\multirow{4}{*}{0.109\tablenotemark{13}}&	 	 \multirow{4}{*}{HBL}& 	\multirow{4}{*}{10.1} & \multirow{4}{*}{21}& 4747/48/49/89/90 & \multirow{4}{*}{SHBL} \\
    & & & & & & & 4831/32/33 & \\	
        & & & & & & & 4847/49/51/79 & \\
            & & & & & & & 4880/82/91 & \\
            \hline 
2FGL J0423.3+5612	& 04 23 27.0	& +56 12 24 &	-	& 	 - &	1.6 & 28 & 5855/5926 & 2FGL\\
\hline	
1FGL J0423.8+4148	& 04 23 56.1&	 +41 50 03&	 -	& 	 - &	1.0 & 15 & 5599 & 1FGL\\	
\hline
\multirow{4}{*}{1ES 0446+449}	& \multirow{4}{*}{04 50 07.3}&	  \multirow{4}{*}{+45 03 12}&	 \multirow{4}{*}{0.203:\tablenotemark{4}}& 	 \multirow{4}{*}{IBL\tablenotemark{38}}& 	 \multirow{4}{*}{7.1} &  \multirow{4}{*}{22}& 4734/42/61/62&  \multirow{4}{*}{S96}\\	
& & & & & & & 4763/64/65 & \\
& & & & & & & 5209/13/34/35 & \\
& & & & & & & 5838 & \\
\hline
\multirow{3}{*}{RGB J0505+612}	& \multirow{3}{*}{05 05 58.7}	& \multirow{3}{*}{+61 13 36}&	 \multirow{3}{*}{-} &	 	 \multirow{3}{*}{-} &	\multirow{3}{*}{9.2} & \multirow{3}{*}{32} & 5502/03/29/30& \multirow{3}{*}{1FGL}\\
& & & & & & & 5535/36/40/41 & \\
& & & & & & & 5543/44/57/58/72 & \\
\hline	
\multirow{2}{*}{1FGL J0515.9+1528}&	 \multirow{2}{*}{05 15 47.3} &	 \multirow{2}{*}{+15 27 17}&	 \multirow{2}{*}{-}&	  \multirow{2}{*}{-}&	\multirow{2}{*}{3.9} & \multirow{2}{*}{18} & 5480/81/82& \multirow{2}{*}{1FGL}\\
& & & & & & & 5558/59 & \\
\hline	
2FGL J0540.4+5822	& 05 40 26.0 &	 +58 22 54 &  -  & -  &	1.3 & 30 & 5856/5926/27& 2FGL\\	
\hline
RGB J0643+422 & 06 43 26.8 &	+42 14 19 &	0.080\tablenotemark{14}&	 	HBL &	1.2 & 23  & 4439/4790/4892  &  B97\\	
\hline
\multirow{2}{*}{RGB J0656+426}&	\multirow{2}{*}{06 56 10.7}&	 \multirow{2}{*}{+42 37 02}&	\multirow{2}{*}{0.061\tablenotemark{15}}&	 	\multirow{2}{*}{IBL\tablenotemark{37}}&	\multirow{2}{*}{9.4}	 & \multirow{2}{*}{16} & 4746/66/67/77 & \multirow{3}{*}{P00}\\
& & & & & & & 4778/79/90/4800 & \\
\hline	
\multirow{2}{*}{1ES 0735+178}	& \multirow{2}{*}{07 38 07.4}&	 \multirow{2}{*}{+17 42 19}&	\multirow{2}{*}{0.424\tablenotemark{16}}&	 	\multirow{2}{*}{IBL\tablenotemark{44}}&	\multirow{2}{*}{5.2}	 & \multirow{2}{*}{18} & 5531/32/58/59& \multirow{2}{*}{1FGL}\\
& & & & & & & 5574/87/5602 & \\
\hline
BZB J0809+3455&	 08 09 38.9&	 +34 55 37&	0.082\tablenotemark{8}&	 	 HBL&	1.6 & 13 & 5928/29/30& 1FGL\\	
\hline
PKS 0829+046	& 08 31 48.9 &	 +04 29 39 &	0.174\tablenotemark{17}&		IBL\tablenotemark{37}& 2.4 &30  & 4822/4921/5181 & M01\\
\hline	
\multirow{2}{*}{Mrk 1218}	& \multirow{2}{*}{08 38 10.9}&	 \multirow{2}{*}{+24 53 43}&	\multirow{2}{*}{0.028\tablenotemark{18}}&	FSRQ/&	\multirow{2}{*}{5.9} & \multirow{2}{*}{13} & 4423/25/39/40 &\multirow{2}{*}{F04}\\
& & & & Sy1\tablenotemark{45}& & & 4448/49/50/52 & \\
\hline			
\multirow{4}{*}{OJ 287}	& \multirow{4}{*}{08 54 48.9}&	 \multirow{4}{*}{+20 06 31}&	\multirow{4}{*}{0.306\tablenotemark{19}}&	 	 \multirow{4}{*}{LBL\tablenotemark{46}}&	\multirow{4}{*}{10.2} & \multirow{4}{*}{19} & 4438/39/40/48 & \multirow{3}{*}{CG02, M01}\\
& & & & & & & 4449/50/52/65/66& \\
& & & & & & & 5233/35/37/66& ToO \\
& & & & & & & 5302 & \\	
\hline
\multirow{3}{*}{B2 0912+29}&	 \multirow{3}{*}{09 15 52.4}&	 \multirow{3}{*}{+29 33 24}&	\multirow{3}{*}{0.36:\tablenotemark{7}}&	 	 \multirow{3}{*}{HBL}&	\multirow{3}{*}{11.7}	 & \multirow{3}{*}{10} & 5571/72/74/75& \multirow{3}{*}{1FGL}\\
& & & & & & & 5576/87/88/89/90& \\
& & & & & & & 5630/44/72/73 & \\
\hline	
\multirow{4}{*}{1ES 0927+500}&	 \multirow{4}{*}{09 30 37.6}&	 \multirow{4}{*}{+49 50 26}&	\multirow{4}{*}{0.188\tablenotemark{4}}&	 	 \multirow{4}{*}{HBL}&	\multirow{4}{*}{11.7} & \multirow{4}{*}{23} & 4466/4770/71 & \multirow{4}{*}{SHBL, ROXA} \\
& & & & & & & 4800/01/02/03/06 & \\
& & & & & & & 4807/20/21/22 & \\
& & & & & & & 5247/75 & \\
\hline	
RBS 0831	 &10 08 11.4&	 +47 05 22&	0.343\tablenotemark{20}&	 	 HBL&	1.6 & 18 & 5531/59/89& SHBL, ROXA\\	
\hline
RGB J1012+424	& 10 12 44.3 &	 +42 29 57 &	0.36:\tablenotemark{21}&	 	IBL\tablenotemark{37} &	1.7 & 15 & 5589/5931/86& ROXA\\	
\hline
\multirow{11}{*}{1ES 1028+511}	& \multirow{11}{*}{10 31 18.5}	& \multirow{11}{*}{+50 53 36}&	\multirow{11}{*}{0.36:\tablenotemark{10}}&	 	 \multirow{11}{*}{HBL}& 	\multirow{11}{*}{24.1} & \multirow{11}{*}{23} & 4412/13/15/4530 & \\
& & & & & & & 4828/29/30/31 & \\
& & & & & & & 4832/59/83 & \\	
& & & & & & & 4905/06/11/21 & \\
& & & & & & & 4922/23/27/28 & SHBL, CG02\\
& & & & & & & 5292/93/95/98 & ROXA\\
& & & & & & & 5301/03 &\\
& & & & & & & 5919/23/27/31/45 & \\
& & & & & & & 5946/58/70/79/82 & \\
& & & & & & & 6000/02/09/27 & \\
& & & & & & & 6035/38 & \\
\hline
RGB J1037+571	& 10 37 44.3&	 +57 11 56&	>0.62:\tablenotemark{7}&	 	 IBL\tablenotemark{37}&	3.7 & 26 & 5241/42/43/46 & 1FGL\\
\hline	
\multirow{3}{*}{RGB J1053+494}	 & \multirow{3}{*}{10 53 44.1}&	 \multirow{3}{*}{+49 29 56}&	\multirow{3}{*}{0.140\tablenotemark{22}}&	 	 \multirow{3}{*}{HBL}	& \multirow{3}{*}{7.8} & \multirow{3}{*}{24} & 4879/81/82/88/91 & \multirow{3}{*}{1FGL} \\
& & & & & & & 4921/22 & \\
& & & & & & & 5157/58/59 & \\
\hline	
RBS 0921	 &10 56 06.6&	 +02 52 14&	0.236\tablenotemark{23}&	 	 HBL	& 2.7 & 30 & 4821/22/51 &  SHBL\\	
\hline
 RBS 0929	 & 11 00 21.1	& +40 19 28 &	- &	 HBL & 4.4  &  16 &  5333/5587/88/89 & SHBL, ROXA\\
\hline	
1ES 1106+244	 &11 09 16.2&	 +24 11 20&	0.482\tablenotemark{24}&	 	 HBL&	1.0 & 17 & 5981/82& C07\\	
\hline
\multirow{3}{*}{RX J1117.1+2014}	& \multirow{3}{*}{11 17 06.3}&	 \multirow{3}{*}{+20 14 07}&	\multirow{3}{*}{0.139\tablenotemark{15}}&	 	 \multirow{3}{*}{HBL}&	\multirow{3}{*}{9.1} & \multirow{3}{*}{16} & 4940/41/42/5538 & \\ 
& & & & & & & 5540/41/42/43/44 & SHBL, CG02\\
& & & & & & & 5543/63/64/65/71 & 1FGL, ToO\\
\hline	
\multirow{2}{*}{1ES 1118+424}	& \multirow{2}{*}{11 20 48.1}	& \multirow{2}{*}{+42 12 12}&	\multirow{2}{*}{0.230\tablenotemark{12}}&	 	 \multirow{2}{*}{HBL}	& \multirow{2}{*}{6.4}	 & \multirow{2}{*}{17} & 5212/32/74/75&\multirow{2}{*}{SHBL, S96} \\
& & & & & & & 5276/89/90/91 & \\	
\hline
\multirow{2}{*}{S4 1150+497}	 & \multirow{2}{*}{11 53 24.5}	&  \multirow{2}{*}{+49 31 09} &	\multirow{2}{*}{0.334\tablenotemark{25}} &	 	 \multirow{2}{*}{FSRQ\tablenotemark{38}}&	\multirow{2}{*}{3.8} & \multirow{2}{*}{24} & 5701/02/03/04/05 & \multirow{2}{*}{ROXA, ToO}\\	
& & & & & & & 5706/07/08/09/10 & \\
\hline
\multirow{2}{*}{RGB J1231+287}&	 \multirow{2}{*}{12 31 43.6}&	 \multirow{2}{*}{+28 47 50}&	\multirow{2}{*}{1.03\tablenotemark{26}}&	 	 \multirow{2}{*}{HBL}&	\multirow{2}{*}{5.1} & \multirow{2}{*}{17} & 5239/66/91/98 & \multirow{2}{*}{1FGL}\\	
& & & & & & & 5300/03 & \\
\hline
1ES 1239+069&	 12 41 48.3&	 +06 36 01&	0.150\tablenotemark{4}&	  HBL&	1.9 & 26 & 4979/80 & S96\\	
\hline
\multirow{3}{*}{PG 1246+586}	& \multirow{3}{*}{12 48 18.8}&	 \multirow{3}{*}{+58 20 29}&	\multirow{3}{*}{>0.73:\tablenotemark{10}}&	 	 \multirow{3}{*}{IBL\tablenotemark{37}}&	\multirow{3}{*}{9.6} & \multirow{3}{*}{29} & 5595/ 5601/03/05 & \multirow{3}{*}{1FGL}\\
 & & & & & & & 5621/24/25/29 & \\
& & & & & & & 5630/31 & \\
\hline
\multirow{11}{*}{1ES 1255+244}	 & \multirow{11}{*}{12 57 31.9}&	 \multirow{11}{*}{+24 12 40}&	\multirow{11}{*}{0.141\tablenotemark{27}}&	 	 \multirow{11}{*}{HBL}&	\multirow{11}{*}{26.0}	 & \multirow{11}{*}{16} & 4531/34/68/80 & \multirow{11}{*}{SHBL, S96}\\
& & & & & & & 4581/82/83/84 & \\
& & & & & & & 4585/86/87/91 & \\
& & & & & & & 4907/11/23/50 & \\
& & & & & & & 4970/79/80 & \\
& & & & & & & 5207/08/36 & \\
& & & & & & & 5591/5617 & \\
 & & & & & & & 5947/49/53/59 & \\
& & & & & & & 5972/78/89 & \\
& & & & & & & 6016/42/44/45 & \\
& & & & & & & 6072/73/74 & \\
\hline
\multirow{3}{*}{BZB J1309+4305}&	 \multirow{3}{*}{13 09 25.5}&	 \multirow{3}{*}{+43 05 06}&	 \multirow{3}{*}{0.691\tablenotemark{6}}	 &	 \multirow{3}{*}{HBL}&	 \multirow{3}{*}{9.4} & \multirow{3}{*}{15}& 5594/96/98 & \multirow{3}{*}{1FGL}\\
& & & & & & & 5600/02/04/06 & \\	
& & & & & & & 5620/22/23/25 & \\
\hline
\multirow{4}{*}{1FGL J1323.1+2942}&	 \multirow{4}{*}{13 23 02.4}&	 \multirow{4}{*}{+29 41 35}&	 \multirow{4}{*}{-}&		 \multirow{4}{*}{FSRQ\tablenotemark{47}}&		 \multirow{4}{*}{8.4} & \multirow{4}{*}{14} & 4832/34/92/93 & \multirow{4}{*}{1FGL, ToO}\\
& & & & & & & 4909/14/79 & \\
& & & & & & & 5596/97/99 & \\
& & & & & & & 5602/05/48/49 & \\
\hline	
RX J1326.2+2933	& 13 26 15.0&	 +29 33 31&	0.431\tablenotemark{14}&	 	 HBL	&8.4	 & 14 & Same as above & ROXA, C07\\	
\hline
\multirow{2}{*}{RGB J1341+399}	& \multirow{2}{*}{13 41 05.2}&	 \multirow{2}{*}{+39 59 46}&	\multirow{2}{*}{0.169\tablenotemark{14}}&	 	 \multirow{2}{*}{HBL}	& \multirow{2}{*}{2.7} & \multirow{2}{*}{26} & 4938/78 & \multirow{2}{*}{ROXA, N06}\\
& & & & & & & 5972/6045 & \\
\hline	
\multirow{3}{*}{RGB J1351+112	}& \multirow{3}{*}{13 51 20.8}&	 \multirow{3}{*}{+11 14 53}&	 \multirow{3}{*}{>0.619\tablenotemark{28}}	&	 \multirow{3}{*}{HBL}	& \multirow{3}{*}{6.2}	 & \multirow{3}{*}{24} & 5210/21/39/40 & \multirow{3}{*}{1FGL}\\
& & & & & & & 5241/43/46 & \\
& & & & & & & 5293/97/98 & \\
\hline	
RX J1353.4+5601	& 13 53 28.1 &	 +53 00 57 &	0.370\tablenotemark{14}&	 	HBL &	 2.5 & 26 & 5589/90/6002 & ROXA, N06\\	
\hline
\multirow{2}{*}{RBS 1350}	& \multirow{2}{*}{14 06 59.2}&	 \multirow{2}{*}{+16 42 06} &	\multirow{2}{*}{>0.623\tablenotemark{13}}	&	 \multirow{2}{*}{HBL}	&\multirow{2}{*}{4.5}	 & \multirow{2}{*}{22} & 5269/97/98& \multirow{2}{*}{1FGL}\\
& & & & & & & 5300/01 & \\
\hline	
\multirow{4}{*}{RBS 1366}	 & \multirow{4}{*}{14 17 56.7}&	 \multirow{4}{*}{+25 43 56}&	\multirow{4}{*}{0.237\tablenotemark{12}}	& 	 \multirow{4}{*}{HBL}	& \multirow{4}{*}{10.0} & \multirow{4}{*}{17} & 4591/92/93/94 &\\	
& & & & & & & 4596/99 & SHBL, CG02\\
& & & & & & & 4611/12/13/15/16 & ROXA\\
& & & & & & & 6046 & \\
\hline
\multirow{2}{*}{1ES 1421+582}	 & \multirow{2}{*}{14 22 38.9} &	 \multirow{2}{*}{+58 01 56} &	\multirow{2}{*}{0.683\tablenotemark{14}}&	 	 \multirow{2}{*}{HBL}&	 \multirow{2}{*}{3.4} & \multirow{2}{*}{28} & 5324/25/26/28/30 & \multirow{2}{*}{SHBL}\\
& & & & & & & 5333/35/50/51 & \\
\hline	
RGB J1439+395 &	14 39 17.5&	+39 32 43&	0.344\tablenotemark{13}&	 HBL &	1.5 & 12 & 5620/6002 & SHBL, ROXA\\
\hline
1RXS J144053.2+061013 &	14 40 52.9 &	+06 10 16 &	0.396\tablenotemark{28}	& 	IBL\tablenotemark{48}&	 2.5 & 27 & 5731/32/34/35/36& 1FGL\\
\hline	
RBS 1452	& 15 01 01.8&	 +22 38 06&	0.235\tablenotemark{29}	& 	IBL\tablenotemark{49}&	 4.1	 & 21 & 5297/98/99 & 1FGL\\
\hline	
\multirow{2}{*}{RGB J1532+302}	& \multirow{2}{*}{15 32 02.3}&	 \multirow{2}{*}{+30 16 29}&	\multirow{2}{*}{0.065\tablenotemark{3}}&	 	 \multirow{2}{*}{HBL} &	 \multirow{2}{*}{6.5} & \multirow{2}{*}{17}& 4939/40/67/68 & \multirow{2}{*}{P00}\\
& & & & & & & 4970/75/76/77 & \\
\hline
\multirow{2}{*}{RGB J1533+189} &	 \multirow{2}{*}{15 33 11.3} &	 \multirow{2}{*}{+18 54 29}	& \multirow{2}{*}{0.305\tablenotemark{13}}&	 	 \multirow{2}{*}{HBL}	& \multirow{2}{*}{2.9} & \multirow{2}{*}{20} & 5648/77 & SHBL, ROXA\\
& & & & & & & 5705/06/20 & N06\\
\hline	
1ES 1533+535 & 15 35 00.9	& +53 20 37 &	0.89:\tablenotemark{10}	 &  HBL	& 1.0 & 23 & 4256/6002 & SHBL \\
 \hline
\multirow{2}{*}{RGB J1610+671B}	& \multirow{2}{*}{16 10 04.1} &	 \multirow{2}{*}{+67 10 26}&	\multirow{2}{*}{0.067\tablenotemark{14}}	 &	 \multirow{2}{*}{HBL}	 & \multirow{2}{*}{6.6}	 & \multirow{2}{*}{36} & 4908/38/5268 & \multirow{2}{*}{P00}\\
& & & & & & & 5292/93/94/95 & \\
\hline	
\multirow{5}{*}{1ES 1627+402}	& \multirow{5}{*}{16 29 01.3}	 & \multirow{5}{*}{+40 08 00} &	\multirow{5}{*}{0.272\tablenotemark{3}}&	 	&	 \multirow{5}{*}{13.1} & \multirow{5}{*}{16} & 4229/35/36 & \multirow{5}{*}{P02, F04}\\	
& & & & & & & 4537/38/39/40/57 & \\
& & & & HBL/& & & 4559/60/61/62/63 & \\
& & & & NLS1\tablenotemark{50,51}& & & 4564/65/69/83 & \\
& & & & & & & 4954& \\
\hline
GB6 J1700+6830&	 17 00 09.3	& +68 30 07&	0.301\tablenotemark{30}&	 	 FSRQ\tablenotemark{30}&	 0.8 & 37 & 4914/16& 1FGL, ToO \\	
\hline
\multirow{2}{*}{PKS 1717+177}	& \multirow{2}{*}{17 19 13.0}&	 \multirow{2}{*}{+17 45 06}&	\multirow{2}{*}{>0.58\tablenotemark{31}}&	 	 \multirow{2}{*}{LBL\tablenotemark{52}}&	 \multirow{2}{*}{5.1} & \multirow{2}{*}{18} & 4909/11/17/18 & \multirow{2}{*}{1FGL}\\
& & & & & & & 4920/21/22 & \\	
\hline
PKS 1725+045	& 17 28 25.0&	 +04 27 05&	0.2966\tablenotemark{32}&	 	 FSRQ\tablenotemark{53}&	 0.3 & 27 & 4412 & M01\\	
\hline
PKS 1749+096	 &17 51 32.8&	 +09 39 01&	0.32\tablenotemark{33}	 &	 LBL\tablenotemark{54}&	 0.3 & 22 & 6072 & ToO\\	
\hline
RGB J1838+480&	 18 39 49.2&	 +48 02 34&	0.30:\tablenotemark{21}&	 	IBL\tablenotemark{37}&	 0.5 & 26 & 6090 & 2FGL\\	
\hline
RGB J1903+556&	 19 03 11.6&	 +55 40 39&	>0.58:\tablenotemark{7}&	 	IBL\tablenotemark{41}&	 1.0 & 27 & 5099 & 1FGL\\	
\hline
1FGL J1926.8+6153&	 19 26 41.9&	 +61 54 41&	 -	 &	 -&	 1.3 & 31 & 5706/07 & 1FGL\\	
\hline
PKS 2233-148&	 22 36 34.1&	 -14 33 22&	>0.49:\tablenotemark{34}&	 	 LBL\tablenotemark{55}&	 0.3 & 49 & 6100 & ToO \\						
\hline
3C 454.3	& 22 53 57.7&	 +16 08 54&	 0.859\tablenotemark{35}&	  FSRQ\tablenotemark{56}&	1.0 & 24 & 5504/31 & ToO\\		
\hline
\multirow{2}{*}{RGB J2322+346}&	 \multirow{2}{*}{23 22 44.0} &	 \multirow{2}{*}{+34 36 14} &	\multirow{2}{*}{0.098\tablenotemark{3}}&	 	\multirow{2}{*}{IBL\tablenotemark{37}}&	\multirow{2}{*}{2.9} & \multirow{2}{*}{10} & 4731/36/39 & \multirow{2}{*}{P00} \\	
 & & & & & & &4745/46/47 & \\
\hline
\multirow{3}{*}{1ES 2321+419} &	 \multirow{3}{*}{23 23 52.1} 	& \multirow{3}{*}{+42 10 59} &	\multirow{3}{*}{>0.45\tablenotemark{36}}&	 	 \multirow{3}{*}{HBL}&	 \multirow{3}{*}{4.2} & \multirow{3}{*}{21} & 4773/76 & \multirow{3}{*}{S96}\\	
& & & & & & &4802/03/30/31 & \\
& & & & & & & 5091 & \\
\hline
B3 2322+396&	 23 25 17.9&	 +39 57 37&	>1.05\tablenotemark{32}&		 LBL\tablenotemark{54}&	 1.0 & 16 & 5118/30& 1FGL\\	
\hline
\multirow{2}{*}{1FGL J2329.2+3755}&	 \multirow{2}{*}{23 29 14.2} 	& \multirow{2}{*}{+37 54 15} &	\multirow{2}{*}{-}  &		 \multirow{2}{*}{-}&	 \multirow{2}{*}{3.3}  & \multirow{2}{*}{11} & 5470/71/72/75/76 & \multirow{2}{*}{1FGL}\\	
& & & & & & & 5477/78/80 & \\
\hline
1RXS J234332.5+343957&	 23 43 33.8&	 +34 40 04&	0.366\tablenotemark{13}&	 	 HBL&	 1.5 & 17 & 5912 & SHBL \\
\hline
\end{longtable*}
\end{ThreePartTable}

\begin{ThreePartTable}
\renewcommand\TPTminimum{\textwidth}
  \begin{TableNotes}
    \vspace{0.2cm}
  \item[a] If the redshift value is uncertain it is followed by a colon ($:$). Redshift references: \tablenotemark{1} {private communication from Perlman, see \citet{Falomo99}}; \tablenotemark{2} {\citet{Laurent98}}; \tablenotemark{3} {\citet{Cao99}};  \tablenotemark{4} {\citet{Bade94}};  \tablenotemark{5} {\citet{Burbidge66}}; \tablenotemark{6} {\citet{Marziani96}}; \tablenotemark{7} {\citet{Thompson90}};  \tablenotemark{8} {\citet{Landoni14}}; \tablenotemark{9} {\citet{Magic2001}}; \tablenotemark{10} {\citet{Meisner10}}
    \vspace{0.2cm}
\item[b] Selection references: see Table \ref{Sourcelist}
  \vspace{0.2cm}
\item[c] VHE detection references (see as well for the blazar subclass classification): \tablenotemark{1} {\citet{1ES0033MAGIC}}; \tablenotemark{2} {\citet{0152HESS}};  \tablenotemark{3} {\citet{Magic0847}}; \tablenotemark{4} {\citet{1136Magic}}; \tablenotemark{5a} {\citet{1222Magic}}; \tablenotemark{5b} {\citet{1222VERITAS}};   \tablenotemark{6a} {\citet{3C279Magic}};  \tablenotemark{6b} {\citet{3C279Magicbis}}; \tablenotemark{7a} {\citet{1510hess}}; \tablenotemark{7b} {\citet{1510Magic}}; \tablenotemark{8} {\citet{Magic1725}}; \tablenotemark{9} {\citet{Magic2001}};  \tablenotemark{10} {\citet{2243veritas}};    \tablenotemark{11} {\citet{2247magic}}
  \vspace{0.5cm}
 \end{TableNotes}
\begin{longtable*}{|c|c|c|c|c|c|c|c|c|c|}
\tablecaption{List of known VHE sources observed but not detected by VERITAS in 2007-2012  \label{knownVHElist}} 
\tablehead{
 Source name & R.A. & Dec & $z^a$ & Type & Exposure & $\hat{\theta}_{zenith}$ & MJDs & Selection$^b$ & VHE detection$^c$ \\
 \hline
    & [hr min sec] & [deg min sec] &  & & [hr] & [deg] & [-50000] & &
    }
     \insertTableNotes
    \endlastfoot
\hline 
\multirow{7}{*}{1ES 0033+595} &	 \multirow{7}{*}{00 35 52.6}&	 \multirow{7}{*}{+59 50 05}&	\multirow{7}{*}{0.086:$^1$}&	 \multirow{7}{*}{HBL}& 	\multirow{7}{*}{22.6}  & \multirow{7}{*}{31} &  4411/18/19/20 &  \multirow{7}{*}{CG02, P00} & \multirow{7}{*}{1}\\
& & & & & & & 4421/38/40/48 & & \\
& & & & & & & 4464/66/76 & &\\
& & & & & & & 4734/36/37/74  &  & \\
& & & & & & & 4775/76/77 & &\\
& & & & & & & 4803/04/06 & &\\
& & & & & & & 5866/67& &\\
\hline
\multirow{4}{*}{RGB J0152+017}	& \multirow{4}{*}{01 52 33.5}&	\multirow{4}{*}{+01 46 40}	& \multirow{4}{*}{0.080$^2$} &	 \multirow{4}{*}{HBL} &	\multirow{4}{*}{8.4} & \multirow{4}{*}{35} & 4421/22/37/38 & \multirow{4}{*}{CG02} & \multirow{4}{*}{2} \\
& & & & & & & 4439/40/47/48 &  &\\
& & & & & & & 4449/50/64/65/78 &  &\\
& & & & & & & 5537/5832/33/68 & & \\
\hline
\multirow{4}{*}{RGB J0847+115} 	& \multirow{4}{*}{08 47 13.0}&	 \multirow{4}{*}{+11 33 50}&	\multirow{4}{*}{0.198$^3$} &		 \multirow{4}{*}{HBL}	& \multirow{4}{*}{12.1} &  \multirow{4}{*}{25} & 4499/4505/07/08& \multirow{4}{*}{SHBL} & \multirow{4}{*}{3}\\
& & & & & & & 4522/23/24/25/26 & &\\
& & & & & & & 5303 & &\\
& & & & & & & 5502/03/31/59 & &\\
\hline
\multirow{4}{*}{RX J1136.5+6737}	& \multirow{4}{*}{11 36 30.1} &	 \multirow{4}{*}{+67 37 04}&	\multirow{4}{*}{0.134$^4$} &	 	 \multirow{4}{*}{HBL} &	 \multirow{4}{*}{7.7} & \multirow{4}{*}{37} & 4860/61/91/92 &  & \multirow{4}{*}{4}\\
& & & & & & & 4918/21 & CG02, SHBL & \\
& & & & & & & 5292/94/99 & ROXA & \\
& & & & & & & 5303 & & \\
\hline
\multirow{6}{*}{PKS 1222+216} & \multirow{6}{*}{12 24 54.4} &	\multirow{6}{*}{+21 22 47} &	\multirow{6}{*}{0.432$^5$} & \multirow{6}{*}{FSRQ} & \multirow{6}{*}{25.8} & \multirow{6}{*}{16} & 4939/5182& \multirow{6}{*}{ToO} & \multirow{6}{*}{5a, 5b} \\
& & & & & & & 5318/19/20/21 & &\\
& & & & & & & 5322/24/25/26 & &\\
& & & & & & & 5327/28/30/33 & &\\
& & & & & & & 5622/24/25/26 & &\\
& & & & & & & 5631/33/34 &  &\\
\hline
\multirow{5}{*}{3C 279} & \multirow{5}{*}{12 56 11.1} &	\multirow{5}{*}{-05 47 22} &	\multirow{5}{*}{0.536$^6$} & \multirow{5}{*}{FSRQ} & \multirow{5}{*}{8.3} & \multirow{5}{*}{40} & 5623 & \multirow{5}{*}{ToO} & \multirow{5}{*}{6a, 6b}\\
& & & & & & & 5707/08/09/10 & & \\
& & & & & & & 5711/15/17 & & \\
& & & & & & & 5923/24/25/26/27 & & \\
& & & & & & & 6016 & & \\
\hline
\multirow{4}{*}{PKS 1510-089}&	 \multirow{4}{*}{15 12 52.2}&	\multirow{4}{*}{-09 06 22}&	\multirow{4}{*}{0.361$^7$}&	 	\multirow{4}{*}{FSRQ}&	\multirow{4}{*}{14.9} & \multirow{4}{*}{42} & 4909/11/48  & \multirow{4}{*}{ToO} & \multirow{4}{*}{7a, 7b}\\
& & & & & & & 5976/77/78/79 & & \\
 & & & & & & & 5980/81/82/83 & & \\
& & & & & & & 5984/87 &  &\\	
\hline
\multirow{3}{*}{RGB J1725+118}&	 \multirow{3}{*}{17 25 04.3}&	 \multirow{3}{*}{+11 52 16} &	\multirow{3}{*}{>0.35:$^8$}&	 	 \multirow{3}{*}{HBL}	& \multirow{3}{*}{10.0} & \multirow{3}{*}{23} & 4593/94 & \multirow{3}{*}{CG02} & \multirow{3}{*}{8}\\
& & & & & & & 4615/16/17/18 & & \\
& & & & & & & 4619/20/21/22 & &\\	
\hline
\multirow{2}{*}{0FGL J2001.0+4352}&	 \multirow{2}{*}{20 01 13.5}&	\multirow{2}{*}{+43 53 03}&	\multirow{2}{*}{0.18:$^9$}&		\multirow{2}{*}{ HBL}&	 \multirow{2}{*}{4.9} & \multirow{2}{*}{25} & 5143/44/46/51/52 &  \multirow{2}{*}{1FGL} & \multirow{2}{*}{9}\\
& & & & & & & 5326/52/57 & & \\
\hline	
\multirow{2}{*}{RGB J2243+203}	& \multirow{2}{*}{22 43 54.7}&	 \multirow{2}{*}{+20 21 04} &	\multirow{2}{*}{>0.39:$^{10}$} &		\multirow{2}{*}{IBL}&	 \multirow{2}{*}{4.1}	 & \multirow{2}{*}{17} & 5094/98/99 & \multirow{2}{*}{1FGL}  & \multirow{2}{*}{10}\\
& & & & & & & 5101/16/28/29 & & \\
\hline
\multirow{2}{*}{B3 2247+381}&	 \multirow{2}{*}{22 50 06.6} &	\multirow{2}{*}{+38 25 58} &	\multirow{2}{*}{0.118$^{2}$} &	 \multirow{2}{*}{HBL} &\multirow{2}{*}{6.0}  & \multirow{2}{*}{13} &  5092/93/95/96/97& \multirow{2}{*}{1FGL} & \multirow{2}{*}{11}\\
& & & & & & & 5832/89 &  &\\
\hline
\end{longtable*}
\end{ThreePartTable}

\begin{ThreePartTable}
\renewcommand\TPTminimum{\textwidth}
  \begin{TableNotes}
    \vspace{0.2cm}
  \item[a] Coordinates are provided for the counterpart. If the \fermilat\ source is not associated with any lower-energy blazar, coordinates from the 2FGL catalog are given instead.
    \vspace{0.2cm}
\item[b] Unconstrained redshifts are indicated with a hyphen ($-$). \\ Redshift references:  \tablenotemark{1} {\citet{Shaw12}}; \tablenotemark{2} {\citet{Shaw13}}; \tablenotemark{3} {\citet{Kraus75}};  \tablenotemark{4} {\citet{Halpern86}}; \tablenotemark{5} {\citet{Plotkin10}}; \tablenotemark{6} {\citet{Healey08}}; \tablenotemark{7} {\citet{Hewitt93}};  \tablenotemark{8} {\citet{White88}};  \tablenotemark{9} {\citet{Glikman07}};  \tablenotemark{10} {\citet{Afanas05}}; \tablenotemark{11} {\citet{Sowards05}}.
  \vspace{0.2cm}
\item[c] Blazars of unknown type are indicated with a hyphen ($-$). \\ Blazar type references: \tablenotemark{12} {\citet{Shaw12}}; \tablenotemark{13} {\citet{Kraus75}}; \tablenotemark{14} {\citet{Ajello14}}; \tablenotemark{15} {\citet{Ghisellini11}}; \tablenotemark{16} {\citet{LaurentIBL}};  \tablenotemark{17} {\citet{Hewitt93}}; \tablenotemark{18} {\citet{Plotkin10}}; \tablenotemark{19} {\citet{Glikman07}}; \tablenotemark{20} {\citet{Afanas05}}; \tablenotemark{21} {\citet{Maselli10}}.
  \vspace{0.2cm}
\item[d] See Tables 1 and 2 for information on the exposure and the zenith angle of the observations. \vspace{0.5cm}
  \vspace{0.2cm}
\end{TableNotes}
\begin{longtable*}{|c|c|c|c|c|c|c|}
\tablecolumns{9} 
\tablewidth{0pc} 
\tablecaption{List of 2FGL sources in the \veritas\ field of view of sources listed in Tables 1 and 2\label{2FGLlist}} 
\tablehead{
 Source name & Counterpart & R.A.$^a$ & Dec$^a$ & $z^b$ & Type$^c$ & in FoV of$^d$  \\
 \hline
    & & [hr min sec] & [deg min sec] &  & &  
    }
     \insertTableNotes
    \endlastfoot
\hline 
2FGL J0047.9+2232	&      BWE 0045+2218 &	00 48 02.5&	+22 34 53&	 1.161$^{1}$&	 FSRQ$^{12}$&	RGB J0045+214	\\
2FGL J0148.6+0127	&	PMN J0148+0129  &	01 48 33.8&	+01 29 01&	 0.940$^{2}$&	 -&	RGB J0152+017	\\
2FGL J0158.4+0107	&	 -&	 01 58 25.4&	 +01 07 31&	 -&	 -&	RGB J0152+017	\\	
2FGL J0205.4+3211	&	1Jy 0202+319	&02 05 04.9&	+32 12 30 &	1.466$^{3}$&	FSRQ$^{13}$&	B2 0200+30	\\	
2FGL J0212.1+5318	&	 -	& 02 12 09.4&	 +53 18 19 &	 -	& -&	 RGB J0214+517	\\
2FGL J0213.1+2245	&	1RXS J021252.2+224510 &	02 12 52.8&	+22 44 52&	 0.459$^{2}$	&HBL$^{14}$&	RBS 0298	\\	
2FGL J0326.1+2226	&	TXS 0322+222 &	03 25 36.8	&+22 24 00&	2.06$^{4}$&	FSRQ$^{15}$ & RGB J0321+236			\\	
2FGL J0440.4+1433	&	TXS 0437+145 &	04 40 21.1	&+14 37 57	& -	& - & 1RXS J044127.8+150455			\\	
2FGL J0856.3+2058	&	TXS 0853+211 &	08 56 39.7&	+20 57 43	&>0.388$^{5}$&	 -	& OJ 287	\\		
2FGL J0929.5+5009	&	QSO J0929+5013 &	09 29 15.4 &	+50 13 36&	0.370$^{6}$&	 IBL$^{16}$	& 1ES 0927+500 \\
2FGL J1058.4+0133	&	4C 01.28	& 10 58 29.6 &	+01 33 58&	0.888$^{7}$&	 FSRQ$^{13}$				& RBS 0921\\
2FGL J1059.0+0222	&	PMN J1059+0225 &	10 59 06.0 &	+02 25 12&	 -&	 -			& RBS 0921\\
2FGL J1141.0+6803	&	1RXS J114118.3+680433 &	11 41 18.0	 & +68 04 33&	 -&	 -			& RX J1136.5+6737\\
2FGL J1239.5+0728	&	PKS 1236+077	 & 12 38 24.6&	+07 30 17&	0.400$^{8}$&	FSRQ$^{17}$		& 1ES 1239+069\\	
2FGL J1245.1+5708	&	GB6 J1245+5710  &	12 45 10.0&	+57 09 54&	>0.521$^{5}$&	 LBL$^{18}$			& PG 1246+586\\	
2FGL J1303.1+2435	&	VIPS J13030+2433	&13 03 03.2&	+24 33 56&	0.993$^{9}$&	 LBL$^{19}$				& 1ES 1255+244\\
2FGL J1359.4+5541	&	VIPS J13590+5544	& 13 59 05.7&	+55 44 29&	1.014$^{1}$&	 FSRQ$^{12}$			& RX J1353.4+5601\\
2FGL J1722.7+1013	&	TXS 1720+102&	17 22 44.6&	+10 13 36&	0.732$^{10}$&	FSRQ$^{20}$&	RGB J1725+118		\\	
2FGL J1727.9+1220	&      PKS 1725+123&	17 28 07.1&	+12 15 39&	0.583$^{11}$&	FSRQ$^{20}$&	RGB J1725+118\\
2FGL J1927.5+6117	&	S4 1926+611&	19 27 30.4&	+61 17 33&	 -&	LBL$^{21}$&	1FGL J1926.8+6153	\\
2FGL J1959.9+4212 &      1RXS J195956.1+421339& 19 59 56.1 & +42 13 39& - & - & 0FGL J2001.0+4352	\\
\hline
\end{longtable*}
\end{ThreePartTable}

The results presented in this paper have been obtained using a set of $\gamma$-hadron separation cuts specifically optimized for the detection of soft spectrum sources (differential spectrum parametrized by a power-law function $dN/dE=N(E/E_0)^{-\Gamma}$ with $\Gamma=3.5$). The spectral index assumed is in line with the typical value of $\Gamma$ observed for VHE blazars \citep[see][]{Gunes13}.\\

All the observations presented in this paper were made using the \textit{`wobble'} observing strategy \citep{Fomin94}. Here, the telescopes point 0.5$^\circ$ away from the target, alternatively in each of the four cardinal directions, to enable background estimation from the same field of view. This procedure ensures a similar acceptance for both the source (ON) and the background (OFF) regions. Regions overlapping bright stars are excluded from background estimates. The ratio of the ON over the OFF region size defines the background normalization parameter $\alpha$. The dead time of the telescope array is explicitly calculated and is approximately 10$\%$ for the observations described here. The exposure values provided in Table 1 are all corrected for dead time, and represent the effective live-time of \veritas\ observations.\\

The \veritas\ observations here have an average length of twenty minutes (referred to as a \textit{run}), before switching targets or wobble directions. For quality assurance all the runs with a length lower than ten minutes were excluded, often being associated with technical problems, resulting in the early termination of the run. Additionally, all observations characterized by non-optimal weather conditions or malfunctioning hardware were excluded from the run selection. On certain occasions one of the \veritas\ telescopes can be non-operational due to technical problems; all the runs analyzed in this paper have at least three telescopes in operation. Runs with all four telescopes in operation represent the bulk ($92\%$) of the data.\\

In the standard configuration, \veritas\ observations are not performed under bright-moonlight conditions (Moon illumination $> 35\%$ of full Moon). Since 2012, the \veritas\ collaboration has started a new observing program in order to extend the duty cycle of the observatory and perform observations also under bright moonlight \citep{1727Veritas}. None of the data presented in this paper were taken under bright moonlight conditions. Observations performed under moderate moonlight (Moon illumination $< 35\%$) are included, and analyzed in the same manner as dark-time observations, with appropriate instrument response functions to account for the increased night-sky background.\\

The significance at the source location is computed using Equation 17 in \citet{LiMaa}. The upper limit on the VHE flux is estimated according to \citet{Rolke05} at the $99 \%$ confidence level. It is first calculated assuming three different values of the spectral index ($\Gamma=2.5$, $3.5$ and $4.5$) in order to estimate the decorrelation energy $E_{dec}$ (the energy at which the upper limit estimate depends the least on the spectral index). The upper limit is then recomputed at the reference energy $E_{dec}$ assuming a spectral index $\Gamma=3.5$.  The threshold of the analysis (which depends mainly on the zenith angle of the observations) is also calculated. For every source we verified that, not only the overall significance is lower than $5$ standard deviations ($\sigma$), but that no flares have been detected, i.e. that none of the sources was detected at more than $4 \sigma$ during any single run.\\

For sources which are detected by \fermilat\ (76\% of the sample), the flux is extrapolated into the \veritas\ energy band, taking into account the absorption from the EBL using the model by \citet{Franceschini08}, which is in agreement with the most recent observational constraints \citep{HESSEBL}. The extrapolated flux is then compared to the \veritas\ upper limit. If the \veritas\ measurement is lower than the extrapolation, it means that an additional cut-off should be present in the $\gamma$-ray component between the \fermilat\ and the \veritas\ energy bands. \\

 The results of the analysis are reported in Tables \ref{ULtable} and  \ref{knownVHEtable} (for known VHE sources) and  \ref{2FGLtable} (for 2FGL sources in the \veritas\ field of view). For every target, we provide the significance, the number of ON and OFF counts, the value of the $\alpha$ parameter (ratio of the ON over OFF region size), the threshold of the analysis $E_{th}$, the decorrelation energy $E_{dec}$, the differential flux upper limit at $E_{dec}$, the integral flux upper limit \citep[above $E_{th}$, provided in Crab units, following][]{WhippleCrab}\footnote{The best-fit of the VHE emission from the Crab Nebula as measured with the Whipple 10-m telescope and presented in \citet{WhippleCrab} is a power-law function with index $\Gamma=2.49$ and normalization $K= 3.2\times10^{-11}$ cm$^{\textrm{-2}}$s$^{\textrm{-1}}$TeV$^{\textrm{-1}}$ at 1 TeV. The integral upper limits are computed from the differential ones and provided here as a reference. They can easily be recomputed for different values of $E_{th}$, or for other definitions of the Crab unit. For example, using as a reference the \magic\ spectrum of the Crab nebula \citep{Crabmagic}, the Crab unit above $200$ GeV is 74$\%$ of the Whipple Crab unit above the same threshold.}, the ratio between the \veritas\ differential upper limit and the extrapolation of the \fermilat\ detection ($\Phi_{HE}$, evaluated at $E_{dec}$). Note that the values of the decorrelation and threshold energies are provided with three decimal values to ease any extrapolation to other energy bands, but they are known only to the second decimal value.\\

All the results presented in this paper have been cross-checked using a separate analysis, which provided consistent results for the single upper limit values, significance distributions (Section \ref{sigdist}) and stacked analysis (Section \ref{section6}).\\

\subsection{Notes on individual sources}
\label{section5}
Among the blazars targeted by \veritas\ between 2007 and 2012, eleven of them were later identified as VHE emitters. They are listed in Table \ref{knownVHElist}. The \veritas\ upper limits are useful in these cases to constrain the properties of the VHE emission during low-flux states, as well as the variability properties of the source. The discussion of these blazar observations, in order of R.A., follows:
\begin{itemize}
\item 1ES 0033+595 (HBL, $z=0.086:$)\\
VHE emission from 1ES 0033+595 was discovered by \magic\ \citep{1ES0033MAGIC}. The flux, measured from 24 hours of observations taken from August to October 2009, is $0.9\%$ Crab above 290 GeV\footnote{In order to ease the comparison between the results from different instruments, the integral fluxes provided in this section have been recomputed above the \veritas\ threshold, when the spectral information is available, and are expressed in Crab units as defined in \citet{WhippleCrab}.}. The observed spectral index during the \magic\ observations is $3.8\pm0.7$. The \veritas\ upper limit ($5.4\%$ Crab above 290 GeV) is fully consistent with the \magic\ measurement. The VHE variability of the source has been demonstrated by more recent \veritas\ observations, which detected a bright VHE flare (with integral flux higher than 10\% Crab) from the source during September 2013, with simultaneous X-ray and ultraviolet coverage by \textit{Swift} \citep{wystan}. A paper presenting the results of this multi-wavelength campaign is currently in preparation.
\item RGB J0152+017 (HBL, $z=0.08$)\\
This source has been known as a VHE emitter since 2008 \citep{0152HESS}, when it was detected by \hess\ at a flux of $2.6\%$ Crab above 240 GeV, during $15$ hours of observation. \veritas\ observed the blazar in three different seasons: 2007-2008 (covering the \hess\ period), 2010-2011 and 2011-2012. The \veritas\ upper limit ($3.6\%$ Crab above 240 GeV) is fully consistent with the \hess\ detection. 
\item RGB J0847+115 (HBL, $z=0.198$)\\
VHE emission from this blazar was announced by \magic\ in 2014, at a flux corresponding to $2.5\%$ Crab above 200 GeV \citep{Magic0847}. No spectral information is currently available, but the \magic\ collaboration reported a preliminary classification of the source as an extreme-HBL, with synchrotron peak-frequency in hard-X-rays, and inverse-Compton peak-frequency at TeV energies. Evidence of VHE and optical variability was also claimed by \magic. The \veritas\ upper limit ($2.0\%$ of the Crab Nebula flux above 180 GeV) is marginally consistent with the preliminary flux estimate by \magic, and could be related to variability of the VHE emission.
\item RX J1136.5+6737 (HBL, $z=0.134$)\\
The \magic\ collaboration recently reported the detection of this source at a flux of $1.5\%$ Crab above 200 GeV, in $20$ hours of observations between January and April 2014 \citep{1136Magic}. No spectral information is available at the present time. The \veritas\ upper limit ($5.2\%$ Crab above 290 GeV) is fully consistent with the preliminary flux estimate by \magic.
\item PKS 1222+216 (FSRQ, $z=0.432$)\\
VHE emission from this FSRQ was detected by \magic\ in 2010 at a flux of the order of the Crab Nebula flux \citep{1222Magic}. The detection of this VHE flare is of paramount importance for blazar physics: the rapid variability, and the fact that VHE photons can escape the bright photon field present in FSRQs was used to put constraints on the location of the $\gamma$-ray emitting region in blazars. The \veritas\ non-detection constrains the low-flux state at a level of $2.2\%$ Crab above 180 GeV.\\ During May 2014, PKS 1222+216 underwent another $\gamma$-ray flare, and \veritas\ detected VHE emission at a flux of $3\%$ Crab \citep{1222VERITAS}. A paper describing the \veritas\ detection in 2014 is currently in preparation.
\item 3C 279 (FSRQ, $z=0.536$)\\
This quasar is the first of its class detected as a VHE emitter \citep{3C279Magic}. \veritas\ observations during 2011 were triggered by flaring activity observed at lower wavelengths (optical, X-rays and HE $\gamma$-rays). The same flare triggered observations with the \magic\ telescopes \citep{3C279Magicbis}, which resulted as well in no VHE detection (flux upper limit equal to $1.7\%$ Crab above 260 GeV). The \veritas\ flux upper limit ($2.1\%$ Crab above 260 GeV) is similar to the one measured with the \magic\ telescopes.
\item PKS 1510-089 (FSRQ, $z=0.361$)\\
Two VHE flares from this quasar have been reported so far: the first during March-April 2009, seen by \hess\ \citep[around 0.6\% of the Crab Nebula flux above 260 GeV, see][]{1510hess}, the second during February 2012, seen by \magic\ \citep[around 1\% of the Crab Nebula flux above 260 GeV, see][]{1510Magic}. The \veritas\ observations presented in this paper are quasi-simultaneous with both flares (see Table \ref{knownVHElist} for details). For the 2009 flare, \veritas\ observations were taken a few days before the VHE flare seen by \hess. For the 2012 flare, \veritas\ observations were taken every night from February 19 to February 27, covering the \fermilat\ flare. The \veritas\ upper limit is 2.9\% of the Crab Nebula flux above 260 GeV. 
\item RGB J1725+118 (HBL, $z > 0.35$)\\
The discovery of VHE emission from this blazar was recently reported by \magic\ \citep{Magic1725} at a flux of $2 \%$ Crab above 140 GeV during observations in May 2013 triggered by an elevated optical state. No spectral information is available at the present time. The \veritas\ upper limit ($3.2\%$ Crab above 200 GeV)  is consistent with the \magic\ measurement.
\item 0FGL J2001.0+4352 (HBL, $z = 0.18:$)\\
Early results from \fermilat\ indicated that this source was a good candidate for IACTs, in particular due to its hard GeV spectrum \citep{Fermi0FGL}. The \magic\ collaboration detected the source at a flux of $\sim 10\%$ Crab during a single night \citep[$1.4$ hours on July 16, 2010, see][]{Magic2001}. The \veritas\ upper limit clearly indicates that this blazar is variable at VHE, and that its baseline flux is below $5.2\%$ Crab above 200 GeV. 
\item RGB J2243+203 (HBL, $z>0.39$)\\
VHE emission from this source was detected by \veritas\ during December 2014, following a trigger from a high \fermilat\ flux \citep{2243veritas, 2243icrc}. Preliminary analysis indicates that the flux from the flaring blazar was at $\sim 4\%$ Crab above 180 GeV. The upper limits computed from 2009 observations indicate that the VHE emission from this source is variable, being significantly lower (2.1\% Crab above 170 GeV) than the 2014 detection. 
\item B3 2247+381 (HBL, $z=0.119$)\\
This source was detected by \magic\ in 14 hours of observations from September to October 2010, at a flux of $2.2\%$ Crab above 170 GeV \citep{2247magic}. The \magic\ observations were triggered by a high optical state, and there is evidence of variability in the simultaneous X-ray light curve. \veritas\ observations do not cover the \magic\ detection, nor the high optical flux state measured by the \textit{Tuorla} observatory. The non-detection by \veritas\ (flux upper limit equal to $1.8\%$ Crab above 170 GeV) constrains the low-state flux of this blazar to be lower than the \magic\ detection, suggesting that it may have been related to a VHE high-flux state.\\
\end{itemize}

In addition to these known VHE emitters, we discuss a few other targets with noteworthy histories:
\begin{itemize}
\item 1ES 0037+405 (HBL, $z$ unknown)\\
\veritas\ observations of this target were taken as a self-triggered ToO. During observations of the Andromeda galaxy \citep[M31, see][]{Bird15}, a 4$\sigma$ hotspot coincident with this blazar was observed in the reconstructed sky-map. However, further observations did not confirm the hotspot, and the cumulative significance is 1.5$\sigma$ in 36 hours.

\item OJ 287 (LBL, $z=0.306$)\\ 
This blazar is one of the most studied objects of its kind due to a clear periodicity in its optical lightcurve, with a period of about twelve years.  The \veritas\ observations presented in this work cover the last active phase in Fall 2007 (from December 4, 2007 to January 1, 2008), with additional observations during 2010. The VHE upper limits are comparable to the ones measured with the \magic\ telescopes and presented by \citet{OJ287}.
 
\item 1FGL J1323.1+2942 (FSRQ, $z$ unknown) and\\ RX~J1326.2+2933 (HBL, $z=0.431$)\\
Although the angular distance between the two sources is only 43$^\prime$, they are not the same blazar. \veritas\ can resolve the two objects, and they have been targeted by \veritas\ independently (see last column of Table \ref{Sourcelist}). Since they are well within the \veritas\ FoV, the exposures on these two objects have been merged into a single dataset. 

\item B2 0912+29 (HBL, $z=0.36$)\\
 This blazar shows the highest significance in our dataset ($3.5 \sigma$ in 11.7 hours). This excess was confirmed at the same significance level by the cross-check analysis chain. Further observations were taken during the 2013 and 2014 observing seasons, but the initial excess did not increase. While VHE blazars are known to be variable, and one could interpret the lack of a detection in 2013-2014 as due to variability, we also note that the probability of a $3.5 \sigma$ excess reduces to only $2.0 \sigma$ when 103 trials (the sources from Tables \ref{Sourcelist} and \ref{2FGLlist}) are taken into account, and it is thus not enough to make any claim. \\
   
\end{itemize}
  
\subsection{Significance distributions}
\label{sigdist}
\begin{figure*}[t!]
\begin{center}
\includegraphics[width=217pt]{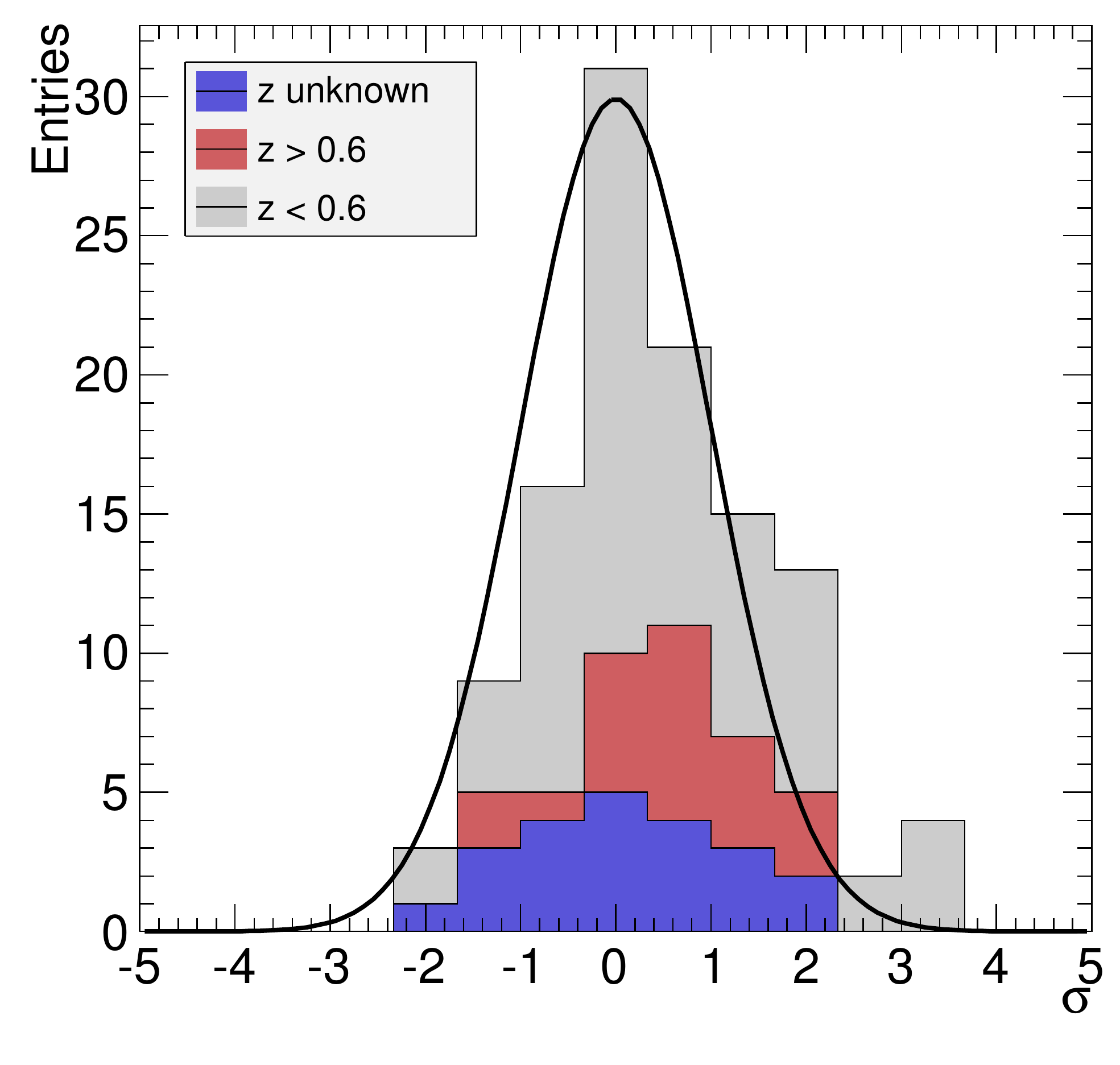}
\includegraphics[width=217pt]{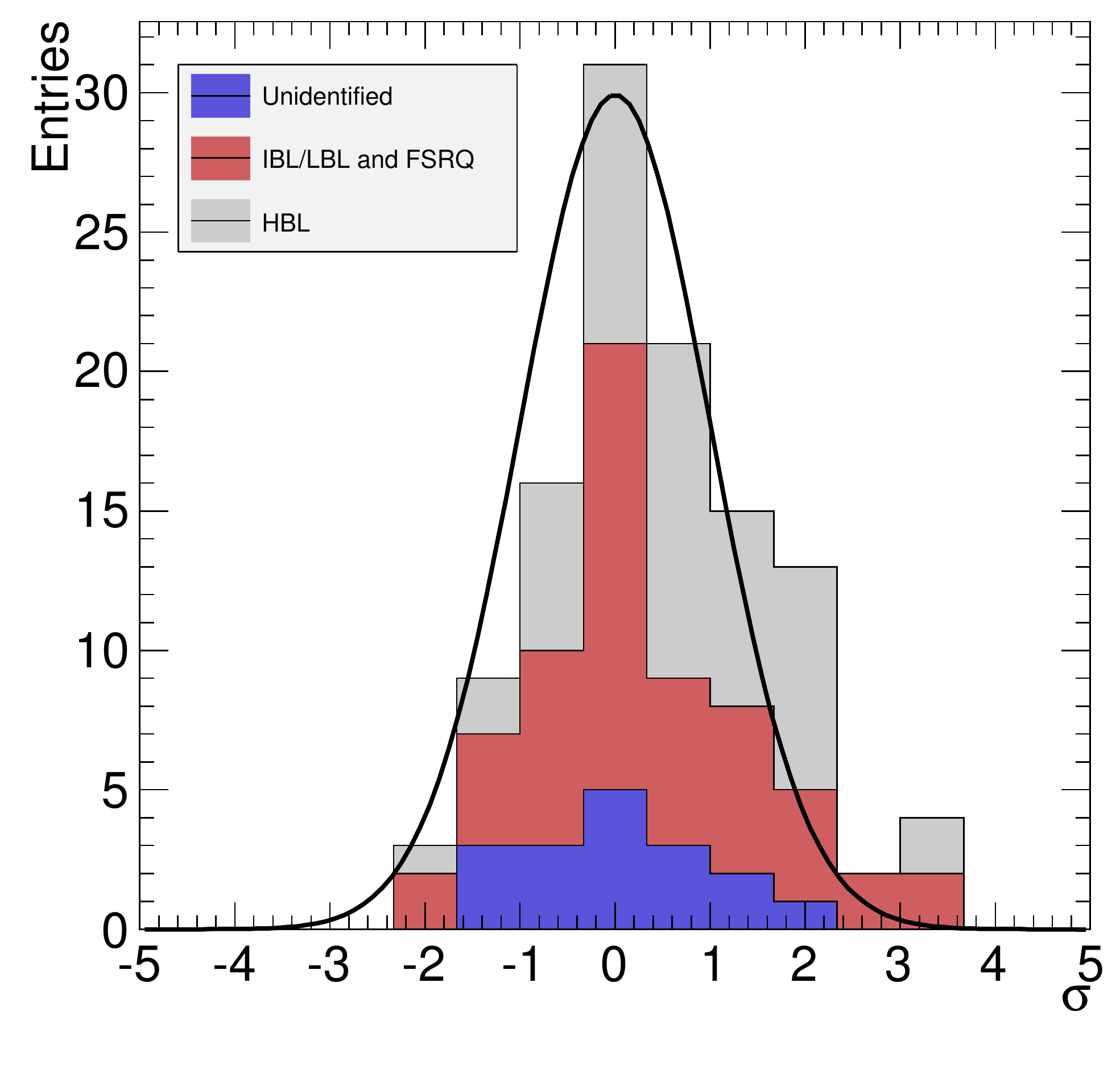}

\caption{\textit{Left}: stacked significance distribution of the sources included in our sample, classified according to their redshift. Sources with unknown z are in blue, sources with $z>0.6$ are in red and sources with $z<0.6$ are in grey. The Gaussian function represents the expectation from a randomly distributed sample, with mean equal to zero, and variance equal to 1. \textit{Right}: same as left panel, but for sources classified according to the AGN type. Unidentified sources are in blue, IBL/LBL/FSRQ in red and HBL in grey.} \label{figone}
\end{center}
\end{figure*}

In Fig.\ \ref{figone} we present the distribution of the significances for all the sources presented in our work. Given that the blazar population at VHE is not homogeneous (see the Introduction), and depends on both the blazar sub-class (which is correlated with the energy of the high-energy SED peak) and the blazar redshift (which implies a different level of EBL absorption), significance distributions are produced as a function of these two parameters. For the redshift division (left plot of Fig.\ \ref{figone}), redshifts lower or higher than 0.6 were considered, along with unknown redshift. Concerning the division of blazar sub-classes (right plot of Fig.\ \ref{figone}), the sources were categorized as HBLs, IBLs/LBLs/FSRQs, and blazars of unknown type. 
 The Gaussian distribution expected from a sample with average $\hat{X}=0$ and $\sigma=1$ is overlaid on the significance distribution. A fit of the histogram with a Gaussian function instead yields $\hat{X}
=0.3 \pm 0.1$ and $\sigma=1.2 \pm 0.1$.\\

 \newpage
 \vspace{1cm}
\LongTables
\begin{deluxetable*}{|c||c|c|c|c||c| c |c | c|c|}
\tablecaption{Analysis results and flux upper limits for the non-detected AGN observed by \veritas  \label{ULtable}} 
\tablehead{
 Source name & $\sigma$ & ON & OFF & $\alpha$ & E$_{\rm{th}}$ & E$_{\rm{dec}}$ & UL$_{@E_{\rm{dec}}}$ &   UL$_{>E_{\rm{th}}}$  & UL/ $\Phi_{HE}$ \\
  & &  & & &   \scriptsize{[TeV]} & \scriptsize{[TeV]} & \scriptsize{[10$^{-12}$\ cm$^{\textrm{-2}}$s$^{\textrm{-1}}$TeV$^{\textrm{-1}}$]} & \scriptsize{[\% C.U.]} &  
}
\startdata
\hline
 RBS 0042 &	0.02& 1239 & 6455 & 0.192 & 	  0.182& 0.345 & \phn 8.7& 2.2& 0.2 \footnotesize{(z=0.1)}\\		
 RBS 0082	 &  0.16 & 1680 & 9062 & 0.185 & 	 0.166& 0.264 & 16.9& 1.8 & 45.0 \\		
1ES 0037+405	 &	 1.50 & 7515 & 64132 & 0.115 &  0.166 & 0.322 & 29.3	& 6.3 & \nodata\\ 	
1RXS J0045.3+2127&	  2.04 & 224 & 1116 & 0.172 & 0.166 & 0.297 &	 54.6& 8.9 & 3.0/14.8 \footnotesize{(z=0.1/0.5)} \\			
RGB J0110+418&	 -0.02 & 801 & 4810 & 0.167 & 0.182& 0.299 &	 22.4		 & 3.4 & \nodata \\		 
1ES 0120+340	&	 1.47 & 1174 & 5215 & 0.214 & 0.166& 0.283 & 24.6		& 3.4 & 1.4\\	
QSO 0133+476	&	 1.24 & 114 & 496& 0.202& 0.417& 0.728 &	 11.8	 		& 17.4 \phn &1.9e4 \\	
B2 0200+30&	 	 1.38 & 495 & 2775 & 0.167 & 0.151& 0.273 & 	 51.9	 			 & 6.9 & 59.6\\	
CGRaBS J0211+1051&	 1.01 & 977 & 5659 & 0.167 & 0.200& 0.318 &	 20.8	 	 & 3.6 & 3.4\\	
RGB J0214+517	&  	 0.33 & 1113& 5877 & 0.187 & 0.182& 0.336 &	 15.9	 	& 3.7 & \nodata\\	
RBS 0298	& 	 1.76 & 606 & 3245 & 0.173 &	0.240& 0.435 & 21.5	 	& 9.2 & \nodata \\	
RBS 0319	&	 -0.52 & 61 & 393 & 0.167 &	 0.219&  0.351 & 38.4	 	& 8.6 & 44.9 \\	
AO 0235+16	&	0.63 & 704 & 4116 & 0.167 & 0.182& 0.311 &	 18.7	 	&  3.2 & 9.6\\	
RGB J0250+172&			 -0.06 & 1274 & 5637 & 0.226 & 0.166& 0.316 &	 13.5	 	& 2.7 & 1.4\\	
 2FGL J0312.8+2013&	 		 -0.36 & 2124 & 9046 & 0.238 & 0.166& 0.257 &	 10.0		& 1.0 & 0.5/0.6 \footnotesize{(z=0.1/0.5)} \\	
RGB J0314+247&	 			 1.07 & 691 & 3967 & 0.167 & 0.240& 0.460 &	 16.4	 	& 8.5 & \nodata \\	
RGB J0314+063&		 1.18  & 76 & 326 &  0.200 & 0.182& 0.294 &	 76.7						& 11.0 \phn & \nodata \\	
RGB J0321+236&		 1.24 & 3065 & 17948 & 0.167 & 0.138&0.233 & 	 31.2	 	& 2.6 & 2.0/2.9 \footnotesize{(z=0.1/0.5)} \\	
B2 0321+33 &	 	 -0.03 & 2190 & 8223 & 0.267 & 0.166& 0.272 &	 \phn 9.7	& 1.2 & 15.9\\		
1FGL J0333.7+2919&	  	 0.37 & 158 & 606 & 0.223 & 0.138 & 0.226  &	 101.0 \phn		& 7.6 & 2.7/7.4 \footnotesize{(z=0.1/0.5)}\\	
1RXS J044127.8+150455&	 	1.83 & 2351 & 10636& 0.212& 0.182 & 0.338 &	 17.5	 	& 4.1 & \nodata \\	 
2FGL J0423.3+5612	&  		 0.67 & 283 & 1522 & 0.178 & 0.240  & 0.457 &	 16.4	& 8.3 & 1.9/32.6 \footnotesize{(z=0.1/0.5)}\\	
1FGL J0423.8+4148	&		 -0.22 & 240 & 1462 & 0.167 & 0.166 & 0.274 &	 46.4			& 5.7 & 0.6/2.2 \footnotesize{(z=0.1/0.5)}\\	
1ES 0446+449	& -1.49 & 1482 &  10866 & 0.142 & 	0.219 & 0.363& \phn 4.7 & 1.2 & \nodata \\	
RGB J0505+612	&    -1.50 &  2167 & 9896 & 0.227 & 0.219& 0.377 &	 \phn 4.1		&1.2  & 0.6/5.7 \footnotesize{(z=0.1/0.5)}\\\	
1FGL J0515.9+1528& 	 -0.54 & 1149 & 6734 & 0.173& 0.151& 0.283 &	 16.7 & 2.5 & 0.9/3.8 \footnotesize{(z=0.1/0.5)}\\\	
2FGL J0540.4+5822	& 	 0.43 & 226 & 1315 & 0.167 & 	 0.240&  0.459& 16.5	 &8.5 & 2.8/49.4 \footnotesize{(z=0.1/0.5)}\\\		
RGB J0643+422&		 0.46 & 240 & 1282 & 0.181 & 0.200& 0.369 &	 32.7 &  9.5 & \nodata \\	
RGB J0656+426&	 1.16 & 1960 & 9607 & 0.198 & 0.200  & 0.322 &	 20.2&  3.6 & \nodata \\	
1ES 0735+178	& 	 -1.20 & 1259 &  6865 & 0.190 & 0.166 & 0.260 &\phn 9.0 		& 0.9 & 0.5\\
BZB J0809+3455&		 -0.24 & 252 & 1537 & 0.167 & 0.151 & 0.251 &	 39.8	 	& 4.0 & \nodata \\		
PKS 0829+046	&  -0.77 & 465 & 2465 & 0.196  & 0.240 &	 0.379 & 10.1	 & 2.7 & 0.6 \\	
Mrk 1218	& 		 2.44 & 1589 & 8994 & 0.165 & 0.166 & 0.280 &	 43.1 & 5.7 & \nodata \\			
OJ 287	& 	 0.97 & 2197  & 12966 & 0.166 & 0.182& 0.296  & 	 17.4	& 2.6 & 3.1 \\	
B2 0912+29&	 3.49 & 3466 & 19492 & 0.167 & 0.138& 0.228 &45.9	& 3.6 & 1.6\\	
1ES 0927+500&	  -0.18 & 2378 & 11404 & 0.209 & 0.182& 0.346 &	   11.0	 & 2.8 & \nodata \\	
RBS 0831	 &	 -0.31 & 394 & 2403 & 0.167 & 0.166& 0.297 &	 29.5	& 4.8 & \nodata \\	
RGB J1012+424	& 	 0.18 & 270 & 1324 & 0.167 & 0.219& 0.316&	 43.5	 & 6.7 & 22.2 \\	
1ES 1028+511	& 	  1.16 & 4610  & 27154 &  0.167& 0.182& 0.305&	 12.4	 & 2.0 & 1.4\\	
RGB J1037+571 & 	  -1.53 & 790 & 3798 & 0.221 & 0.200& 0.331 & \phn	 5.7	 	& 1.1 & 2.5 (z=0.6)\\	
RGB J1053+494	 &  -0.76 & 1397 & 8567 & 0.167 & 0.200& 0.386& \phn	 6.4	& 2.2 & 0.5 \\	
RBS 0921	 &	0.84 & 633 & 3534 & 0.173 & 0.240& 0.354 &	20.1& 4.2 & \nodata \\		
RBS 0929	 & -0.62 & 923 & 4678 & 0.202 & 0.166& 0.319	 & 11.3	& 2.4 & 0.5/2.7 \footnotesize{(z=0.1/0.5)} \\	
1ES 1106+244	 &	 -1.65 & 200 & 1151 & 0.197 & 0.151& 0.257 &	 14.3	 & 1.5 & 3.3 \\	
RX J1117.1+2014	&  0.16 & 2545 & 12950 & 0.190 & 0.151& 0.281 &	 12.5	 & 1.8 &  0.18\\	
1ES 1118+424	&	 0.39 & 1685 & 9703 & 0.172 & 0.151& 0.267 &	 22.4	 	& 2.8 & 0.39 \\	
S4 1150+497	 &	 -0.53 & 749 & 4589 & 0.167 & 0.182& 0.315&	 12.9 & 2.3 & 70.0\\	
RGB J1231+287&	 0.74 & 1258 & 6664 & 0.185 & 0.138& 0.243 &	 36.6	 & 3.5 & 12.2\\	
1ES 1239+069& -0.86 & 224 & 1429 & 0.167 & 0.240& 0.369 & \phn	 8.7	 	& 2.1& \nodata \\	
PG 1246+586	& 	 0.23 & 2123 & 12670 & 0.167 & 0.200& 0.363&	\phn  9.5	 & 2.4&  14.5 \footnotesize{(z=0.73)}\\
1ES 1255+244	 &  2.24 & 5127 & 29732 & 0.167& 0.166& 0.315	& 12.4	 & 2.5&  \nodata \\	
BZB J1309+4305&	 0.54  & 2359 & 14020 & 0.167 & 0.151&  0.298&	 17.8	 	& 3.2&  7.9 \\	
1FGL J1323.1+2942& 0.24 & 1781& 14622& 0.121& 0.151& 0.243&	 18.0	 & 1.6& 1.2/3.7 \footnotesize{(z=0.1/0.5)} \\	
RX J1326.2+2933	&  1.36 & 1771 & 14150 & 0.127 & 0.166& 0.256 &	 17.8 		& 1.7& \nodata \\	
RGB J1341+399	&  0.00 & 381 & 2286 & 0.167 & 0.200& 0.405&	  12.7	 	& 5.1& \nodata \\	
RGB J1351+112	&  1.44 & 1715 & 8994 & 0.184 & 0.151&  0.248 &	 30.0	& 2.8 & 2.9 (z=0.62) \\	
RX J1353.4+5601	&  0.65 & 569 & 3163 & 0.175 & 0.200& 0.339&	 24.8	 & 5.3 & \nodata \\	
RBS 1350	& 0.55 & 1387 & 8190 & 0.167& 0.151 & 0.248 &	 27.9	 	& 2.6 & \nodata \\	
RBS 1366	 &	 1.89 & 1789 & 9843 & 0.173 & 0.200& 0.327 &	 17.1	 	 & 3.3 & 7.9\\	
1ES 1421+582	 & 0.17 & 674 & 4016 & 0.167 & 0.219& 0.378&	 13.2	 	& 3.8 & \nodata \\	
 RGB J1439+395&	 	 0.80 & 404 & 2321 & 0.167 & 0.151& 0.246 &	 62.4	 & 5.7 & 5.0\\
1RXS J144053.2+061013&	 -0.09 & 424 & 2556 & 0.167 & 0.200& 0.343 &	 13.8	 		& 3.1 & 16.7\\
RBS 1452	&  0.03 & 1232 & 7474 & 0.165 & 0.151& 0.254 &	 25.9	 & 2.7 & 0.5\\	
RGB J1532+302	& 		 -0.56 & 812 & 3648 & 0.227 & 0.166& 0.348&	 10.7	 	& 3.0 & \nodata \\	
RGB J1533+189&	 -1.44 & 653 & 4460 & 0.167 & 0.151& 0.293& \phn	 8.5	 	& 1.5 & \nodata \\	
1ES 1533+535	&  0.54 & 191 & 973 & 0.188 &  0.182& 0.324 &	41.7 			& 8.9 & \nodata \\	
RGB J1610+671B	& -1.75 & 1318 & 6510 & 0.214 & 0.263& 0.516 & \phn	 1.6	 	&1.1 & \nodata \\	
1ES 1627+402	&  -0.33 & 2191 & 13244 & 0.167 & 0.182& 0.360 &	\phn  7.3	 	& 2.1 & \nodata \\	
GB6 J1700+6830&	 -1.98 & 81 & 610  & 0.167 & 0.316& 0.528 & \phn	 3.5	 	& 2.2 & 161\\	
PKS 1717+177	&  0.42 & 934 & 4343 & 0.212 & 0.182& 0.290 &	 19.3	 	& 2.6 & 3.1(z=0.58)  \\	
PKS 1725+045	&  -0.58 & 41 & 271 & 0.167 & 0.200& 0.329 &	  37.5	 	& 7.3 & 127\\	
PKS 1749+096	 & -1.00 & 64 & 438 & 0.167 & 0.166& 0.257 &	 43.6					& 4.3 & 1.6\\	
RGB J1838+480& 0.27 & 87 & 329 & 0.167 & 0.200& 0.346 &	 52.3	 				& 12.1 \phn & 4.2\\	
RGB J1903+556& 0.19 & 164 & 968 & 0.167 & 0.240& 0.398 &	 22.4	 & 7.0 & 22.2 (z=0.58) \\	
1FGL J1926.8+6153& 0.62  & 231 & 1326 & 0.167 & 0.240& 0.408 &	 21.6	 & 7.4 & 1.1/12.5 \footnotesize{(z=0.1/0.5)}  \\	
PKS 2233-148& 0.06 & 32 & 190 & 0.167 & 0.501 & 0.829 &	\phn 7.8	 	& 15.0 \phn & 1.4e3 (z=0.49) \\						
3C 454.3	&  -1.21 & 220 & 981 & 0.25 & 0.138& 0.250 &	 23.0			& 2.5 & 0.6 \\		
RGB J2322+346&	 -0.42 & 518  & 2926 & 0.181 & 0.182& 0.296 &	 19.2	 	&2.8 & 1.4 \\	
1ES 2321+419& 2.05 & 992 & 3686 & 0.250 & 0.219& 0.414 &	 23.7	 & 9.4 & 15.9 (z=0.5) \\	
B3 2322+396&	 -0.63 & 131 & 705 & 0.197 & 0.166& 0.256 &	 49.2		& 4.8 & 80 (z=1.05) \\	
1FGL J2329.2+3755&	 -0.13 & 847 & 5106 & 0.167 & 0.166& 0.254 &	 27.3	 	& 2.6 & 1.0/3.4 \footnotesize{(z=0.1/0.5)} \\	
1RXS J234332.5+343957& 	 0.80 & 341& 1952 & 0.167 & 0.151& 0.250 & 	 53.5		&  5.2 & 3.2
\enddata
\end{deluxetable*}

\begin{deluxetable*}{|c||c|c|c|c||c| c |c | c|c|}
\tablecolumns{8} 
\tablewidth{0pc} 
\tablecaption{Results and upper limits for the known VHE sources \label{knownVHEtable}} 
\tablehead{
 Source name & $\sigma$ & ON & OFF & $\alpha$ & E$_{\rm{th}}$ & E$_{\rm{dec}}$ & UL$_{@E_{\rm{dec}}}$ &   UL$_{>E_{\rm{th}}}$  & UL/ $\Phi_{HE}$ \\
  & &  & & &   \scriptsize{[TeV]} & \scriptsize{[TeV]} & \scriptsize{[10$^{-12}$\ cm$^{\textrm{-2}}$s$^{\textrm{-1}}$TeV$^{\textrm{-1}}$]} & \scriptsize{[\% C.U.]} &  
}
\startdata
\hline
1ES 0033+595&	 3.23 & 4560 & 20321& 0.214&  0.288 & 0.490 &	 10.0 & 5.4 & 0.8 (z=0.086) \\	
RGB J0152+017	&  1.83 & 1720 & 8693 & 0.188 & 0.240& 0.376 &	 14.1		& 3.6 & 1.2\\	
RGB J0847+115	& 	 -0.03 & 3391 & 19889 & 0.171 & 0.182& 0.338  &	 8.63	 & 2.0 & 0.3\\
RX J1136.5+6737	& 1.10 & 1324 & 7271 & 0.177 &	 0.288& 0.519 & 7.86	 & 5.2 & 1.3 \\
PKS 1222+216 & 3.40 & 7482& 39770& 0.180& 0.182&0.257 & 24.1& 2.2 &0.2 \\
3C 279 & 0.10 & 1479 & 8849 & 0.167 & 0.263 & 0.430 & 5.5 & 2.1 & 2.6 \\
PKS 1510-089 & 2.00 & 2691 & 13698 & 0.167 & 0.263& 0.496& 4.7& 2.9 & 1.1 \\
RGB J1725+118& 2.29 & 1979 & 10833 & 0.079 & 0.200& 0.316 &	 18.7	 	& 3.2 & 0.6/2.6 \footnotesize{(z=0.1/0.5)} \\
0FGL J2001.0+4352 & 0.76 & 1102 & 6449 & 0.167 & 0.200 & 0.384 & 15.5 & 5.2 & 0.3 (z=0.2) \\
RGB J2243+203	& -0.12 & 1111 & 6458 & 0.173& 0.166& 0.263 &	 19.3	 & 2.1 & 0.3 (z=0.39) \\	
B3 2247+381 & -0.93 & 1447 & 8042 & 0.185 & 0.166 & 0.315 & 8.8 &1.8  & 0.5 
\enddata
\end{deluxetable*}

\begin{deluxetable*}{|c||c|c|c|c||c| c |c | c|c|}
\tablecolumns{7} 
\tablewidth{0pc} 
\tablecaption{Results and upper limits for the 2FGL sources in the \veritas\ field of view \label{2FGLtable}} 
\tablehead{
 Source name & $\sigma$ & ON & OFF & $\alpha$ & E$_{\rm{th}}$ & E$_{\rm{dec}}$ & UL$_{@E_{\rm{dec}}}$ &   UL$_{>E_{\rm{th}}}$  & UL/ $\Phi_{HE}$ \\
  & &  & & &   \scriptsize{[TeV]} & \scriptsize{[TeV]} & \scriptsize{[10$^{-12}$\ cm$^{\textrm{-2}}$s$^{\textrm{-1}}$TeV$^{\textrm{-1}}$]} & \scriptsize{[\% C.U.]} &  
}
\startdata
\hline
2FGL J0047.9+2232	&	0.32 & 123 & 2058 & 0.0770 &		0.200 &  0.314 & 	70.5 & 11.6 & 1.6e4\\		
2FGL J0148.6+0127	&	1.15 & 1239 & 17272 & 0.063  &		0.240 &  0.412 &	25.1 & 8.9  & 6.7e3\\		
2FGL J0158.4+0107	&	-1.62 & 582 & 13194& 0.0625 &		0.240 &  0.435 &	16.6& 7.1  & 150/2.2e3 \footnotesize{(z=0.1/0.5)} \\			 
2FGL J0205.4+3211	&	1.76 & 232 & 3743& 0.0524&		0.166  & 0.290 &	102	& 15.4 & 5.4e5\\	
2FGL J0212.1+5318	&	-1.22 & 389 & 8089 & 0.0512&		0.200  & 0.377 & 	37.8	& 11.9 & 1.0/9.1 \footnotesize{(z=0.1/0.5)} \\	
2FGL J0213.1+2245	&	0.57 & 501 & 7460 & 0.0655 &		0.182 & 0.295 & 	63.3	& 9.2 & 29.3\\	
2FGL J0326.1+2226	&	0.22 & 1342 & 25891 & 0.0515 &		0.151&  0.249 & 	92.9	& 9.0 & 2.5e4\\	
2FGL J0440.4+1433	&	0.20 & 1028 & 19458 & 0.0525 &		0.182 &  0.340 & 	50.3	& 12.1 & 60/416 \footnotesize{(z=0.1/0.5)}\\	
2FGL J0856.3+2058	 &	0.19 & 1686 & 21961& 0.0764&		0.182 &  0.297 & 	37.1	& 5.6 & 53 \footnotesize{(z=0.5)} \\	
2FGL J0929.5+5009	&        3.28 & 1712 & 17578 & 0.0884&		0.200 &  0.349 & 	40.2		& 9.6 & 28\\	
2FGL J1058.4+0133	&	1.19 & 319 & 5882 & 0.0506&		0.240 &  0.373 & 	51.7	& 12.9 & 985 \\	
2FGL J1059.0+0222	 &      -0.37& 533 & 7097 & 0.0764 & 		0.240&  0.359 & 	25.8			& 5.7 & 30/240 \footnotesize{(z=0.1/0.5)} \\	
2FGL J1141.0+6803	&	0.17 & 1195 & 12124 & 0.0981 &		0.288 &  0.525 & 	7.5		& 5.1 & 0.7/18.9 \footnotesize{(z=0.1/0.5)} \\	
2FGL J1239.5+0728 & 	1.41 & 198 & 2766 & 0.0644&		0.240&  0.368 & 	55.1		& 13.1 & 67\\	
2FGL J1245.1+5708 &	2.34 & 1418 & 24171 & 0.0545&	        0.219 &  0.369 & 	40.7	& 10.8 & 276 \footnotesize{(z=0.52)} \\	
2FGL J1303.1+2435 &	0.33 & 3021 & 57951 & 0.0518 &		0.166 &  0.322 & 	36.2		& 7.8 & 117 \\	
2FGL J1359.4+5541&	1.74 & 507 & 5493 & 0.0850 &		0.219 &  0.347 & 	50.5		& 10.8 & 1.4e5\\	
2FGL J1722.7+1013 &	1.73 & 693 & 14010& 0.0462 &	         0.200 &  0.332 & 	72.2		&  14.5 & 433\\	
2FGL J1727.9+1220	&	-0.56 & 1816 & 23431 & 0.0789&	         0.200 &  0.315 & 	15.3				& 2.5 & 41\\											
2FGL J1927.5+6117	& -1.94 & 127 & 1964 & 0.077 &      	0.240 &  0.412 & 	12.1	&  4.3 & 5.1/62 \footnotesize{(z=0.1/0.5)} \\	
2FGL J1959.9+4212	&	1.82 & 458 & 9118 & 0.037&	0.219		& 0.347 & 64.4 & 13.8 &  24/176 \footnotesize{(z=0.1/0.5)} 
\enddata
\end{deluxetable*}

\clearpage
\newpage
\phantom{text}

\section{Stacked analysis}
\label{section6}

Motivated by the skew in the significance distribution and in order to study if there is any evidence of emission from a population of blazars below the \veritas\ sensitivity level, a stacked analysis of the data-set is performed. For every source the $\gamma$-ray excess (ON - $\alpha$ OFF) and its uncertainty (the excess divided by the significance) are calculated. We then compute the sum of the excesses, and its uncertainty (the square root of the sum of the squared uncertainties), whose ratio provides the significance of the stacked excess. Sources known as VHE emitters are excluded from the stacked analysis, which only includes sources listed in Tables \ref{Sourcelist} and \ref{2FGLlist}.\\

The stacked analysis indicates that there is evidence of VHE emission at a level of $4.0\sigma$, corresponding to an excess of 1990 $\gamma$-rays.
The same study is then performed for sub-samples of the overall data-set. The majority of the excess ($3.0\sigma$) comes from nearby ($z<0.6$) HBLs. On the other hand, the stacked analysis including only non-HBL sources located at an unknown distance or $z>0.6$ results in a stacked significance of $1.1\sigma$. However, because nearby HBLs are considered the most likely VHE candidates, there are more of them and they often have deeper exposures. So this study has more sensitivity to the nearby HBLs. Indeed, the \veritas\ exposure on HBLs located at $z<0.6$ is about 196 hours. By assuming that the $4.0\sigma$ stacked excess comes from a constant signal from all sources in 570 hours, one would expect a $2.3\sigma$ excess in 196 hours. The excess from the $z<0.6$  HBLs is thus compatible with this expectation, and it is not possible to claim that the stacked excess is dominated by a particular blazar population.\\

The \magic\ collaboration has also reported evidence for VHE emission from a stacked sample of IBL/HBL sources \citep{MAGICUL}, detecting a signal at a significance level of $4.9\sigma$ from an exposure of 394 hours. The following sources included in the present work are also part of the \magic\ sample: 1ES~0120+340, 1RXS~J044127.8+150455, 1ES~0927+500, 1ES~1028+511, RX~J1117.1+2014, RX~1136.5+6737, and RBS~1366. The four sources with the highest significance in the \magic\ publication (1ES~0033+595, 1ES~1011+496, B2~1215+30, and 1ES~1741+196) were, notably, later confirmed as VHE emitters, either during flaring activity, or by increasing the integration time. \\

\section{Conclusions}
\label{section7}

The results from the analysis of the observations of non-detected blazars targeted by \veritas\ from 2007 to 2012 have been presented. In addition, $\gamma$-ray sources from the 2FGL catalog which were within the field of view of these \veritas\ observations were included in this study. For all the 114 sources included in this data-set we provided the \veritas\ upper limit at VHE.
Given that the redshift estimate of blazars is particularly important for VHE extragalactic astronomy, due to the $\gamma$-ray absorption on the EBL, we also presented the results from optical spectroscopy of 18 of these targets, determining the redshift for three of them, and providing a lower limit for the redshift of one of the sources.\\

We have presented the results from a stacked analysis of the data-set, showing that there is some evidence of signal with a significance level of $4\sigma$.\\

In the near future, the sensitivity of VHE astronomy will be significantly increased thanks to the Cherenkov Telescope Array (CTA), which will be capable of detecting sources with VHE fluxes of the order of 0.001 Crab units, about a factor of ten better than current IACTs \citep{CTA}. Among the scientific goals of CTA, an important endeavor will be to increase the number of known VHE blazars in order to perform population studies. Among the \veritas\ targets presented in this work, the sources with the highest significance could be considered as primary candidates for observations with CTA, which may be able to detect many of them on the basis of the extrapolation of their \fermilat\ spectra to higher energies. The non-detection of a number of later detected VHE blazars emphasizes the variable nature of these sources, highlighting the importance of monitoring observations in order to increase the likelihood of catching the sources at detectable VHE states. \\

\acknowledgments{This research is supported by grants from the U.S. Department of Energy Office of Science, the U.S. National Science Foundation and the Smithsonian Institution, and by NSERC in Canada. We acknowledge the excellent work of the technical support staff at the Fred Lawrence Whipple Observatory and at the collaborating institutions in the construction and operation of the instrument. The VERITAS Collaboration is grateful to Trevor Weekes for his seminal contributions and leadership in the field of VHE gamma-ray astrophysics, which made this study possible. MF acknowledges support by the Science and Technology Facilities Council [grant number  ST/L00075X/1].}\\

 \bibliographystyle{apj}
 \bibliography{UL_biblio}  

\begin{thebibliography}{149}
\expandafter\ifx\csname natexlab\endcsname\relax\def\natexlab#1{#1}\fi

\bibitem[{{Abdo} {et~al.}(2009{\natexlab{a}}){Abdo}, {Ackermann}, {Ajello},
  {Atwood}, {Axelsson}, {Baldini}, {Ballet}, {Band}, {Barbiellini}, {Bastieri},
  \& et~al.}]{Fermi0FGL}
{Abdo}, A.~A., {Ackermann}, M., {Ajello}, M., {et~al.} 2009{\natexlab{a}},
  \apjs, 183, 46

\bibitem[{{Abdo} {et~al.}(2009{\natexlab{b}}){Abdo}, {Ackermann}, {Ajello},
  {Baldini}, {Ballet}, {Barbiellini}, {Bastieri}, {Bechtol}, {Bellazzini},
  {Berenji}, {Bloom}, {Bonamente}, {Borgland}, {Bregeon}, {Brez}, {Brigida},
  {Bruel}, {Burnett}, {Caliandro}, {Cameron}, {Caraveo}, {Casandjian},
  {Cecchi}, {{\c C}elik}, {Chekhtman}, {Cheung}, {Chiang}, {Ciprini}, {Claus},
  {Cohen-Tanugi}, {Conrad}, {Cutini}, {Dermer}, {de Palma}, {Silva}, {Drell},
  {Dubois}, {Dumora}, {Farnier}, {Favuzzi}, {Fegan}, {Focke}, {Foschini},
  {Frailis}, {Fukazawa}, {Fusco}, {Gargano}, {Gehrels}, {Germani}, {Giebels},
  {Giglietto}, {Giordano}, {Giroletti}, {Glanzman}, {Godfrey}, {Grenier},
  {Grove}, {Guillemot}, {Guiriec}, {Hayashida}, {Hays}, {Horan}, {Hughes},
  {J{\'o}hannesson}, {Johnson}, {Johnson}, {Kadler}, {Kamae}, {Katagiri},
  {Kataoka}, {Kerr}, {Kn{\"o}dlseder}, {Kuss}, {Lande}, {Latronico}, {Longo},
  {Loparco}, {Lott}, {Lovellette}, {Lubrano}, {Makeev}, {Mazziotta},
  {McConville}, {McEnery}, {Meurer}, {Michelson}, {Mitthumsiri}, {Mizuno},
  {Monte}, {Monzani}, {Morselli}, {Moskalenko}, {Murgia}, {Nolan}, {Norris},
  {Nuss}, {Ohsugi}, {Omodei}, {Orlando}, {Ormes}, {Pelassa}, {Pepe}, {Persic},
  {Pesce-Rollins}, {Piron}, {Porter}, {Rain{\`o}}, {Rando}, {Razzano},
  {Rochester}, {Rodriguez}, {Ryde}, {Sadrozinski}, {Sambruna}, {Sander}, {Saz
  Parkinson}, {Scargle}, {Sgr{\`o}}, {Smith}, {Spandre}, {Spinelli},
  {Strickman}, {Suson}, {Tagliaferri}, {Takahashi}, {Takahashi}, {Tanaka},
  {Thayer}, {Thayer}, {Thompson}, {Tibaldo}, {Tibolla}, {Torres}, {Tosti},
  {Tramacere}, {Uchiyama}, {Usher}, {Vasileiou}, {Vilchez}, {Vitale}, {Waite},
  {Wang}, {Winer}, {Wood}, {Ylinen}, {Ziegler}, {Fermi/LAT Collaboration},
  {Ghisellini}, {Maraschi}, \& {Tavecchio}}]{FermiNLS1}
---. 2009{\natexlab{b}}, \apjl, 707, L142

\bibitem[{{Abdo} {et~al.}(2010{\natexlab{a}}){Abdo}, {Ackermann}, {Ajello},
  {Allafort}, {Antolini}, {Atwood}, {Axelsson}, {Baldini}, {Ballet},
  {Barbiellini}, \& et~al.}]{1FGL}
---. 2010{\natexlab{a}}, \apjs, 188, 405

\bibitem[{{Abdo} {et~al.}(2010{\natexlab{b}}){Abdo}, {Ackermann}, {Agudo},
  {Ajello}, {Aller}, {Aller}, {Angelakis}, {Arkharov}, {Axelsson}, {Bach}, \&
  et~al.}]{Abdo10}
{Abdo}, A.~A., {Ackermann}, M., {Agudo}, I., {et~al.} 2010{\natexlab{b}}, \apj,
  716, 30

\bibitem[{{Abeysekara}(2015)}]{2243icrc}
{Abeysekara}, A.~U. 2015, to appear in Proceedings of the International Cosmic
  Ray Conference, ArXiv e-prints 1508.06334

\bibitem[{Abeysekara {et~al.}(2015)Abeysekara, Archambault, Archer, Aune,
  Barnacka, Benbow, Bird, Biteau, Buckley, Bugaev, Cardenzana, Cerruti, Chen,
  Christiansen, Ciupik, Connolly, Coppi, Cui, Dickinson, Dumm, Eisch, Errando,
  Falcone, Feng, Finley, Fleischhack, Flinders, Fortin, Fortson, Furniss,
  Gillanders, Griffin, Grube, Gyuk, H{\"u}tten, H{\aa}kansson, Hanna, Holder,
  Humensky, Johnson, Kaaret, Kar, Kelley-Hoskins, Khassen, Kieda, Krause,
  Krennrich, Kumar, Lang, Maier, McArthur, McCann, Meagher, Moriarty,
  Mukherjee, Nieto, de~Bhr{\'o}ithe, Ong, Otte, Park, Perkins, Petrashyk, Pohl,
  Popkow, Pueschel, Quinn, Ragan, Ratliff, Reynolds, Richards, Roache,
  Rousselle, Santander, Sembroski, Shahinyan, Smith, Staszak, Telezhinsky,
  Todd, Tucci, Tyler, Vassiliev, Vincent, Wakely, Weiner, Weinstein, Wilhelm,
  Williams, Zitzer, VERITAS, Smith, SPOL, Holoien, Prieto, Kochanek, Stanek,
  Shappee, ASAS-SN, Hovatta, Max-Moerbeck, Pearson, Reeves, Richards, Readhead,
  OVRO, Madejski, NuSTAR, Djorgovski, Drake, Graham, Mahabal, \&
  CRTS}]{1441veritas}
Abeysekara, A.~U., Archambault, S., Archer, A., {et~al.} 2015, \apjl, 815, L22

\bibitem[{{Abramowski} {et~al.}(2013{\natexlab{a}}){Abramowski}, {Acero},
  {Aharonian}, {Akhperjanian}, {Anton}, {Balenderan}, {Balzer}, {Barnacka},
  {Becherini}, {Becker Tjus}, {Behera}, {Bernl{\"o}hr}, {Birsin}, {Biteau},
  {Bochow}, {Boisson}, {Bolmont}, {Bordas}, {Brucker}, {Brun}, {Brun}, {Bulik},
  {Carrigan}, {Casanova}, {Cerruti}, {Chadwick}, {Chaves}, {Cheesebrough},
  {Colafrancesco}, {Cologna}, {Conrad}, {Couturier}, {Dalton}, {Daniel},
  {Davids}, {Degrange}, {Deil}, {deWilt}, {Dickinson}, {Djannati-Ata{\"i}},
  {Domainko}, {O'C.~Drury}, {Dubus}, {Dutson}, {Dyks}, {Dyrda}, {Egberts},
  {Eger}, {Espigat}, {Fallon}, {Farnier}, {Fegan}, {Feinstein}, {Fernandes},
  {Fernandez}, {Fiasson}, {Fontaine}, {F{\"o}rster}, {F{\"u}{\ss}ling},
  {Gajdus}, {Gallant}, {Garrigoux}, {Gast}, {Giebels}, {Glicenstein},
  {Gl{\"u}ck}, {G{\"o}ring}, {Grondin}, {Grudzi{\'n}ska}, {H{\"a}ffner},
  {Hague}, {Hahn}, {Hampf}, {Harris}, {Hauser}, {Heinz}, {Heinzelmann},
  {Henri}, {Hermann}, {Hillert}, {Hinton}, {Hofmann}, {Hofverberg}, {Holler},
  {Horns}, {Jacholkowska}, {Jahn}, {Jamrozy}, {Jung}, {Kastendieck},
  {Katarzy{\'n}ski}, {Katz}, {Kaufmann}, {Kh{\'e}lifi}, {Klepser}, {Klochkov},
  {Klu{\'z}niak}, {Kneiske}, {Kolitzus}, {Komin}, {Kosack}, {Kossakowski},
  {Krayzel}, {Kr{\"u}ger}, {Laffon}, {Lamanna}, {Lefaucheur},
  {Lemoine-Goumard}, {Lenain}, {Lennarz}, {Lohse}, {Lopatin}, {Lu}, {Marandon},
  {Marcowith}, {Masbou}, {Maurin}, {Maxted}, {Mayer}, {McComb}, {Medina},
  {M{\'e}hault}, {Menzler}, {Moderski}, {Mohamed}, {Moulin}, {Naumann},
  {Naumann-Godo}, {de Naurois}, {Nedbal}, {Nguyen}, {Niemiec}, {Nolan}, {Ohm},
  {de O{\~n}a Wilhelmi}, {Opitz}, {Ostrowski}, {Oya}, {Panter}, {Parsons}, {Paz
  Arribas}, {Pekeur}, {Pelletier}, {Perez}, {Petrucci}, {Peyaud}, {Pita},
  {P{\"u}hlhofer}, {Punch}, {Quirrenbach}, {Raab}, {Raue}, {Reimer}, {Reimer},
  {Renaud}, {de los Reyes}, {Rieger}, {Ripken}, {Rob}, {Rosier-Lees}, {Rowell},
  {Rudak}, {Rulten}, {Sahakian}, {Sanchez}, {Santangelo}, {Schlickeiser},
  {Schulz}, {Schwanke}, {Schwarzburg}, {Schwemmer}, {Sheidaei}, {Skilton},
  {Sol}, {Spengler}, {Stawarz}, {Steenkamp}, {Stegmann}, {Stinzing}, {Stycz},
  {Sushch}, {Szostek}, {Tavernet}, {Terrier}, {Tluczykont}, {Trichard},
  {Valerius}, {van Eldik}, {Vasileiadis}, {Venter}, {Viana}, {Vincent},
  {V{\"o}lk}, {Volpe}, {Vorobiov}, {Vorster}, {Wagner}, {Ward}, {White},
  {Wierzcholska}, {Wouters}, {Zacharias}, {Zajczyk}, {Zdziarski}, {Zech}, \&
  {Zechlin}}]{1510hess}
{Abramowski}, A., {Acero}, F., {Aharonian}, F., {et~al.} 2013{\natexlab{a}},
  \aap, 554, A107

\bibitem[{{Abramowski} {et~al.}(2013{\natexlab{b}}){Abramowski}, {Acero},
  {Aharonian}, {Akhperjanian}, {Anton}, {Balenderan}, {Balzer}, {Barnacka},
  {Becherini}, {Becker Tjus}, {Bernl{\"o}hr}, {Birsin}, {Biteau}, {Bochow},
  {Boisson}, {Bolmont}, {Bordas}, {Brucker}, {Brun}, {Brun}, {Bulik},
  {Carrigan}, {Casanova}, {Cerruti}, {Chadwick}, {Charbonnier}, {Chaves},
  {Cheesebrough}, {Cologna}, {Conrad}, {Couturier}, {Dalton}, {Daniel},
  {Davids}, {Degrange}, {Deil}, {deWilt}, {Dickinson}, {Djannati-Ata{\"i}},
  {Domainko}, {O'C.~Drury}, {Dubus}, {Dutson}, {Dyks}, {Dyrda}, {Egberts},
  {Eger}, {Espigat}, {Fallon}, {Farnier}, {Fegan}, {Feinstein}, {Fernandes},
  {Fernandez}, {Fiasson}, {Fontaine}, {F{\"o}rster}, {F{\"u}{\ss}ling},
  {Gajdus}, {Gallant}, {Garrigoux}, {Gast}, {Giebels}, {Glicenstein},
  {Gl{\"u}ck}, {G{\"o}ring}, {Grondin}, {H{\"a}ffner}, {Hague}, {Hahn},
  {Hampf}, {Harris}, {Heinz}, {Heinzelmann}, {Henri}, {Hermann}, {Hillert},
  {Hinton}, {Hofmann}, {Hofverberg}, {Holler}, {Horns}, {Jacholkowska}, {Jahn},
  {Jamrozy}, {Jung}, {Kastendieck}, {Katarzy{\'n}ski}, {Katz}, {Kaufmann},
  {Kh{\'e}lifi}, {Klochkov}, {Klu{\'z}niak}, {Kneiske}, {Komin}, {Kosack},
  {Kossakowski}, {Krayzel}, {Laffon}, {Lamanna}, {Lenain}, {Lennarz}, {Lohse},
  {Lopatin}, {Lu}, {Marandon}, {Marcowith}, {Masbou}, {Maurin}, {Maxted},
  {Mayer}, {McComb}, {Medina}, {M{\'e}hault}, {Menzler}, {Moderski}, {Mohamed},
  {Moulin}, {Naumann}, {Naumann-Godo}, {de Naurois}, {Nedbal}, {Nguyen},
  {Niemiec}, {Nolan}, {Ohm}, {de O{\~n}a Wilhelmi}, {Opitz}, {Ostrowski},
  {Oya}, {Panter}, {Parsons}, {Paz Arribas}, {Pekeur}, {Pelletier}, {Perez},
  {Petrucci}, {Peyaud}, {Pita}, {P{\"u}hlhofer}, {Punch}, {Quirrenbach},
  {Raue}, {Reimer}, {Reimer}, {Renaud}, {de los Reyes}, {Rieger}, {Ripken},
  {Rob}, {Rosier-Lees}, {Rowell}, {Rudak}, {Rulten}, {Sahakian}, {Sanchez},
  {Santangelo}, {Schlickeiser}, {Schulz}, {Schwanke}, {Schwarzburg},
  {Schwemmer}, {Sheidaei}, {Skilton}, {Sol}, {Spengler}, {Stawarz},
  {Steenkamp}, {Stegmann}, {Stinzing}, {Stycz}, {Sushch}, {Szostek},
  {Tavernet}, {Terrier}, {Tluczykont}, {Valerius}, {van Eldik}, {Vasileiadis},
  {Venter}, {Viana}, {Vincent}, {V{\"o}lk}, {Volpe}, {Vorobiov}, {Vorster},
  {Wagner}, {Ward}, {White}, {Wierzcholska}, {Wouters}, {Zacharias}, {Zajczyk},
  {Zdziarski}, {Zech}, \& {Zechlin}}]{HESSEBL}
---. 2013{\natexlab{b}}, \aap, 550, A4

\bibitem[{{Abramowski} {et~al.}(2014){Abramowski}, {Aharonian}, {Ait Benkhali},
  {Akhperjanian}, {Ang{\"u}ner}, {Anton}, {Balenderan}, {Balzer}, {Barnacka},
  \& et~al.}]{HESSUL2014}
{Abramowski}, A., {Aharonian}, F., {Ait Benkhali}, F., {et~al.} 2014, \aap,
  564, A9

\bibitem[{{Acciari} {et~al.}(2010){Acciari}, {Aliu}, {Arlen}, {Aune},
  {Bautista}, {Beilicke}, {Benbow}, {B{\"o}ttcher}, {Boltuch}, {Bradbury}, \&
  et~al.}]{PKS1424Veritas}
{Acciari}, V.~A., {Aliu}, E., {Arlen}, T., {et~al.} 2010, \apjl, 708, L100

\bibitem[{{Ackermann} {et~al.}(2012{\natexlab{a}}){Ackermann}, {Ajello},
  {Allafort}, {Antolini}, {Baldini}, {Ballet}, {Barbiellini}, {Bastieri},
  {Bellazzini}, {Berenji}, {Blandford}, {Bloom}, {Bonamente}, {Borgland},
  {Bouvier}, {Brandt}, {Bregeon}, {Brigida}, {Bruel}, {Buehler}, {Burnett},
  {Buson}, {Caliandro}, {Cameron}, {Caraveo}, {Casandjian}, {Cavazzuti},
  {Cecchi}, {{\c C}elik}, {Charles}, {Chekhtman}, {Chen}, {Cheung}, {Chiang},
  {Ciprini}, {Claus}, {Cohen-Tanugi}, {Conrad}, {Cutini}, {de Angelis},
  {DeCesar}, {De Luca}, {de Palma}, {Dermer}, {Silva}, {Drell},
  {Drlica-Wagner}, {Dubois}, {Enoto}, {Favuzzi}, {Fegan}, {Ferrara}, {Focke},
  {Fortin}, {Fukazawa}, {Funk}, {Fusco}, {Gargano}, {Gasparrini}, {Gehrels},
  {Germani}, {Giglietto}, {Giordano}, {Giroletti}, {Glanzman}, {Godfrey},
  {Grenier}, {Grondin}, {Grove}, {Guillemot}, {Guiriec}, {Gustafsson},
  {Hadasch}, {Hanabata}, {Harding}, {Hayashida}, {Hays}, {Healey}, {Hill},
  {Horan}, {Hou}, {J{\'o}hannesson}, {Johnson}, {Johnson}, {Kamae}, {Katagiri},
  {Kataoka}, {Kerr}, {Kn{\"o}dlseder}, {Kuss}, {Lande}, {Latronico}, {Lee},
  {Lemoine-Goumard}, {Longo}, {Loparco}, {Lott}, {Lovellette}, {Lubrano},
  {Madejski}, {Mazziotta}, {McEnery}, {Mehault}, {Michelson}, {Mignani},
  {Mitthumsiri}, {Mizuno}, {Monte}, {Monzani}, {Morselli}, {Moskalenko},
  {Murgia}, {Nakamori}, {Naumann-Godo}, {Nolan}, {Norris}, {Nuss}, {Ohsugi},
  {Okumura}, {Omodei}, {Orlando}, {Ormes}, {Ozaki}, {Paneque}, {Panetta},
  {Parent}, {Pelassa}, {Pesce-Rollins}, {Pierbattista}, {Piron}, {Pivato},
  {Porter}, {Rain{\`o}}, {Rando}, {Ray}, {Razzano}, {Reimer}, {Reimer},
  {Reposeur}, {Romani}, {Sadrozinski}, {Salvetti}, {Saz Parkinson}, {Schalk},
  {Sgr{\`o}}, {Shaw}, {Siskind}, {Smith}, {Spandre}, {Spinelli}, {Suson},
  {Takahashi}, {Tanaka}, {Thayer}, {Thayer}, {Thompson}, {Tibaldo}, {Tibolla},
  {Torres}, {Tosti}, {Tramacere}, {Troja}, {Usher}, {Vandenbroucke},
  {Vasileiou}, {Vianello}, {Vilchez}, {Vitale}, {Waite}, {Wallace}, {Wang},
  {Winer}, {Wolff}, {Wood}, {Wood}, {Yang}, \& {Zimmer}}]{Ackermann12}
{Ackermann}, M., {Ajello}, M., {Allafort}, A., {et~al.} 2012{\natexlab{a}},
  \apj, 753, 83

\bibitem[{{Ackermann} {et~al.}(2012{\natexlab{b}}){Ackermann}, {Ajello},
  {Ballet}, {Barbiellini}, {Bastieri}, {Bellazzini}, {Blandford}, {Bloom},
  {Bonamente}, {Borgland}, \& et~al.}]{0235Fermi}
{Ackermann}, M., {Ajello}, M., {Ballet}, J., {et~al.} 2012{\natexlab{b}}, \apj,
  751, 159

\bibitem[{{Actis} {et~al.}(2011){Actis}, {Agnetta}, {Aharonian},
  {Akhperjanian}, {Aleksi{\'c}}, {Aliu}, {Allan}, {Allekotte}, {Antico},
  {Antonelli}, \& et~al.}]{CTA}
{Actis}, M., {Agnetta}, G., {Aharonian}, F., {et~al.} 2011, Experimental
  Astronomy, 32, 193

\bibitem[{{Afanas'Ev} {et~al.}(2005){Afanas'Ev}, {Dodonov}, {Moiseev},
  {Gorshkov}, {Konnikova}, \& {Mingaliev}}]{Afanas05}
{Afanas'Ev}, V.~L., {Dodonov}, S.~N., {Moiseev}, A.~V., {et~al.} 2005,
  Astronomy Reports, 49, 374

\bibitem[{{Aharonian} {et~al.}(2004){Aharonian}, {Akhperjanian}, {Beilicke},
  {Bernl{\"o}hr}, {B{\"o}rst}, {Bojahr}, {Bolz}, {Coarasa}, {Contreras},
  {Cortina}, {Denninghoff}, {Fonseca}, {Girma}, {G{\"o}tting}, {Heinzelmann},
  {Hermann}, {Heusler}, {Hofmann}, {Horns}, {Jung}, {Kankanyan}, {Kestel},
  {Konopelko}, {Kornmeyer}, {Kranich}, {Lampeitl}, {Lopez}, {Lorenz},
  {Lucarelli}, {Mang}, {Mazin}, {Meyer}, {Mirzoyan}, {Moralejo},
  {Ona-Wilhelmi}, {Panter}, {Plyasheshnikov}, {P{\"u}hlhofer}, {de los Reyes},
  {Rhode}, {Ripken}, {Rowell}, {Sahakian}, {Samorski}, {Schilling}, {Siems},
  {Sobzynska}, {Stamm}, {Tluczykont}, {Vitale}, {V{\"o}lk}, {Wiedner}, \&
  {Wittek}}]{HegraUL}
{Aharonian}, F., {Akhperjanian}, A., {Beilicke}, M., {et~al.} 2004, \aap, 421,
  529

\bibitem[{{Aharonian} {et~al.}(2005){Aharonian}, {Akhperjanian}, {Bazer-Bachi},
  {Beilicke}, {Benbow}, {Berge}, {Bernl{\"o}hr}, {Boisson}, {Bolz}, {Borrel},
  {Braun}, {Breitling}, {Brown}, {Chadwick}, {Chounet}, {Cornils},
  {Costamante}, {Degrange}, {Dickinson}, {Djannati-Ata{\"i}}, {O'C.~Drury},
  {Dubus}, {Emmanoulopoulos}, {Espigat}, {Feinstein}, {Fontaine}, {Fuchs},
  {Funk}, {Gallant}, {Giebels}, {Gillessen}, {Glicenstein}, {Goret},
  {Hadjichristidis}, {Hauser}, {Heinzelmann}, {Henri}, {Hermann}, {Hinton},
  {Hofmann}, {Holleran}, {Horns}, {Jacholkowska}, {de Jager}, {Kh{\'e}lifi},
  {Komin}, {Konopelko}, {Latham}, {Le Gallou}, {Lemi{\`e}re},
  {Lemoine-Goumard}, {Leroy}, {Lohse}, {Martin}, {Martineau-Huynh},
  {Marcowith}, {Masterson}, {McComb}, {de Naurois}, {Nolan}, {Noutsos},
  {Orford}, {Osborne}, {Ouchrif}, {Panter}, {Pelletier}, {Pita},
  {P{\"u}hlhofer}, {Punch}, {Raubenheimer}, {Raue}, {Raux}, {Rayner}, {Reimer},
  {Reimer}, {Ripken}, {Rob}, {Rolland}, {Rowell}, {Sahakian}, {Saug{\'e}},
  {Schlenker}, {Schlickeiser}, {Schuster}, {Schwanke}, {Siewert}, {Sol},
  {Spangler}, {Steenkamp}, {Stegmann}, {Tavernet}, {Terrier}, {Th{\'e}oret},
  {Tluczykont}, {Vasileiadis}, {Venter}, {Vincent}, {V{\"o}lk}, \&
  {Wagner}}]{HESSUL2005}
{Aharonian}, F., {Akhperjanian}, A.~G., {Bazer-Bachi}, A.~R., {et~al.} 2005,
  \aap, 441, 465

\bibitem[{{Aharonian} {et~al.}(2008{\natexlab{a}}){Aharonian}, {Akhperjanian},
  {Barres de Almeida}, {Bazer-Bachi}, {Behera}, {Beilicke}, {Benbow},
  {Bernl{\"o}hr}, {Boisson}, {Borrel}, {Braun}, {Brion}, {Brucker},
  {B{\"u}hler}, {Bulik}, {B{\"u}sching}, {Boutelier}, {Carrigan}, {Chadwick},
  {Chaves}, {Chounet}, {Clapson}, {Coignet}, {Cornils}, {Costamante}, {Dalton},
  {Degrange}, {Dickinson}, {Djannati-Ata{\"i}}, {Domainko}, {O'C.~Drury},
  {Dubois}, {Dubus}, {Dyks}, {Egberts}, {Emmanoulopoulos}, {Espigat},
  {Farnier}, {Feinstein}, {Fiasson}, {F{\"o}rster}, {Fontaine},
  {F{\"u}{\ss}ling}, {Gabici}, {Gallant}, {Giebels}, {Glicenstein},
  {Gl{\"u}ck}, {Goret}, {Hadjichristidis}, {Hauser}, {Hauser}, {Heinzelmann},
  {Henri}, {Hermann}, {Hinton}, {Hoffmann}, {Hofmann}, {Holleran}, {Hoppe},
  {Horns}, {Jacholkowska}, {de Jager}, {Jung}, {Katarzy{\'n}ski}, {Kaufmann},
  {Kendziorra}, {Kerschhaggl}, {Khangulyan}, {Kh{\'e}lifi}, {Keogh}, {Komin},
  {Kosack}, {Lamanna}, {Latham}, {Lenain}, {Lohse}, {Martin},
  {Martineau-Huynh}, {Marcowith}, {Masterson}, {Maurin}, {McComb}, {Moderski},
  {Moulin}, {Naumann-Godo}, {de Naurois}, {Nedbal}, {Nekrassov}, {Nolan},
  {Ohm}, {Olive}, {de O{\~n}a Wilhelmi}, {Orford}, {Osborne}, {Ostrowski},
  {Panter}, {Pedaletti}, {Pelletier}, {Petrucci}, {Pita}, {P{\"u}hlhofer},
  {Punch}, {Quirrenbach}, {Raubenheimer}, {Raue}, {Rayner}, {Renaud}, {Rieger},
  {Ripken}, {Rob}, {Rosier-Lees}, {Rowell}, {Rudak}, {Ruppel}, {Sahakian},
  {Santangelo}, {Schlickeiser}, {Sch{\"o}ck}, {Schr{\"o}der}, {Schwanke},
  {Schwarzburg}, {Schwemmer}, {Shalchi}, {Sol}, {Spangler}, {Stawarz},
  {Steenkamp}, {Stegmann}, {Superina}, {Tam}, {Tavernet}, {Terrier}, {van
  Eldik}, {Vasileiadis}, {Venter}, {Vialle}, {Vincent}, {Vivier}, {V{\"o}lk},
  {Volpe}, {Wagner}, {Ward}, {Zdziarski}, \& {Zech}}]{0152HESS}
{Aharonian}, F., {Akhperjanian}, A.~G., {Barres de Almeida}, U., {et~al.}
  2008{\natexlab{a}}, \aap, 481, L103

\bibitem[{{Aharonian} {et~al.}(2008{\natexlab{b}}){Aharonian}, {Akhperjanian},
  {Barres de Almeida}, {Bazer-Bachi}, {Behera}, {Beilicke}, {Benbow},
  {Bernl{\"o}hr}, {Boisson}, {Bolz}, {Borrel}, {Braun}, {Brion}, {Brown},
  {B{\"u}hler}, {Bulik}, {B{\"u}sching}, {Boutelier}, {Carrigan}, {Chadwick},
  {Chounet}, {Clapson}, {Coignet}, {Cornils}, {Costamante}, {Dalton},
  {Degrange}, {Dickinson}, {Djannati-Ata{\"i}}, {Domainko}, {O'C.~Drury},
  {Dubois}, {Dubus}, {Dyks}, {Egberts}, {Emmanoulopoulos}, {Espigat},
  {Farnier}, {Feinstein}, {Fiasson}, {F{\"o}rster}, {Fontaine}, {Funk},
  {F{\"u}{\ss}ling}, {Gallant}, {Giebels}, {Glicenstein}, {Gl{\"u}ck}, {Goret},
  {Hadjichristidis}, {Hauser}, {Hauser}, {Heinzelmann}, {Henri}, {Hermann},
  {Hinton}, {Hoffmann}, {Hofmann}, {Holleran}, {Hoppe}, {Horns},
  {Jacholkowska}, {de Jager}, {Jung}, {Katarzy{\'n}ski}, {Kendziorra},
  {Kerschhaggl}, {Kh{\'e}lifi}, {Keogh}, {Komin}, {Kosack}, {Lamanna},
  {Latham}, {Lemi{\`e}re}, {Lemoine-Goumard}, {Lenain}, {Lohse}, {Martin},
  {Martineau-Huynh}, {Marcowith}, {Masterson}, {Maurin}, {Maurin}, {McComb},
  {Moderski}, {Moulin}, {de Naurois}, {Nedbal}, {Nolan}, {Ohm}, {Olive}, {de
  O{\~n}a Wilhelmi}, {Orford}, {Osborne}, {Ostrowski}, {Panter}, {Pedaletti},
  {Pelletier}, {Petrucci}, {Pita}, {P{\"u}hlhofer}, {Punch}, {Ranchon},
  {Raubenheimer}, {Raue}, {Rayner}, {Renaud}, {Ripken}, {Rob}, {Rolland},
  {Rosier-Lees}, {Rowell}, {Rudak}, {Ruppel}, {Sahakian}, {Santangelo},
  {Schlickeiser}, {Sch{\"o}ck}, {Schr{\"o}der}, {Schwanke}, {Schwarzburg},
  {Schwemmer}, {Shalchi}, {Sol}, {Spangler}, {Stawarz}, {Steenkamp},
  {Stegmann}, {Superina}, {Tam}, {Tavernet}, {Terrier}, {van Eldik},
  {Vasileiadis}, {Venter}, {Vialle}, {Vincent}, {Vivier}, {V{\"o}lk}, {Volpe},
  {Wagner}, {Ward}, {Zdziarski}, \& {Zech}}]{HESSUL2008}
---. 2008{\natexlab{b}}, \aap, 478, 387

\bibitem[{Ahnen {et~al.}(2015)Ahnen, Ansoldi, Antonelli, Antoranz, Babic,
  Banerjee, Bangale, de~Almeida, Barrio, Bednarek, Bernardini, Biasuzzi,
  Biland, Blanch, Bonnefoy, Bonnoli, Borracci, Bretz, Carmona, Carosi,
  Chatterjee, Clavero, Colin, Colombo, Contreras, Cortina, Covino, Vela, Dazzi,
  Angelis, Lotto, de~O{\~n}a~Wilhelmi, Mendez, Pierro, Prester, Dorner, Doro,
  Einecke, Glawion, Elsaesser, Fern{\'a}ndez-Barral, Fidalgo, Fonseca, Font,
  Frantzen, Fruck, Galindo, L{\'o}pez, Garczarczyk, Terrats, Gaug, Giammaria,
  Godinovi{\'c}, Mu{\~n}oz, Guberman, Hahn, Hanabata, Hayashida, Herrera, Hose,
  Hrupec, Hughes, Idec, Kodani, Konno, Kubo, Kushida, Barbera, Lelas, Lindfors,
  Lombardi, L{\'o}pez, L{\'o}pez-Coto, L{\'o}pez-Oramas, Lorenz, Majumdar,
  Makariev, Mallot, Maneva, Manganaro, Mannheim, Maraschi, Marcote, Mariotti,
  Mart{\'\i}nez, Mazin, Menzel, Miranda, Mirzoyan, Moralejo, Moretti, Nakajima,
  Neustroev, Niedzwiecki, Rosillo, Nilsson, Nishijima, Noda, Orito,
  Overkemping, Paiano, Palacio, Palatiello, Paneque, Paoletti, Paredes,
  Paredes-Fortuny, Persic, Poutanen, Moroni, Prandini, Puljak, Rhode, Rib{\'o},
  Rico, Garcia, Saito, Satalecka, Schultz, Schweizer, Shore, Sillanp{\"a}{\"a},
  Sitarek, Snidaric, Sobczynska, Stamerra, Steinbring, Strzys, Takalo, Takami,
  Tavecchio, Temnikov, Terzi{\'c}, Tescaro, Teshima, Thaele, Torres, Toyama,
  Treves, Verguilov, Vovk, Ward, Will, Wu, Zanin, Collaboration, Ajello,
  Baldini, Barbiellini, Bastieri, Gonz{\'a}lez, Bellazzini, Bissaldi,
  Blandford, Bonino, Bregeon, Bruel, Buson, Caliandro, Cameron, Caragiulo,
  Caraveo, Cavazzuti, Chiang, Chiaro, Ciprini, D'Ammando, de~Palma, Desiante,
  Venere, Dom{\'\i}nguez, Fusco, Gargano, Gasparrini, Giglietto, Giordano,
  Giroletti, Grenier, Guiriec, Hays, Hewitt, Jogler, Kuss, Larsson, Li, Li,
  Longo, Loparco, Lovellette, Lubrano, Maldera, Mayer, Mazziotta, McEnery,
  Mirabal, Mizuno, Monzani, Morselli, Moskalenko, Nuss, Ojha, Ohsugi, Omodei,
  Orlando, Perkins, Pesce-Rollins, Piron, Pivato, Porter, Raino, Rando,
  Razzano, Reimer, Reimer, Sgro, Siskind, Spada, Spandre, Spinelli, Tajima,
  Takahashi, Thayer, Thompson, Troja, Wood, Collaboration, Balokovic,
  Berdyugin, Carraminana, Carrasco, Chavushyan, Ramazani, Feige, Haarto,
  Haeusner, Hovatta, Kania, Klamt, L{\"a}hteenm{\"a}ki, Leon-Tavares, Lorey,
  Pacciani, Porras, Recillas, Reinthal, Tornikoski, Wolfert, \&
  Zottmann}]{1441magic}
Ahnen, M.~L., Ansoldi, S., Antonelli, L.~A., {et~al.} 2015, \apjl, 815, L23

\bibitem[{{Ajello} {et~al.}(2014){Ajello}, {Romani}, {Gasparrini}, {Shaw},
  {Bolmer}, {Cotter}, {Finke}, {Greiner}, {Healey}, {King}, {Max-Moerbeck},
  {Michelson}, {Potter}, {Rau}, {Readhead}, {Richards}, \& {Schady}}]{Ajello14}
{Ajello}, M., {Romani}, R.~W., {Gasparrini}, D., {et~al.} 2014, \apj, 780, 73

\bibitem[{{Albert} {et~al.}(2008){Albert}, {Aliu}, {Anderhub}, {Antonelli},
  {Antoranz}, {Backes}, {Baixeras}, {Barrio}, {Bartko}, {Bastieri}, {Becker},
  {Bednarek}, {Berger}, {Bernardini}, {Bigongiari}, {Biland}, {Bock},
  {Bonnoli}, {Bordas}, {Bosch-Ramon}, {Bretz}, {Britvitch}, {Camara},
  {Carmona}, {Chilingarian}, {Commichau}, {Contreras}, {Cortina}, {Costado},
  {Covino}, {Curtef}, {Dazzi}, {De Angelis}, {Cea del Pozo}, {de los Reyes},
  {De Lotto}, {De Maria}, {De Sabata}, {Mendez}, {Dominguez}, {Dorner}, {Doro},
  {Errando}, {Fagiolini}, {Ferenc}, {Fern{\'a}ndez}, {Firpo}, {Fonseca},
  {Font}, {Galante}, {Garc{\'{\i}}a L{\'o}pez}, {Garczarczyk}, {Gaug},
  {Goebel}, {Hayashida}, {Herrero}, {H{\"o}hne}, {Hose}, {Hsu}, {Huber},
  {Jogler}, {Kneiske}, {Kranich}, {La Barbera}, {Laille}, {Leonardo},
  {Lindfors}, {Lombardi}, {Longo}, {L{\'o}pez}, {Lorenz}, {Majumdar}, {Maneva},
  {Mankuzhiyil}, {Mannheim}, {Maraschi}, {Mariotti}, {Mart{\'{\i}}nez},
  {Mazin}, {Meucci}, {Meyer}, {Miranda}, {Mirzoyan}, {Mizobuchi}, {Moles},
  {Moralejo}, {Nieto}, {Nilsson}, {Ninkovic}, {Otte}, {Oya}, {Panniello},
  {Paoletti}, {Paredes}, {Pasanen}, {Pascoli}, {Pauss}, {Pegna},
  {Perez-Torres}, {Persic}, {Peruzzo}, {Piccioli}, {Prada}, {Prandini},
  {Puchades}, {Raymers}, {Rhode}, {Rib{\'o}}, {Rico}, {Rissi}, {Robert},
  {R{\"u}gamer}, {Saggion}, {Saito}, {Salvati}, {Sanchez-Conde}, {Sartori},
  {Satalecka}, {Scalzotto}, {Scapin}, {Schmitt}, {Schweizer}, {Shayduk},
  {Shinozaki}, {Shore}, {Sidro}, {Sierpowska-Bartosik}, {Sillanp{\"a}{\"a}},
  {Sobczynska}, {Spanier}, {Stamerra}, {Stark}, {Takalo}, {Tavecchio},
  {Temnikov}, {Tescaro}, {Teshima}, {Tluczykont}, {Torres}, {Turini}, {Vankov},
  {Venturini}, {Vitale}, {Wagner}, {Wittek}, {Zabalza}, {Zandanel}, {Zanin}, \&
  {Zapatero}}]{3C279Magic}
{Albert}, J., {Aliu}, E., {Anderhub}, H., {et~al.} 2008, Science, 320, 1752

\bibitem[{{Aleksi{\'c}} {et~al.}(2011{\natexlab{a}}){Aleksi{\'c}}, {Antonelli},
  {Antoranz}, {Backes}, {Baixeras}, {Barrio}, {Bastieri}, {Becerra
  Gonz{\'a}lez}, {Bednarek}, {Berdyugin}, {Berger}, {Bernardini}, {Biland},
  {Blanch}, {Bock}, {Bonnoli}, {Bordas}, {Borla Tridon}, {Bosch-Ramon}, {Bose},
  {Braun}, {Bretz}, {Britzger}, {Camara}, {Carmona}, {Carosi}, {Colin},
  {Commichau}, {Contreras}, {Cortina}, {Costado}, {Covino}, {Dazzi}, {De
  Angelis}, {De Cea del Pozo}, {De los Reyes}, {De Lotto}, {De Maria}, {De
  Sabata}, {Delgado Mendez}, {Doert}, {Dom{\'{\i}}nguez}, {Dominis Prester},
  {Dorner}, {Doro}, {Elsaesser}, {Errando}, {Ferenc}, {Fonseca}, {Font},
  {Garc{\'{\i}}a L{\'o}pez}, {Garczarczyk}, {Gaug}, {Godinovic}, {Hadasch},
  {Herrero}, {Hildebrand}, {H{\"o}hne-M{\"o}nch}, {Hose}, {Hrupec}, {Hsu},
  {Jogler}, {Klepser}, {Kr{\"a}henb{\"u}hl}, {Kranich}, {La Barbera}, {Laille},
  {Leonardo}, {Lindfors}, {Lombardi}, {Longo}, {L{\'o}pez}, {Lorenz},
  {Majumdar}, {Maneva}, {Mankuzhiyil}, {Mannheim}, {Maraschi}, {Mariotti},
  {Mart{\'{\i}}nez}, {Mazin}, {Meucci}, {Miranda}, {Mirzoyan}, {Miyamoto},
  {Mold{\'o}n}, {Moles}, {Moralejo}, {Nieto}, {Nilsson}, {Ninkovic}, {Orito},
  {Oya}, {Paiano}, {Paoletti}, {Paredes}, {Partini}, {Pasanen}, {Pascoli},
  {Pauss}, {Pegna}, {Perez-Torres}, {Persic}, {Peruzzo}, {Prada}, {Prandini},
  {Puchades}, {Puljak}, {Reichardt}, {Rhode}, {Rib{\'o}}, {Rico}, {Rissi},
  {R{\"u}gamer}, {Saggion}, {Saito}, {Salvati}, {S{\'a}nchez-Conde},
  {Satalecka}, {Scalzotto}, {Scapin}, {Schultz}, {Schweizer}, {Shayduk},
  {Shore}, {Sierpowska-Bartosik}, {Sillanp{\"a}{\"a}}, {Sitarek}, {Sobczynska},
  {Spanier}, {Spiro}, {Stamerra}, {Steinke}, {Struebig}, {Suric}, {Takalo},
  {Tavecchio}, {Temnikov}, {Terzic}, {Tescaro}, {Teshima}, {Tibolla}, {Torres},
  {Vankov}, {Wagner}, {Weitzel}, {Zabalza}, {Zandanel}, \& {Zanin}}]{MAGICUL}
{Aleksi{\'c}}, J., {Antonelli}, L.~A., {Antoranz}, P., {et~al.}
  2011{\natexlab{a}}, \apj, 729, 115

\bibitem[{{Aleksi{\'c}} {et~al.}(2011{\natexlab{b}}){Aleksi{\'c}}, {Antonelli},
  {Antoranz}, {Backes}, {Barrio}, {Bastieri}, {Becerra Gonz{\'a}lez},
  {Bednarek}, {Berdyugin}, {Berger}, {Bernardini}, {Biland}, {Blanch}, {Bock},
  {Boller}, {Bonnoli}, {Borla Tridon}, {Braun}, {Bretz}, {Ca{\~n}ellas},
  {Carmona}, {Carosi}, {Colin}, {Colombo}, {Contreras}, {Cortina}, {Cossio},
  {Covino}, {Dazzi}, {De Angelis}, {De Cea del Pozo}, {De Lotto}, {Delgado
  Mendez}, {Diago Ortega}, {Doert}, {Dom{\'{\i}}nguez}, {Dominis Prester},
  {Dorner}, {Doro}, {Elsaesser}, {Ferenc}, {Fonseca}, {Font}, {Fruck},
  {Garc{\'{\i}}a L{\'o}pez}, {Garczarczyk}, {Garrido}, {Giavitto},
  {Godinovi{\'c}}, {Hadasch}, {H{\"a}fner}, {Herrero}, {Hildebrand},
  {H{\"o}hne-M{\"o}nch}, {Hose}, {Hrupec}, {Huber}, {Jogler}, {Klepser},
  {Kr{\"a}henb{\"u}hl}, {Krause}, {La Barbera}, {Lelas}, {Leonardo},
  {Lindfors}, {Lombardi}, {L{\'o}pez}, {Lorenz}, {Makariev}, {Maneva},
  {Mankuzhiyil}, {Mannheim}, {Maraschi}, {Mariotti}, {Mart{\'{\i}}nez},
  {Mazin}, {Meucci}, {Miranda}, {Mirzoyan}, {Miyamoto}, {Mold{\'o}n},
  {Moralejo}, {Nieto}, {Nilsson}, {Orito}, {Oya}, {Paneque}, {Paoletti},
  {Pardo}, {Paredes}, {Partini}, {Pasanen}, {Pauss}, {Perez-Torres}, {Persic},
  {Peruzzo}, {Pilia}, {Pochon}, {Prada}, {Prada Moroni}, {Prandini}, {Puljak},
  {Reichardt}, {Reinthal}, {Rhode}, {Rib{\'o}}, {Rico}, {R{\"u}gamer},
  {Saggion}, {Saito}, {Saito}, {Salvati}, {Satalecka}, {Scalzotto}, {Scapin},
  {Schultz}, {Schweizer}, {Shayduk}, {Shore}, {Sillanp{\"a}{\"a}}, {Sitarek},
  {Sobczynska}, {Spanier}, {Spiro}, {Stamerra}, {Steinke}, {Storz}, {Strah},
  {Suri{\'c}}, {Takalo}, {Tavecchio}, {Temnikov}, {Terzi{\'c}}, {Tescaro},
  {Teshima}, {Thom}, {Tibolla}, {Torres}, {Treves}, {Vankov}, {Vogler},
  {Wagner}, {Weitzel}, {Zabalza}, {Zandanel}, {Zanin}, {MAGIC Collaboration},
  {Tanaka}, {Wood}, \& {Buson}}]{1222Magic}
---. 2011{\natexlab{b}}, \apjl, 730, L8

\bibitem[{{Aleksi{\'c}} {et~al.}(2012){Aleksi{\'c}}, {Alvarez}, {Antonelli},
  {Antoranz}, {Asensio}, {Backes}, {Barrio}, {Bastieri}, {Becerra
  Gonz{\'a}lez}, {Bednarek}, {Berdyugin}, {Berger}, {Bernardini}, {Biland},
  {Blanch}, {Bock}, {Boller}, {Bonnoli}, {Borla Tridon}, {Braun}, {Bretz},
  {Ca{\~n}ellas}, {Carmona}, {Carosi}, {Colin}, {Colombo}, {Contreras},
  {Cortina}, {Cossio}, {Covino}, {Dazzi}, {de Angelis}, {de Caneva}, {de Cea
  Del Pozo}, {de Lotto}, {Delgado Mendez}, {Diago Ortega}, {Doert},
  {Dom{\'{\i}}nguez}, {Dominis Prester}, {Dorner}, {Doro}, {Elsaesser},
  {Ferenc}, {Fonseca}, {Font}, {Fruck}, {Garc{\'{\i}}a L{\'o}pez},
  {Garczarczyk}, {Garrido}, {Giavitto}, {Godinovi{\'c}}, {Hadasch},
  {H{\"a}fner}, {Herrero}, {Hildebrand}, {H{\"o}hne-M{\"o}nch}, {Hose},
  {Hrupec}, {Huber}, {Jogler}, {Kellermann}, {Klepser}, {Kr{\"a}henb{\"u}hl},
  {Krause}, {La Barbera}, {Lelas}, {Leonardo}, {Lindfors}, {Lombardi},
  {L{\'o}pez}, {L{\'o}pez-Oramas}, {Lorenz}, {Makariev}, {Maneva},
  {Mankuzhiyil}, {Mannheim}, {Maraschi}, {Mariotti}, {Mart{\'{\i}}nez},
  {Mazin}, {Meucci}, {Miranda}, {Mirzoyan}, {Miyamoto}, {Mold{\'o}n},
  {Moralejo}, {Munar-Adrover}, {Nieto}, {Nilsson}, {Orito}, {Oya}, {Paneque},
  {Paoletti}, {Pardo}, {Paredes}, {Partini}, {Pasanen}, {Pauss},
  {Perez-Torres}, {Persic}, {Peruzzo}, {Pilia}, {Pochon}, {Prada}, {Prada
  Moroni}, {Prandini}, {Puljak}, {Reichardt}, {Reinthal}, {Rhode}, {Rib{\'o}},
  {Rico}, {R{\"u}gamer}, {Saggion}, {Saito}, {Saito}, {Salvati}, {Satalecka},
  {Scalzotto}, {Scapin}, {Schultz}, {Schweizer}, {Shayduk}, {Shore},
  {Sillanp{\"a}{\"a}}, {Sitarek}, {Snidaric}, {Sobczynska}, {Spanier}, {Spiro},
  {Stamatescu}, {Stamerra}, {Steinke}, {Storz}, {Strah}, {Suri{\'c}}, {Takalo},
  {Takami}, {Tavecchio}, {Temnikov}, {Terzi{\'c}}, {Tescaro}, {Teshima},
  {Tibolla}, {Torres}, {Treves}, {Uellenbeck}, {Vankov}, {Vogler}, {Wagner},
  {Weitzel}, {Zabalza}, {Zandanel}, {Zanin}, {Kadenius}, {Weidinger}, \&
  {Buson}}]{2247magic}
{Aleksi{\'c}}, J., {Alvarez}, E.~A., {Antonelli}, L.~A., {et~al.} 2012, \aap,
  539, A118

\bibitem[{{Aleksi{\'c}} {et~al.}(2014{\natexlab{a}}){Aleksi{\'c}}, {Ansoldi},
  {Antonelli}, {Antoranz}, {Babic}, {Bangale}, {Barres de Almeida}, {Barrio},
  {Becerra Gonz{\'a}lez}, {Bednarek}, {Bernardini}, {Biland}, {Blanch},
  {Bonnefoy}, {Bonnoli}, {Borracci}, {Bretz}, {Carmona}, {Carosi}, {Carreto
  Fidalgo}, {Colin}, {Colombo}, {Contreras}, {Cortina}, {Covino}, {da Vela},
  {Dazzi}, {de Angelis}, {de Caneva}, {de Lotto}, {Delgado Mendez}, {Doert},
  {Dom{\'{\i}}nguez}, {Dominis Prester}, {Dorner}, {Doro}, {Einecke},
  {Eisenacher}, {Elsaesser}, {Farina}, {Ferenc}, {Fonseca}, {Font}, {Frantzen},
  {Fruck}, {Garc{\'{\i}}a L{\'o}pez}, {Garczarczyk}, {Garrido Terrats}, {Gaug},
  {Godinovi{\'c}}, {Gonz{\'a}lez Mu{\~n}oz}, {Gozzini}, {Hadasch}, {Hayashida},
  {Herrera}, {Herrero}, {Hildebrand}, {Hose}, {Hrupec}, {Idec}, {Kadenius},
  {Kellermann}, {Kodani}, {Konno}, {Krause}, {Kubo}, {Kushida}, {La Barbera},
  {Lelas}, {Lewandowska}, {Lindfors}, {Lombardi}, {L{\'o}pez},
  {L{\'o}pez-Coto}, {L{\'o}pez-Oramas}, {Lorenz}, {Lozano}, {Makariev},
  {Mallot}, {Maneva}, {Mankuzhiyil}, {Mannheim}, {Maraschi}, {Marcote},
  {Mariotti}, {Mart{\'{\i}}nez}, {Mazin}, {Menzel}, {Meucci}, {Miranda},
  {Mirzoyan}, {Moralejo}, {Munar-Adrover}, {Nakajima}, {Niedzwiecki},
  {Nilsson}, {Nishijima}, {Noda}, {Nowak}, {Orito}, {Overkemping}, {Paiano},
  {Palatiello}, {Paneque}, {Paoletti}, {Paredes}, {Paredes-Fortuny}, {Partini},
  {Persic}, {Prada}, {Prada Moroni}, {Prandini}, {Preziuso}, {Puljak},
  {Reinthal}, {Rhode}, {Rib{\'o}}, {Rico}, {Rodriguez Garcia}, {R{\"u}gamer},
  {Saggion}, {Saito}, {Saito}, {Satalecka}, {Scalzotto}, {Scapin}, {Schultz},
  {Schweizer}, {Shore}, {Sillanp{\"a}{\"a}}, {Sitarek}, {Snidaric},
  {Sobczynska}, {Spanier}, {Stamatescu}, {Stamerra}, {Steinbring}, {Storz},
  {Strzys}, {Sun}, {Suri{\'c}}, {Takalo}, {Takami}, {Tavecchio}, {Temnikov},
  {Terzi{\'c}}, {Tescaro}, {Teshima}, {Thaele}, {Tibolla}, {Torres}, {Toyama},
  {Treves}, {Uellenbeck}, {Vogler}, {Wagner}, {Zandanel}, {Zanin}, \& {MAGIC
  Collaboration}}]{Magic2001}
{Aleksi{\'c}}, J., {Ansoldi}, S., {Antonelli}, L.~A., {et~al.}
  2014{\natexlab{a}}, \aap, 572, A121

\bibitem[{{Aleksi{\'c}} {et~al.}(2014{\natexlab{b}}){Aleksi{\'c}}, {Ansoldi},
  {Antonelli}, {Antoranz}, {Babic}, {Bangale}, {Barres de Almeida}, {Barrio},
  {Becerra Gonz{\'a}lez}, {Bednarek}, {Bernardini}, {Biland}, {Blanch},
  {Bonnefoy}, {Bonnoli}, {Borracci}, {Bretz}, {Carmona}, {Carosi}, {Carreto
  Fidalgo}, {Colin}, {Colombo}, {Contreras}, {Cortina}, {Covino}, {Da Vela},
  {Dazzi}, {De Angelis}, {De Caneva}, {De Lotto}, {Delgado Mendez}, {Doert},
  {Dom{\'{\i}}nguez}, {Dominis Prester}, {Dorner}, {Doro}, {Einecke},
  {Eisenacher}, {Elsaesser}, {Farina}, {Ferenc}, {Fonseca}, {Font}, {Frantzen},
  {Fruck}, {Garc{\'{\i}}a L{\'o}pez}, {Garczarczyk}, {Garrido Terrats}, {Gaug},
  {Godinovi{\'c}}, {Gonz{\'a}lez Mu{\~n}oz}, {Gozzini}, {Hadasch}, {Hayashida},
  {Herrera}, {Herrero}, {Hildebrand}, {Hose}, {Hrupec}, {Idec}, {Kadenius},
  {Kellermann}, {Kodani}, {Konno}, {Krause}, {Kubo}, {Kushida}, {La Barbera},
  {Lelas}, {Lewandowska}, {Lindfors}, {Lombardi}, {L{\'o}pez},
  {L{\'o}pez-Coto}, {L{\'o}pez-Oramas}, {Lorenz}, {Lozano}, {Makariev},
  {Mallot}, {Maneva}, {Mankuzhiyil}, {Mannheim}, {Maraschi}, {Marcote},
  {Mariotti}, {Mart{\'{\i}}nez}, {Mazin}, {Menzel}, {Meucci}, {Miranda},
  {Mirzoyan}, {Moralejo}, {Munar-Adrover}, {Nakajima}, {Niedzwiecki},
  {Nilsson}, {Nishijima}, {Noda}, {Nowak}, {Orito}, {Overkemping}, {Paiano},
  {Palatiello}, {Paneque}, {Paoletti}, {Paredes}, {Paredes-Fortuny}, {Partini},
  {Persic}, {Prada}, {Prada Moroni}, {Prandini}, {Preziuso}, {Puljak},
  {Reinthal}, {Rhode}, {Rib{\'o}}, {Rico}, {Rodriguez Garcia}, {R{\"u}gamer},
  {Saggion}, {Saito}, {Saito}, {Satalecka}, {Scalzotto}, {Scapin}, {Schultz},
  {Schweizer}, {Shore}, {Sillanp{\"a}{\"a}}, {Sitarek}, {Snidaric},
  {Sobczynska}, {Spanier}, {Stamatescu}, {Stamerra}, {Steinbring}, {Storz},
  {Strzys}, {Sun}, {Suri{\'c}}, {Takalo}, {Takami}, {Tavecchio}, {Temnikov},
  {Terzi{\'c}}, {Tescaro}, {Teshima}, {Thaele}, {Tibolla}, {Torres}, {Toyama},
  {Treves}, {Uellenbeck}, {Vogler}, {Wagner}, {Zandanel}, \&
  {Zanin}}]{1510Magic}
---. 2014{\natexlab{b}}, \aap, 569, A46

\bibitem[{{Aleksi{\'c}} {et~al.}(2014{\natexlab{c}}){Aleksi{\'c}}, {Ansoldi},
  {Antonelli}, {Antoranz}, {Babic}, {Bangale}, {Barres de Almeida}, {Barrio},
  {Becerra Gonz{\'a}lez}, {Bednarek}, {Berger}, {Bernardini}, {Biland},
  {Blanch}, {Bock}, {Bonnefoy}, {Bonnoli}, {Borracci}, {Bretz}, {Carmona},
  {Carosi}, {Carreto Fidalgo}, {Colin}, {Colombo}, {Contreras}, {Cortina},
  {Covino}, {Da Vela}, {Dazzi}, {De Angelis}, {De Caneva}, {De Lotto}, {Delgado
  Mendez}, {Doert}, {Dom{\'{\i}}nguez}, {Dominis Prester}, {Dorner}, {Doro},
  {Einecke}, {Eisenacher}, {Elsaesser}, {Farina}, {Ferenc}, {Fonseca}, {Font},
  {Frantzen}, {Fruck}, {Garc{\'{\i}}a L{\'o}pez}, {Garczarczyk}, {Garrido
  Terrats}, {Gaug}, {Giavitto}, {Godinovi{\'c}}, {Gonz{\'a}lez Mu{\~n}oz},
  {Gozzini}, {Hadasch}, {Herrero}, {Hildebrand}, {Hose}, {Hrupec}, {Idec},
  {Kadenius}, {Kellermann}, {Knoetig}, {Kodani}, {Konno}, {Krause}, {Kubo},
  {Kushida}, {La Barbera}, {Lelas}, {Lewandowska}, {Lindfors}, {Lombardi},
  {L{\'o}pez}, {L{\'o}pez-Coto}, {L{\'o}pez-Oramas}, {Lorenz}, {Lozano},
  {Makariev}, {Mallot}, {Maneva}, {Mankuzhiyil}, {Mannheim}, {Maraschi},
  {Marcote}, {Mariotti}, {Mart{\'{\i}}nez}, {Mazin}, {Menzel}, {Meucci},
  {Miranda}, {Mirzoyan}, {Moralejo}, {Munar-Adrover}, {Nakajima},
  {Niedzwiecki}, {Nilsson}, {Nishijima}, {Nowak}, {Orito}, {Overkemping},
  {Paiano}, {Palatiello}, {Paneque}, {Paoletti}, {Paredes}, {Paredes-Fortuny},
  {Partini}, {Persic}, {Prada}, {Prada Moroni}, {Prandini}, {Preziuso},
  {Puljak}, {Reinthal}, {Rhode}, {Rib{\'o}}, {Rico}, {Rodriguez Garcia},
  {R{\"u}gamer}, {Saggion}, {Saito}, {Saito}, {Salvati}, {Satalecka},
  {Scalzotto}, {Scapin}, {Schultz}, {Schweizer}, {Shore}, {Sillanp{\"a}{\"a}},
  {Sitarek}, {Snidaric}, {Sobczynska}, {Spanier}, {Stamatescu}, {Stamerra},
  {Steinbring}, {Storz}, {Sun}, {Suri{\'c}}, {Takalo}, {Takami}, {Tavecchio},
  {Temnikov}, {Terzi{\'c}}, {Tescaro}, {Teshima}, {Thaele}, {Tibolla},
  {Torres}, {Toyama}, {Treves}, {Vogler}, {Wagner}, {Zandanel}, \&
  {Zanin}}]{3C279Magicbis}
---. 2014{\natexlab{c}}, \aap, 567, A41

\bibitem[{{Aleksi{\'c}} {et~al.}(2015{\natexlab{a}}){Aleksi{\'c}}, {Ansoldi},
  {Antonelli}, {Antoranz}, {Babic}, {Bangale}, {Barres de Almeida}, {Barrio},
  {Becerra Gonz{\'a}lez}, {Bednarek}, {Berger}, {Bernardini}, {Biland},
  {Blanch}, {Bonnefoy}, {Bonnoli}, {Borracci}, {Bretz}, {Carmona}, {Carosi},
  {Carreto Fidalgo}, {Colin}, {Colombo}, {Contreras}, {Cortina}, {Covino}, {da
  Vela}, {Dazzi}, {de Angelis}, {de Caneva}, {de Lotto}, {Delgado Mendez},
  {Doert}, {Dom{\'{\i}}nguez}, {Dominis Prester}, {Dorner}, {Doro}, {Einecke},
  {Eisenacher}, {Elsaesser}, {Farina}, {Ferenc}, {Fonseca}, {Font}, {Frantzen},
  {Fruck}, {Garc{\'{\i}}a L{\'o}pez}, {Garczarczyk}, {Garrido Terrats}, {Gaug},
  {Godinovi{\'c}}, {Gonz{\'a}lez Mu{\~n}oz}, {Gozzini}, {Hadasch}, {Hayashida},
  {Herrera}, {Herrero}, {Hildebrand}, {Hose}, {Hrupec}, {Idec}, {Kadenius},
  {Kellermann}, {Kodani}, {Konno}, {Krause}, {Kubo}, {Kushida}, {La Barbera},
  {Lelas}, {Lewandowska}, {Lindfors}, {Lombardi}, {L{\'o}pez},
  {L{\'o}pez-Coto}, {L{\'o}pez-Oramas}, {Lorenz}, {Lozano}, {Makariev},
  {Mallot}, {Maneva}, {Mankuzhiyil}, {Mannheim}, {Maraschi}, {Marcote},
  {Mariotti}, {Mart{\'{\i}}nez}, {Mazin}, {Menzel}, {Meucci}, {Miranda},
  {Mirzoyan}, {Moralejo}, {Munar-Adrover}, {Nakajima}, {Niedzwiecki},
  {Nilsson}, {Nishijima}, {Noda}, {Nowak}, {Orito}, {Overkemping}, {Paiano},
  {Palatiello}, {Paneque}, {Paoletti}, {Paredes}, {Paredes-Fortuny}, {Partini},
  {Persic}, {Prada}, {Moroni}, {Prandini}, {Preziuso}, {Puljak}, {Reinthal},
  {Rhode}, {Rib{\'o}}, {Rico}, {Rodriguez Garcia}, {R{\"u}gamer}, {Saggion},
  {Saito}, {Saito}, {Satalecka}, {Scalzotto}, {Scapin}, {Schultz}, {Schweizer},
  {Shore}, {Sillanp{\"a}{\"a}}, {Sitarek}, {Snidaric}, {Sobczynska}, {Spanier},
  {Stamatescu}, {Stamerra}, {Steinbring}, {Storz}, {Sun}, {Suri{\'c}},
  {Takalo}, {Takami}, {Tavecchio}, {Temnikov}, {Terzi{\'c}}, {Tescaro},
  {Teshima}, {Thaele}, {Tibolla}, {Torres}, {Toyama}, {Treves}, {Uellenbeck},
  {Vogler}, {Wagner}, {Zandanel}, {Zanin}, {MAGIC Collaboration}, {Tronconi},
  {Buson}, \& {Borghese}}]{1ES0033MAGIC}
---. 2015{\natexlab{a}}, \mnras, 446, 217

\bibitem[{{Aleksi{\'c}} {et~al.}(2015{\natexlab{b}}){Aleksi{\'c}}, {Ansoldi},
  {Antonelli}, {Antoranz}, {Babic}, {Bangale}, {Barrio}, {Becerra
  Gonz{\'a}lez}, {Bednarek}, {Bernardini}, {Biasuzzi}, {Biland}, {Blanch},
  {Bonnefoy}, {Bonnoli}, {Borracci}, {Bretz}, {Carmona}, {Carosi}, {Colin},
  {Colombo}, {Contreras}, {Cortina}, {Covino}, {Da Vela}, {Dazzi}, {De
  Angelis}, {De Caneva}, {De Lotto}, {de O{\~n}a Wilhelmi}, {Delgado Mendez},
  {Doert}, {Dominis Prester}, {Dorner}, {Doro}, {Einecke}, {Eisenacher},
  {Elsaesser}, {Fonseca}, {Font}, {Frantzen}, {Fruck}, {Galindo},
  {Garc{\'{\i}}a L{\'o}pez}, {Garczarczyk}, {Garrido Terrats}, {Gaug},
  {Godinovi{\'c}}, {Gonz{\'a}lez Mu{\~n}oz}, {Gozzini}, {Hadasch}, {Hanabata},
  {Hayashida}, {Herrera}, {Hildebrand}, {Hose}, {Hrupec}, {Idec}, {Kadenius},
  {Kellermann}, {Kodani}, {Konno}, {Krause}, {Kubo}, {Kushida}, {La Barbera},
  {Lelas}, {Lewandowska}, {Lindfors}, {Lombardi}, {L{\'o}pez},
  {L{\'o}pez-Coto}, {L{\'o}pez-Oramas}, {Lorenz}, {Lozano}, {Makariev},
  {Mallot}, {Maneva}, {Mankuzhiyil}, {Mannheim}, {Maraschi}, {Marcote},
  {Mariotti}, {Mart{\'{\i}}nez}, {Mazin}, {Menzel}, {Miranda}, {Mirzoyan},
  {Moralejo}, {Munar-Adrover}, {Nakajima}, {Niedzwiecki}, {Nilsson},
  {Nishijima}, {Noda}, {Nowak}, {Orito}, {Overkemping}, {Paiano}, {Palatiello},
  {Paneque}, {Paoletti}, {Paredes}, {Paredes-Fortuny}, {Persic}, {Prada
  Moroni}, {Prandini}, {Preziuso}, {Puljak}, {Reinthal}, {Rhode}, {Rib{\'o}},
  {Rico}, {Rodriguez Garcia}, {R{\"u}gamer}, {Saggion}, {Saito}, {Saito},
  {Satalecka}, {Scalzotto}, {Scapin}, {Schultz}, {Schweizer}, {Shore},
  {Sillanp{\"a}{\"a}}, {Sitarek}, {Snidaric}, {Sobczynska}, {Spanier},
  {Stamatescu}, {Stamerra}, {Steinbring}, {Storz}, {Strzys}, {Takalo},
  {Takami}, {Tavecchio}, {Temnikov}, {Terzi{\'c}}, {Tescaro}, {Teshima},
  {Thaele}, {Tibolla}, {Torres}, {Toyama}, {Treves}, {Uellenbeck}, {Vogler},
  {Wagner}, {Zanin}, {Horns}, {Mart{\'{\i}}n}, \& {Meyer}}]{Crabmagic}
---. 2015{\natexlab{b}}, Journal of High Energy Astrophysics, 5, 30

\bibitem[{{Aliu} {et~al.}(2012){Aliu}, {Archambault}, {Arlen}, {Aune},
  {Beilicke}, {Benbow}, {B{\"o}ttcher}, {Bouvier}, {Buckley}, {Bugaev},
  {Cesarini}, {Ciupik}, {Collins-Hughes}, {Connolly}, {Cui}, {Dickherber},
  {Duke}, {Dumm}, {Errando}, {Falcone}, {Federici}, {Feng}, {Finley},
  {Finnegan}, {Fortson}, {Furniss}, {Galante}, {Gall}, {Godambe}, {Griffin},
  {Grube}, {Gyuk}, {Hanna}, {Holder}, {Huan}, {Kaaret}, {Karlsson}, {Khassen},
  {Kieda}, {Krawczynski}, {Krennrich}, {Lee}, {Madhavan}, {Maier}, {Majumdar},
  {McArthur}, {McCann}, {Moriarty}, {Mukherjee}, {Nelson}, {O'Faol{\'a}in de
  Bhr{\'o}ithe}, {Ong}, {Orr}, {Otte}, {Park}, {Perkins}, {Pichel}, {Pohl},
  {Prokoph}, {Quinn}, {Ragan}, {Reyes}, {Reynolds}, {Roache}, {Saxon},
  {Sembroski}, {Staszak}, {Telezhinsky}, {Te{\v s}i{\'c}}, {Theiling},
  {Thibadeau}, {Tsurusaki}, {Varlotta}, {Vassiliev}, {Vincent}, {Vivier},
  {Wakely}, {Weekes}, {Weinstein}, {Welsing}, {Williams}, {Zitzer}, {VERITAS
  Collaboration}, {Fortin}, {Horan}, {Fumagalli}, {Kaplan}, \&
  {Prochaska}}]{Fermi6UL}
{Aliu}, E., {Archambault}, S., {Arlen}, T., {et~al.} 2012, \apj, 759, 102

\bibitem[{{Angel} \& {Stockman}(1980)}]{Angel80}
{Angel}, J.~R.~P., \& {Stockman}, H.~S. 1980, \araa, 18, 321

\bibitem[{{Archambault} {et~al.}(2013){Archambault}, {Arlen}, {Aune}, {Behera},
  {Beilicke}, {Benbow}, {Bird}, {Bouvier}, {Buckley}, {Bugaev}, {Byrum},
  {Cesarini}, {Ciupik}, {Connolly}, {Cui}, {Errando}, {Falcone}, {Federici},
  {Feng}, {Finley}, {Fortson}, {Furniss}, {Galante}, {Gall}, {Gillanders},
  {Griffin}, {Grube}, {Gyuk}, {Hanna}, {Holder}, {Hughes}, {Humensky},
  {Kaaret}, {Kertzman}, {Khassen}, {Kieda}, {Krawczynski}, {Krennrich},
  {Kumar}, {Lang}, {Madhavan}, {Maier}, {Majumdar}, {McArthur}, {McCann},
  {Millis}, {Moriarty}, {Mukherjee}, {O'Faol{\'a}in de Bhr{\'o}ithe}, {Ong},
  {Otte}, {Park}, {Perkins}, {Pohl}, {Popkow}, {Prokoph}, {Quinn}, {Ragan},
  {Reyes}, {Reynolds}, {Richards}, {Roache}, {Saxon}, {Sembroski}, {Smith},
  {Staszak}, {Telezhinsky}, {Theiling}, {Varlotta}, {Vassiliev}, {Vincent},
  {Wakely}, {Weekes}, {Weinstein}, {Welsing}, {Williams}, {Zitzer}, {VERITAS
  Collaboration}, {B{\"o}ttcher}, {Fegan}, {Fortin}, {Halpern}, {Kovalev},
  {Lister}, {Liu}, {Pushkarev}, \& {Smith}}]{verj0521}
{Archambault}, S., {Arlen}, T., {Aune}, T., {et~al.} 2013, \apj, 776, 69

\bibitem[{{Archambault} {et~al.}(2014){Archambault}, {Aune}, {Behera},
  {Beilicke}, {Benbow}, {Berger}, {Bird}, {Biteau}, {Bugaev}, {Byrum},
  {Cardenzana}, {Cerruti}, {Chen}, {Ciupik}, {Connolly}, {Cui}, {Dumm},
  {Errando}, {Falcone}, {Federici}, {Feng}, {Finley}, {Fleischhack}, {Fortson},
  {Furniss}, {Galante}, {Gillanders}, {Griffin}, {Griffiths}, {Grube}, {Gyuk},
  {Hanna}, {Holder}, {Hughes}, {Humensky}, {Johnson}, {Kaaret}, {Kertzman},
  {Khassen}, {Kieda}, {Krawczynski}, {Krennrich}, {Kumar}, {Lang}, {Madhavan},
  {Maier}, {McCann}, {Meagher}, {Moriarty}, {Mukherjee}, {Nieto},
  {O'Faol{\'a}in de Bhr{\'o}ithe}, {Ong}, {Otte}, {Park}, {Pohl}, {Popkow},
  {Prokoph}, {Quinn}, {Ragan}, {Rajotte}, {Reyes}, {Reynolds}, {Richards},
  {Roache}, {Sembroski}, {Shahinyan}, {Staszak}, {Telezhinsky}, {Tucci},
  {Tyler}, {Varlotta}, {Vassiliev}, {Vincent}, {Wakely}, {Weinstein},
  {Welsing}, {Wilhelm}, {Williams}, {VERITAS Collaboration}, {Ackermann},
  {Ajello}, {Albert}, {Baldini}, {Bastieri}, {Bellazzini}, {Bissaldi},
  {Bregeon}, {Buehler}, {Buson}, {Caliandro}, {Cameron}, {Caraveo},
  {Cavazzuti}, {Charles}, {Chiang}, {Ciprini}, {Claus}, {Cutini}, {D'Ammando},
  {de Angelis}, {de Palma}, {Dermer}, {Digel}, {Di Venere}, {Drell}, {Favuzzi},
  {Franckowiak}, {Fusco}, {Gargano}, {Gasparrini}, {Giglietto}, {Giordano},
  {Giroletti}, {Grenier}, {Guiriec}, {Jogler}, {Kuss}, {Larsson}, {Latronico},
  {Longo}, {Loparco}, {Lubrano}, {Madejski}, {Mayer}, {Mazziotta}, {Michelson},
  {Mizuno}, {Monzani}, {Morselli}, {Murgia}, {Nuss}, {Ohsugi}, {Ormes},
  {Paneque}, {Perkins}, {Piron}, {Pivato}, {Rain{\`o}}, {Razzano}, {Reimer},
  {Reimer}, {Ritz}, {Schaal}, {Sgr{\`o}}, {Siskind}, {Spinelli}, {Takahashi},
  {Tibaldo}, {Tinivella}, {Troja}, {Vianello}, {Werner}, {Wood}, \& {Fermi LAT
  Collaboration}}]{PKS1424Veritas2}
{Archambault}, S., {Aune}, T., {Behera}, B., {et~al.} 2014, \apjl, 785, L16

\bibitem[{{Archambault} {et~al.}(2015){Archambault}, {Archer}, {Beilicke},
  {Benbow}, {Bird}, {Biteau}, {Bouvier}, {Bugaev}, {Cardenzana}, {Cerruti},
  {Chen}, {Ciupik}, {Connolly}, {Cui}, {Dickinson}, {Dumm}, {Eisch}, {Errando},
  {Falcone}, {Feng}, {Finley}, {Fleischhack}, {Fortin}, {Fortson}, {Furniss},
  {Gillanders}, {Griffin}, {Griffiths}, {Grube}, {Gyuk}, {H{\aa}kansson},
  {Hanna}, {Holder}, {Humensky}, {Johnson}, {Kaaret}, {Kar}, {Kertzman},
  {Khassen}, {Kieda}, {Krause}, {Krennrich}, {Kumar}, {Lang}, {Maier},
  {McArthur}, {McCann}, {Meagher}, {Millis}, {Moriarty}, {Mukherjee}, {Nieto},
  {O'Faol{\'a}in de Bhr{\'o}ithe}, {Ong}, {Otte}, {Park}, {Pohl}, {Popkow},
  {Prokoph}, {Pueschel}, {Quinn}, {Ragan}, {Reyes}, {Reynolds}, {Richards},
  {Roache}, {Santander}, {Sembroski}, {Shahinyan}, {Smith}, {Staszak},
  {Telezhinsky}, {Tucci}, {Tyler}, {Varlotta}, {Vincent}, {Wakely},
  {Weinstein}, {Welsing}, {Wilhelm}, {Williams}, {Zitzer}, {Veritas
  Collaboration}, \& {Hughes}}]{1727Veritas}
{Archambault}, S., {Archer}, A., {Beilicke}, M., {et~al.} 2015, \apj, 808, 110

\bibitem[{{Atwood} {et~al.}(2009){Atwood}, {Abdo}, {Ackermann}, {Althouse},
  {Anderson}, {Axelsson}, {Baldini}, {Ballet}, {Band}, {Barbiellini}, \&
  et~al.}]{FermiLAT}
{Atwood}, W.~B., {Abdo}, A.~A., {Ackermann}, M., {et~al.} 2009, \apj, 697, 1071

\bibitem[{{Bade} {et~al.}(1994){Bade}, {Fink}, \& {Engels}}]{Bade94}
{Bade}, N., {Fink}, H.~H., \& {Engels}, D. 1994, \aap, 286, 381

\bibitem[{{Bauer} {et~al.}(2000){Bauer}, {Condon}, {Thuan}, \&
  {Broderick}}]{Bauer00}
{Bauer}, F.~E., {Condon}, J.~J., {Thuan}, T.~X., \& {Broderick}, J.~J. 2000,
  \apjs, 129, 547

\bibitem[{{Benbow}(2015)}]{wystan}
{Benbow}, W. 2015, to appear in Proceedings of the International Cosmic Ray
  Conference, ArXiv e-prints 1508.07251

\bibitem[{{Bird}(2015)}]{Bird15}
{Bird}, R. 2015, to appear in Proceedings of the International Cosmic Ray
  Conference, ArXiv e-prints 1508.07195

\bibitem[{{B{\"o}hringer} {et~al.}(2000){B{\"o}hringer}, {Voges}, {Huchra},
  {McLean}, {Giacconi}, {Rosati}, {Burg}, {Mader}, {Schuecker}, {Simi{\c c}},
  {Komossa}, {Reiprich}, {Retzlaff}, \& {Tr{\"u}mper}}]{Bohringer00}
{B{\"o}hringer}, H., {Voges}, W., {Huchra}, J.~P., {et~al.} 2000, \apjs, 129,
  435

\bibitem[{{Brinkmann} {et~al.}(1997){Brinkmann}, {Siebert}, {Feigelson},
  {Kollgaard}, {Laurent-Muehleisen}, {Reich}, {Fuerst}, {Reich}, {Voges},
  {Truemper}, \& {McMahon}}]{Brinkmann97}
{Brinkmann}, W., {Siebert}, J., {Feigelson}, E.~D., {et~al.} 1997, \aap, 323,
  739

\bibitem[{{Burbidge} \& {Kinman}(1966)}]{Burbidge66}
{Burbidge}, E.~M., \& {Kinman}, T.~D. 1966, \apj, 145, 654

\bibitem[{{Burbidge} {et~al.}(1977){Burbidge}, {Crowne}, \&
  {Smith}}]{Burbidge77}
{Burbidge}, G.~R., {Crowne}, A.~H., \& {Smith}, H.~E. 1977, \apjs, 33, 113

\bibitem[{{{\c S}ent{\"u}rk} {et~al.}(2013){{\c S}ent{\"u}rk}, {Errando},
  {B{\"o}ttcher}, \& {Mukherjee}}]{Gunes13}
{{\c S}ent{\"u}rk}, G.~D., {Errando}, M., {B{\"o}ttcher}, M., \& {Mukherjee},
  R. 2013, \apj, 764, 119

\bibitem[{{Cao} {et~al.}(1999){Cao}, {Wei}, \& {Hu}}]{Cao99}
{Cao}, L., {Wei}, J.-Y., \& {Hu}, J.-Y. 1999, \aaps, 135, 243

\bibitem[{{Carswell} {et~al.}(1974){Carswell}, {Strittmatter}, {Williams},
  {Kinman}, \& {Serkowski}}]{Carswell74}
{Carswell}, R.~F., {Strittmatter}, P.~A., {Williams}, R.~E., {Kinman}, T.~D.,
  \& {Serkowski}, K. 1974, \apjl, 190, L101

\bibitem[{{Chandra} {et~al.}(2012){Chandra}, {Baliyan}, {Ganesh}, \&
  {Joshi}}]{Chandra12}
{Chandra}, S., {Baliyan}, K.~S., {Ganesh}, S., \& {Joshi}, U.~C. 2012, \apj,
  746, 92

\bibitem[{{Cohen} {et~al.}(1987){Cohen}, {Smith}, {Junkkarinen}, \&
  {Burbidge}}]{Cohen87}
{Cohen}, R.~D., {Smith}, H.~E., {Junkkarinen}, V.~T., \& {Burbidge}, E.~M.
  1987, \apj, 318, 577

\bibitem[{{Cornwell} {et~al.}(1986){Cornwell}, {Saikia}, {Shastri}, {Feretti},
  {Giovannini}, {Parma}, \& {Salter}}]{Cornwell86}
{Cornwell}, T.~J., {Saikia}, D.~J., {Shastri}, P., {et~al.} 1986, Journal of
  Astrophysics and Astronomy, 7, 119

\bibitem[{{Cortina}(2013)}]{Magic1725}
{Cortina}, J. 2013, The Astronomer's Telegram, 5080, 1

\bibitem[{{Costamante}(2007)}]{Costamante07}
{Costamante}, L. 2007, \apss, 309, 487

\bibitem[{{Costamante} \& {Ghisellini}(2002)}]{CG}
{Costamante}, L., \& {Ghisellini}, G. 2002, \aap, 384, 56

\bibitem[{{Doert} \& {Errando}(2014)}]{DoertErrando}
{Doert}, M., \& {Errando}, M. 2014, \apj, 782, 41

\bibitem[{{Drinkwater} {et~al.}(1997){Drinkwater}, {Webster}, {Francis},
  {Condon}, {Ellison}, {Jauncey}, {Lovell}, {Peterson}, \& {Savage}}]{Drink97}
{Drinkwater}, M.~J., {Webster}, R.~L., {Francis}, P.~J., {et~al.} 1997, \mnras,
  284, 85

\bibitem[{{Eracleous} \& {Halpern}(2004)}]{Erac04}
{Eracleous}, M., \& {Halpern}, J.~P. 2004, \apjs, 150, 181

\bibitem[{{Errando}(2011)}]{Errando11}
{Errando}, M. 2011, International Cosmic Ray Conference, 8, 133

\bibitem[{{Falcone} {et~al.}(2004){Falcone}, {Bond}, {Boyle}, {Bradbury},
  {Buckley}, {Carter-Lewis}, {Celik}, {Cui}, {Daniel}, {D'Vali}, {de la Calle
  Perez}, {Duke}, {Fegan}, {Fegan}, {Finley}, {Fortson}, {Gaidos}, {Gammell},
  {Gibbs}, {Gillanders}, {Grube}, {Hall}, {Hall}, {Hanna}, {Hillas}, {Holder},
  {Horan}, {Jarvis}, {Kenny}, {Kertzman}, {Kieda}, {Kildea}, {Knapp}, {Kosack},
  {Krawczynski}, {Krennrich}, {Lang}, {LeBohec}, {Linton}, {Lloyd-Evans},
  {Milovanovic}, {Moriarty}, {Muller}, {Nagai}, {Nolan}, {Ong}, {Pallassini},
  {Petry}, {Pizlo}, {Power-Mooney}, {Quinn}, {Quinn}, {Ragan}, {Rebillot},
  {Reynolds}, {Rose}, {Schroedter}, {Sembroski}, {Swordy}, {Syson}, {Tyler},
  {Vassiliev}, {Wakely}, {Walker}, {Weekes}, \& {Zweerink}}]{Falcone04}
{Falcone}, A.~D., {Bond}, I.~H., {Boyle}, P.~J., {et~al.} 2004, \apj, 613, 710

\bibitem[{{Falomo}(1991)}]{Falomo91}
{Falomo}, R. 1991, \aj, 102, 1991

\bibitem[{{Falomo} \& {Kotilainen}(1999)}]{Falomo99}
{Falomo}, R., \& {Kotilainen}, J.~K. 1999, \aap, 352, 85

\bibitem[{{Fanaroff} \& {Riley}(1974)}]{FR74}
{Fanaroff}, B.~L., \& {Riley}, J.~M. 1974, \mnras, 167, 31P

\bibitem[{{Fischer} {et~al.}(1998){Fischer}, {Hasinger}, {Schwope}, {Brunner},
  {Boller}, {Tr{\"u}mper}, {Voges}, \& {Neizvestnyj}}]{Fischer98}
{Fischer}, J.-U., {Hasinger}, G., {Schwope}, A.~D., {et~al.} 1998,
  Astronomische Nachrichten, 319, 347

\bibitem[{{Fomin} {et~al.}(1994){Fomin}, {Stepanian}, {Lamb}, {Lewis}, {Punch},
  \& {Weekes}}]{Fomin94}
{Fomin}, V.~P., {Stepanian}, A.~A., {Lamb}, R.~C., {et~al.} 1994, Astroparticle
  Physics, 2, 137

\bibitem[{{Franceschini} {et~al.}(2008){Franceschini}, {Rodighiero}, \&
  {Vaccari}}]{Franceschini08}
{Franceschini}, A., {Rodighiero}, G., \& {Vaccari}, M. 2008, \aap, 487, 837

\bibitem[{{Fumagalli} {et~al.}(2012){Fumagalli}, {Dessauges-Zavadsky},
  {Furniss}, {Prochaska}, {Williams}, {Kaplan}, \& {Hogan}}]{Fuma12}
{Fumagalli}, M., {Dessauges-Zavadsky}, M., {Furniss}, A., {et~al.} 2012,
  \mnras, 424, 2276

\bibitem[{{Furniss} {et~al.}(2013){Furniss}, {Williams}, {Danforth},
  {Fumagalli}, {Prochaska}, {Primack}, {Urry}, {Stocke}, {Filippenko}, \&
  {Neely}}]{Furniss13}
{Furniss}, A., {Williams}, D.~A., {Danforth}, C., {et~al.} 2013, \apjl, 768,
  L31

\bibitem[{{Ghisellini} {et~al.}(2011){Ghisellini}, {Tagliaferri}, {Foschini},
  {Ghirlanda}, {Tavecchio}, {Della Ceca}, {Haardt}, {Volonteri}, \&
  {Gehrels}}]{Ghisellini11}
{Ghisellini}, G., {Tagliaferri}, G., {Foschini}, L., {et~al.} 2011, \mnras,
  411, 901

\bibitem[{{Giommi} {et~al.}(2005){Giommi}, {Piranomonte}, {Perri}, \&
  {Padovani}}]{Sedentary}
{Giommi}, P., {Piranomonte}, S., {Perri}, M., \& {Padovani}, P. 2005, \aap,
  434, 385

\bibitem[{{Giommi} {et~al.}(2012){Giommi}, {Polenta}, {L{\"a}hteenm{\"a}ki},
  {Thompson}, {Capalbi}, {Cutini}, {Gasparrini}, {Gonz{\'a}lez-Nuevo},
  {Le{\'o}n-Tavares}, {L{\'o}pez-Caniego}, {Mazziotta}, {Monte}, {Perri},
  {Rain{\`o}}, {Tosti}, {Tramacere}, {Verrecchia}, {Aller}, {Aller},
  {Angelakis}, {Bastieri}, {Berdyugin}, {Bonaldi}, {Bonavera}, {Burigana},
  {Burrows}, {Buson}, {Cavazzuti}, {Chincarini}, {Colafrancesco}, {Costamante},
  {Cuttaia}, {D'Ammando}, {de Zotti}, {Frailis}, {Fuhrmann}, {Galeotta},
  {Gargano}, {Gehrels}, {Giglietto}, {Giordano}, {Giroletti}, {Keih{\"a}nen},
  {King}, {Krichbaum}, {Lasenby}, {Lavonen}, {Lawrence}, {Leto}, {Lindfors},
  {Mandolesi}, {Massardi}, {Max-Moerbeck}, {Michelson}, {Mingaliev}, {Natoli},
  {Nestoras}, {Nieppola}, {Nilsson}, {Partridge}, {Pavlidou}, {Pearson},
  {Procopio}, {Rachen}, {Readhead}, {Reeves}, {Reimer}, {Reinthal},
  {Ricciardi}, {Richards}, {Riquelme}, {Saarinen}, {Sajina}, {Sandri},
  {Savolainen}, {Sievers}, {Sillanp{\"a}{\"a}}, {Sotnikova}, {Stevenson},
  {Tagliaferri}, {Takalo}, {Tammi}, {Tavagnacco}, {Terenzi}, {Toffolatti},
  {Tornikoski}, {Trigilio}, {Turunen}, {Umana}, {Ungerechts}, {Villa}, {Wu},
  {Zacchei}, {Zensus}, \& {Zhou}}]{Giommi12}
{Giommi}, P., {Polenta}, G., {L{\"a}hteenm{\"a}ki}, A., {et~al.} 2012, \aap,
  541, A160

\bibitem[{{Glikman} {et~al.}(2007){Glikman}, {Helfand}, {White}, {Becker},
  {Gregg}, \& {Lacy}}]{Glikman07}
{Glikman}, E., {Helfand}, D.~J., {White}, R.~L., {et~al.} 2007, \apj, 667, 673

\bibitem[{{Halpern} {et~al.}(1986){Halpern}, {Impey}, {Bothun}, {Tapia},
  {Skillman}, {Wilson}, \& {Meurs}}]{Halpern86}
{Halpern}, J.~P., {Impey}, C.~D., {Bothun}, G.~D., {et~al.} 1986, \apj, 302,
  711

\bibitem[{{Healey} {et~al.}(2008){Healey}, {Romani}, {Cotter}, {Michelson},
  {Schlafly}, {Readhead}, {Giommi}, {Chaty}, {Grenier}, \&
  {Weintraub}}]{Healey08}
{Healey}, S.~E., {Romani}, R.~W., {Cotter}, G., {et~al.} 2008, \apjs, 175, 97

\bibitem[{{Henstock} {et~al.}(1997){Henstock}, {Browne}, {Wilkinson}, \&
  {McMahon}}]{Henstock97}
{Henstock}, D.~R., {Browne}, I.~W.~A., {Wilkinson}, P.~N., \& {McMahon}, R.~G.
  1997, \mnras, 290, 380

\bibitem[{{Hewitt} \& {Burbidge}(1993)}]{Hewitt93}
{Hewitt}, A., \& {Burbidge}, G. 1993, \apjs, 87, 451

\bibitem[{{Hillas}(1985)}]{Hillas85}
{Hillas}, A.~M. 1985, International Cosmic Ray Conference, 3, 445

\bibitem[{{Hillas} {et~al.}(1998){Hillas}, {Akerlof}, {Biller}, {Buckley},
  {Carter-Lewis}, {Catanese}, {Cawley}, {Fegan}, {Finley}, {Gaidos},
  {Krennrich}, {Lamb}, {Lang}, {Mohanty}, {Punch}, {Reynolds}, {Rodgers},
  {Rose}, {Rovero}, {Schubnell}, {Sembroski}, {Vacanti}, {Weekes}, {West}, \&
  {Zweerink}}]{WhippleCrab}
{Hillas}, A.~M., {Akerlof}, C.~W., {Biller}, S.~D., {et~al.} 1998, \apj, 503,
  744

\bibitem[{{Holder}(2011)}]{Holder11}
{Holder}, J. 2011, International Cosmic Ray Conference, 12, 137

\bibitem[{{Holder}(2014{\natexlab{a}})}]{1222VERITAS}
---. 2014{\natexlab{a}}, The Astronomer's Telegram, 5981, 1

\bibitem[{{Holder}(2014{\natexlab{b}})}]{2243veritas}
---. 2014{\natexlab{b}}, The Astronomer's Telegram, 6849, 1

\bibitem[{{Holder} {et~al.}(2006){Holder}, {Atkins}, {Badran}, {Blaylock},
  {Bradbury}, {Buckley}, {Byrum}, {Carter-Lewis}, {Celik}, {Chow}, {Cogan},
  {Cui}, {Daniel}, {de la Calle Perez}, {Dowdall}, {Dowkontt}, {Duke},
  {Falcone}, {Fegan}, {Finley}, {Fortin}, {Fortson}, {Gibbs}, {Gillanders},
  {Glidewell}, {Grube}, {Gutierrez}, {Gyuk}, {Hall}, {Hanna}, {Hays}, {Horan},
  {Hughes}, {Humensky}, {Imran}, {Jung}, {Kaaret}, {Kenny}, {Kieda}, {Kildea},
  {Knapp}, {Krawczynski}, {Krennrich}, {Lang}, {LeBohec}, {Linton}, {Little},
  {Maier}, {Manseri}, {Milovanovic}, {Moriarty}, {Mukherjee}, {Ogden}, {Ong},
  {Petry}, {Perkins}, {Pizlo}, {Pohl}, {Quinn}, {Ragan}, {Reynolds}, {Roache},
  {Rose}, {Schroedter}, {Sembroski}, {Sleege}, {Steele}, {Swordy}, {Syson},
  {Toner}, {Valcarcel}, {Vassiliev}, {Wakely}, {Weekes}, {White}, {Williams},
  \& {Wagner}}]{Holder06}
{Holder}, J., {Atkins}, R.~W., {Badran}, H.~M., {et~al.} 2006, Astroparticle
  Physics, 25, 391

\bibitem[{{Horan} {et~al.}(2004){Horan}, {Badran}, {Bond}, {Boyle}, {Bradbury},
  {Buckley}, {Carter-Lewis}, {Catanese}, {Celik}, {Cui}, {Daniel}, {D'Vali},
  {de la Calle Perez}, {Duke}, {Falcone}, {Fegan}, {Fegan}, {Finley},
  {Fortson}, {Gaidos}, {Gammell}, {Gibbs}, {Gillanders}, {Grube}, {Hall},
  {Hall}, {Hanna}, {Hillas}, {Holder}, {Jarvis}, {Jordan}, {Kenny}, {Kertzman},
  {Kieda}, {Kildea}, {Knapp}, {Kosack}, {Krawczynski}, {Krennrich}, {Lang}, {Le
  Bohec}, {Linton}, {Lloyd-Evans}, {Milovanovic}, {Moriarty}, {Muller},
  {Nagai}, {Nolan}, {Ong}, {Pallassini}, {Petry}, {Power-Mooney}, {Quinn},
  {Quinn}, {Ragan}, {Rebillot}, {Reynolds}, {Rose}, {Schroedter}, {Sembroski},
  {Swordy}, {Syson}, {Vassiliev}, {Wakely}, {Walker}, {Weekes}, \&
  {Zweerink}}]{WhippleUL}
{Horan}, D., {Badran}, H.~M., {Bond}, I.~H., {et~al.} 2004, \apj, 603, 51

\bibitem[{{Impey} \& {Neugebauer}(1988)}]{Impey88}
{Impey}, C.~D., \& {Neugebauer}, G. 1988, \aj, 95, 307

\bibitem[{{Jannuzi} {et~al.}(1993){Jannuzi}, {Smith}, \& {Elston}}]{Jannuzi93}
{Jannuzi}, B.~T., {Smith}, P.~S., \& {Elston}, R. 1993, \apjs, 85, 265

\bibitem[{{Kieda}(2013)}]{Kieda13}
{Kieda}, D.~B. 2013, Proceedings of the 33rd International Cosmic Ray
  Conference

\bibitem[{{Komossa} {et~al.}(2006){Komossa}, {Voges}, {Xu}, {Mathur}, {Adorf},
  {Lemson}, {Duschl}, \& {Grupe}}]{Komossa06}
{Komossa}, S., {Voges}, W., {Xu}, D., {et~al.} 2006, \aj, 132, 531

\bibitem[{{Konigl}(1981)}]{Konigl81}
{Konigl}, A. 1981, \apj, 243, 700

\bibitem[{{Kraus} \& {Gearhart}(1975)}]{Kraus75}
{Kraus}, J.~D., \& {Gearhart}, M.~R. 1975, \aj, 80, 1

\bibitem[{{Landoni} {et~al.}(2014){Landoni}, {Falomo}, {Treves}, \&
  {Sbarufatti}}]{Landoni14}
{Landoni}, M., {Falomo}, R., {Treves}, A., \& {Sbarufatti}, B. 2014, \aap, 570,
  A126

\bibitem[{{Laurent-Muehleisen} {et~al.}(1998){Laurent-Muehleisen}, {Kollgaard},
  {Ciardullo}, {Feigelson}, {Brinkmann}, \& {Siebert}}]{Laurent98}
{Laurent-Muehleisen}, S.~A., {Kollgaard}, R.~I., {Ciardullo}, R., {et~al.}
  1998, \apjs, 118, 127

\bibitem[{{Laurent-Muehleisen} {et~al.}(1999){Laurent-Muehleisen}, {Kollgaard},
  {Feigelson}, {Brinkmann}, \& {Siebert}}]{LaurentIBL}
{Laurent-Muehleisen}, S.~A., {Kollgaard}, R.~I., {Feigelson}, E.~D.,
  {Brinkmann}, W., \& {Siebert}, J. 1999, \apj, 525, 127

\bibitem[{{Lavaux} \& {Hudson}(2011)}]{Lavaux11}
{Lavaux}, G., \& {Hudson}, M.~J. 2011, \mnras, 416, 2840

\bibitem[{{Lawrence} {et~al.}(1986){Lawrence}, {Pearson}, {Readhead}, \&
  {Unwin}}]{Lawrence86}
{Lawrence}, C.~R., {Pearson}, T.~J., {Readhead}, A.~C.~S., \& {Unwin}, S.~C.
  1986, \aj, 91, 494

\bibitem[{{Li} {et~al.}(2010){Li}, {Xie}, {Yi}, {Chen}, \& {Dai}}]{Li10}
{Li}, H.~Z., {Xie}, G.~Z., {Yi}, T.~F., {Chen}, L.~E., \& {Dai}, H. 2010, \apj,
  709, 1407

\bibitem[{{Li} \& {Ma}(1983)}]{LiMaa}
{Li}, T.-P., \& {Ma}, Y.-Q. 1983, \apj, 272, 317

\bibitem[{{Lister} {et~al.}(2011){Lister}, {Aller}, {Aller}, {Hovatta},
  {Kellermann}, {Kovalev}, {Meyer}, {Pushkarev}, {Ros}, {MOJAVE Collaboration},
  {Ackermann}, {Antolini}, {Baldini}, {Ballet}, {Barbiellini}, {Bastieri},
  {Bechtol}, {Bellazzini}, {Berenji}, {Blandford}, {Bloom}, {Boeck},
  {Bonamente}, {Borgland}, {Bregeon}, {Brigida}, {Bruel}, {Buehler}, {Buson},
  {Caliandro}, {Cameron}, {Caraveo}, {Casandjian}, {Cavazzuti}, {Cecchi},
  {Chang}, {Charles}, {Chekhtman}, {Cheung}, {Chiang}, {Ciprini}, {Claus},
  {Cohen-Tanugi}, {Conrad}, {Cutini}, {de Palma}, {Dermer}, {Silva}, {Drell},
  {Drlica-Wagner}, {Favuzzi}, {Fegan}, {Ferrara}, {Finke}, {Focke}, {Fortin},
  {Fukazawa}, {Fusco}, {Gargano}, {Gasparrini}, {Gehrels}, {Germani},
  {Giglietto}, {Giordano}, {Giroletti}, {Glanzman}, {Godfrey}, {Grenier},
  {Guiriec}, {Hadasch}, {Hayashida}, {Hays}, {Horan}, {Hughes},
  {J{\'o}hannesson}, {Johnson}, {Kadler}, {Katagiri}, {Kataoka},
  {Kn{\"o}dlseder}, {Kuss}, {Lande}, {Longo}, {Loparco}, {Lott}, {Lovellette},
  {Lubrano}, {Madejski}, {Mazziotta}, {McConville}, {McEnery}, {Mehault},
  {Michelson}, {Mizuno}, {Monte}, {Monzani}, {Morselli}, {Moskalenko},
  {Murgia}, {Naumann-Godo}, {Nishino}, {Nolan}, {Norris}, {Nuss}, {Ohno},
  {Ohsugi}, {Okumura}, {Omodei}, {Orlando}, {Ozaki}, {Paneque}, {Parent},
  {Pesce-Rollins}, {Pierbattista}, {Piron}, {Pivato}, {Rain{\`o}}, {Readhead},
  {Reimer}, {Reimer}, {Richards}, {Ritz}, {Sadrozinski}, {Sgr{\`o}}, {Shaw},
  {Siskind}, {Spandre}, {Spinelli}, {Takahashi}, {Tanaka}, {Thayer}, {Thayer},
  {Thompson}, {Tosti}, {Tramacere}, {Troja}, {Usher}, {Vandenbroucke},
  {Vasileiou}, {Vianello}, {Vitale}, {Waite}, {Wang}, {Winer}, {Wood},
  {Zimmer}, \& {Fermi LAT Collaboration}}]{Lister11}
{Lister}, M.~L., {Aller}, M., {Aller}, H., {et~al.} 2011, \apj, 742, 27

\bibitem[{{Marcha} {et~al.}(1996){Marcha}, {Browne}, {Impey}, \&
  {Smith}}]{Marcha96}
{Marcha}, M.~J.~M., {Browne}, I.~W.~A., {Impey}, C.~D., \& {Smith}, P.~S. 1996,
  \mnras, 281, 425

\bibitem[{{Marziani} {et~al.}(1996){Marziani}, {Sulentic}, {Dultzin-Hacyan},
  {Calvani}, \& {Moles}}]{Marziani96}
{Marziani}, P., {Sulentic}, J.~W., {Dultzin-Hacyan}, D., {Calvani}, M., \&
  {Moles}, M. 1996, \apjs, 104, 37

\bibitem[{{Maselli} {et~al.}(2010){Maselli}, {Massaro}, {Nesci}, {Sclavi},
  {Rossi}, \& {Giommi}}]{Maselli10}
{Maselli}, A., {Massaro}, E., {Nesci}, R., {et~al.} 2010, \aap, 512, A74

\bibitem[{{Massaro} {et~al.}(2009){Massaro}, {Giommi}, {Leto}, {Marchegiani},
  {Maselli}, {Perri}, {Piranomonte}, \& {Sclavi}}]{Massaro09}
{Massaro}, E., {Giommi}, P., {Leto}, C., {et~al.} 2009, \aap, 495, 691

\bibitem[{{Massaro} {et~al.}(2003){Massaro}, {Giommi}, {Perri}, {Tagliaferri},
  {Nesci}, {Tosti}, {Ciprini}, {Montagni}, {Ravasio}, {Ghisellini}, {Frasca},
  {Marilli}, {Valentini}, {Kurtanidze}, \& {Nikolashvili}}]{Massaro03}
{Massaro}, E., {Giommi}, P., {Perri}, M., {et~al.} 2003, \aap, 399, 33

\bibitem[{{Massaro} {et~al.}(2012){Massaro}, {D'Abrusco}, {Tosti}, {Ajello},
  {Gasparrini}, {Grindlay}, \& {Smith}}]{Massaro12}
{Massaro}, F., {D'Abrusco}, R., {Tosti}, G., {et~al.} 2012, \apj, 750, 138

\bibitem[{{Meisner} \& {Romani}(2010)}]{Meisner10}
{Meisner}, A.~M., \& {Romani}, R.~W. 2010, \apj, 712, 14

\bibitem[{{Mirabal} {et~al.}(2012){Mirabal}, {Fr{\'{\i}}as-Martinez}, {Hassan},
  \& {Fr{\'{\i}}as-Martinez}}]{Mirabal12}
{Mirabal}, N., {Fr{\'{\i}}as-Martinez}, V., {Hassan}, T., \&
  {Fr{\'{\i}}as-Martinez}, E. 2012, \mnras, 424, L64

\bibitem[{{Mirzoyan}(2014{\natexlab{a}})}]{1136Magic}
{Mirzoyan}, R. 2014{\natexlab{a}}, The Astronomer's Telegram, 6062, 1

\bibitem[{{Mirzoyan}(2014{\natexlab{b}})}]{S30218}
---. 2014{\natexlab{b}}, The Astronomer's Telegram, 6349, 1

\bibitem[{{Mirzoyan}(2014{\natexlab{c}})}]{Magic0847}
---. 2014{\natexlab{c}}, The Astronomer's Telegram, 5768, 1

\bibitem[{{M{\"u}cke} \& {Protheroe}(2001)}]{Mucke01}
{M{\"u}cke}, A., \& {Protheroe}, R.~J. 2001, Astroparticle Physics, 15, 121

\bibitem[{{Mukherjee}(2001)}]{Egret}
{Mukherjee}, R. 2001, in American Institute of Physics Conference Series, Vol.
  558, American Institute of Physics Conference Series, ed. F.~A. {Aharonian}
  \& H.~J. {V{\"o}lk}, 324--337

\bibitem[{{Muriel} {et~al.}(2015){Muriel}, {Donzelli}, {Rovero}, \&
  {Pichel}}]{Muriel15}
{Muriel}, H., {Donzelli}, C., {Rovero}, A.~C., \& {Pichel}, A. 2015, \aap, 574,
  A101

\bibitem[{{Nieppola} {et~al.}(2006){Nieppola}, {Tornikoski}, \&
  {Valtaoja}}]{Nieppola06}
{Nieppola}, E., {Tornikoski}, M., \& {Valtaoja}, E. 2006, \aap, 445, 441

\bibitem[{{Nilsson} {et~al.}(2003){Nilsson}, {Pursimo}, {Heidt}, {Takalo},
  {Sillanp{\"a}{\"a}}, \& {Brinkmann}}]{Nilsson03}
{Nilsson}, K., {Pursimo}, T., {Heidt}, J., {et~al.} 2003, \aap, 400, 95

\bibitem[{{Nolan} {et~al.}(2012){Nolan}, {Abdo}, {Ackermann}, {Ajello},
  {Allafort}, {Antolini}, {Atwood}, {Axelsson}, {Baldini}, {Ballet}, \&
  et~al.}]{2FGL}
{Nolan}, P.~L., {Abdo}, A.~A., {Ackermann}, M., {et~al.} 2012, \apjs, 199, 31

\bibitem[{{Osterbrock} \& {Dahari}(1983)}]{Osterbrock83}
{Osterbrock}, D.~E., \& {Dahari}, O. 1983, \apj, 273, 478

\bibitem[{{Padovani}(1992)}]{Padovani92}
{Padovani}, P. 1992, \aap, 256, 399

\bibitem[{{Padovani} {et~al.}(2002){Padovani}, {Costamante}, {Ghisellini},
  {Giommi}, \& {Perlman}}]{Padovani02}
{Padovani}, P., {Costamante}, L., {Ghisellini}, G., {Giommi}, P., \& {Perlman},
  E. 2002, \apj, 581, 895

\bibitem[{{Padovani} \& {Giommi}(1995{\natexlab{a}})}]{Padovani95b}
{Padovani}, P., \& {Giommi}, P. 1995{\natexlab{a}}, \mnras, 277, 1477

\bibitem[{{Padovani} \& {Giommi}(1995{\natexlab{b}})}]{Padovani95}
---. 1995{\natexlab{b}}, \apj, 444, 567

\bibitem[{{Perlman}(2000)}]{Perlman00}
{Perlman}, E.~S. 2000, in American Institute of Physics Conference Series, Vol.
  515, American Institute of Physics Conference Series, ed. B.~L. {Dingus},
  M.~H. {Salamon}, \& D.~B. {Kieda}, 53--65

\bibitem[{{Perlman} {et~al.}(1996){Perlman}, {Stocke}, {Schachter}, {Elvis},
  {Ellingson}, {Urry}, {Potter}, {Impey}, \& {Kolchinsky}}]{Perlman96}
{Perlman}, E.~S., {Stocke}, J.~T., {Schachter}, J.~F., {et~al.} 1996, \apjs,
  104, 251

\bibitem[{{Piranomonte} {et~al.}(2007){Piranomonte}, {Perri}, {Giommi},
  {Landt}, \& {Padovani}}]{Piranomonte07}
{Piranomonte}, S., {Perri}, M., {Giommi}, P., {Landt}, H., \& {Padovani}, P.
  2007, \aap, 470, 787

\bibitem[{{Piron}(2000)}]{Piron00}
{Piron}, F. 2000, PhD thesis, Universit{\'e} de Paris-Sud (Paris XI)

\bibitem[{{Plotkin} {et~al.}(2010){Plotkin}, {Anderson}, {Brandt},
  {Diamond-Stanic}, {Fan}, {Hall}, {Kimball}, {Richmond}, {Schneider},
  {Shemmer}, {Voges}, {York}, {Bahcall}, {Snedden}, {Bizyaev}, {Brewington},
  {Malanushenko}, {Malanushenko}, {Oravetz}, {Pan}, \& {Simmons}}]{Plotkin10}
{Plotkin}, R.~M., {Anderson}, S.~F., {Brandt}, W.~N., {et~al.} 2010, \aj, 139,
  390

\bibitem[{{Rani} {et~al.}(2011){Rani}, {Gupta}, {Bachev}, {Strigachev},
  {Semkov}, {D'Ammando}, {Wiita}, {Gurwell}, {Ovcharov}, {Mihov}, {Boeva}, \&
  {Peneva}}]{Rani11}
{Rani}, B., {Gupta}, A.~C., {Bachev}, R., {et~al.} 2011, \mnras, 417, 1881

\bibitem[{{Rau} {et~al.}(2012){Rau}, {Schady}, {Greiner}, {Salvato}, {Ajello},
  {Bottacini}, {Gehrels}, {Afonso}, {Elliott}, {Filgas}, {Kann}, {Klose},
  {Kr{\"u}hler}, {Nardini}, {Nicuesa Guelbenzu}, {Olivares E.}, {Rossi},
  {Sudilovsky}, {Updike}, \& {Hartmann}}]{Rau2012}
{Rau}, A., {Schady}, P., {Greiner}, J., {et~al.} 2012, \aap, 538, A26

\bibitem[{{Rolke} {et~al.}(2005){Rolke}, {L{\'o}pez}, \& {Conrad}}]{Rolke05}
{Rolke}, W.~A., {L{\'o}pez}, A.~M., \& {Conrad}, J. 2005, Nuclear Instruments
  and Methods in Physics Research A, 551, 493

\bibitem[{{Salamon} \& {Stecker}(1998)}]{Salamon98}
{Salamon}, M.~H., \& {Stecker}, F.~W. 1998, \apj, 493, 547

\bibitem[{{Sandrinelli} {et~al.}(2013){Sandrinelli}, {Treves}, {Falomo},
  {Farina}, {Foschini}, {Landoni}, \& {Sbarufatti}}]{Sandrinelli13}
{Sandrinelli}, A., {Treves}, A., {Falomo}, R., {et~al.} 2013, \aj, 146, 163

\bibitem[{{Sbarufatti} {et~al.}(2009){Sbarufatti}, {Treves}, {Decarli},
  {Veronesi}, {Kotilainen}, {Ciprini}, \& {Falomo}}]{Sbarufatti09}
{Sbarufatti}, B., {Treves}, A., {Decarli}, R., {et~al.} 2009, The Astronomer's
  Telegram, 2123, 1

\bibitem[{{Sbarufatti} {et~al.}(2005){Sbarufatti}, {Treves}, {Falomo}, {Heidt},
  {Kotilainen}, \& {Scarpa}}]{Sbarufatti05}
{Sbarufatti}, B., {Treves}, A., {Falomo}, R., {et~al.} 2005, \aj, 129, 559

\bibitem[{{Sbarufatti} {et~al.}(2006){Sbarufatti}, {Treves}, {Falomo}, {Heidt},
  {Kotilainen}, \& {Scarpa}}]{Sbarufatti06}
---. 2006, \aj, 132, 1

\bibitem[{{Seta} {et~al.}(2009){Seta}, {Isobe}, {Tashiro}, {Yaji}, {Arai},
  {Fukuhara}, {Kohno}, {Nakanishi}, {Sasada}, {Shimajiri}, {Tosaki}, {Uemura},
  {Anderhub}, {Antonelli}, {Antoranz}, {Backes}, {Baixeras}, {Balestra},
  {Barrio}, {Bastieri}, {Becerra Gonz{\'a}lez}, {Becker}, {Bednarek}, {Berger},
  {Bernardini}, {Biland}, {Bock}, {Bonnoli}, {Bordas}, {Borla Tridon},
  {Bosch-Ramon}, {Bose}, {Braun}, {Bretz}, {Britvitch}, {Camara}, {Carmona},
  {Commichau}, {Contreras}, {Cortina}, {Costado Dios}, {Covino}, {Curtef},
  {Dazzi}, {de Angelis}, {de Cea Del Pozo}, {de Los Reyes}, {de Lotto}, {de
  Maria}, {de Sabata}, {Delgado M{\'e}ndez}, {Dom{\'{\i}}nguez}, {Dorner},
  {Doro}, {Elsaesser}, {Errando}, {Ferenc}, {Fern{\'a}ndez}, {Firpo},
  {Fonseca}, {Font}, {Galante}, {Garc{\'{\i}}a L{\'o}pez}, {Garczarczyk},
  {Gaug}, {Goebel}, {Hadasch}, {Hayashida}, {Herrero}, {Hildebrand},
  {H{\"o}hne-M{\"o}nch}, {Hose}, {Hsu}, {Jogler}, {Kranich}, {La Barbera},
  {Laille}, {Leonardo}, {Lindfors}, {Lombardi}, {Longo}, {L{\'o}pez}, {Lorenz},
  {Majumdar}, {Maneva}, {Mankuzhiyil}, {Mannheim}, {Maraschi}, {Mariotti},
  {Mart{\'{\i}}nez}, {Mazin}, {Meucci}, {Meyer}, {Miguel Miranda}, {Mirzoyan},
  {Miyamoto}, {Mold{\'o}n}, {Moles}, {Moralejo}, {Nieto}, {Nilsson},
  {Ninkovic}, {Otte}, {Oya}, {Paoletti}, {Paredes}, {Pasanen}, {Pascoli},
  {Pauss}, {Pegna}, {Perez-Torres}, {Persic}, {Peruzzo}, {Prada}, {Prandini},
  {Puchades}, {Reichardt}, {Rhode}, {Rib{\'o}}, {Rico}, {Rissi}, {Robert},
  {R{\"u}gamer}, {Saggion}, {Saito}, {Salvati}, {S{\'a}nchez-Conde},
  {Satalecka}, {Scalzotto}, {Scapin}, {Schweizer}, {Shayduk}, {Shore}, {Sidro},
  {Sierpowska-Bartosik}, {Sillanp{\"a}{\"a}}, {Sitarek}, {Sobczynska},
  {Spanier}, {Stamerra}, {Stark Schneebeli}, {Takalo}, {Tavecchio}, {Temnikov},
  {Tescaro}, {Teshima}, {Tluczykont}, {Torres}, {Turini}, {Vankov}, {Wagner},
  {Wittek}, {Zabalza}, {Zandanel}, {Zanin}, \& {Zapatero}}]{OJ287}
{Seta}, H., {Isobe}, N., {Tashiro}, M.~S., {et~al.} 2009, \pasj, 61, 1011

\bibitem[{{Shaw} {et~al.}(2009){Shaw}, {Romani}, {Healey}, {Cotter},
  {Michelson}, \& {Readhead}}]{Shaw09}
{Shaw}, M.~S., {Romani}, R.~W., {Healey}, S.~E., {et~al.} 2009, \apj, 704, 477

\bibitem[{{Shaw} {et~al.}(2012){Shaw}, {Romani}, {Cotter}, {Healey},
  {Michelson}, {Readhead}, {Richards}, {Max-Moerbeck}, {King}, \&
  {Potter}}]{Shaw12}
{Shaw}, M.~S., {Romani}, R.~W., {Cotter}, G., {et~al.} 2012, \apj, 748, 49

\bibitem[{{Shaw} {et~al.}(2013){Shaw}, {Romani}, {Cotter}, {Healey},
  {Michelson}, {Readhead}, {Richards}, {Max-Moerbeck}, {King}, \&
  {Potter}}]{Shaw13}
---. 2013, \apj, 764, 135

\bibitem[{{Sikora} {et~al.}(1994){Sikora}, {Begelman}, \& {Rees}}]{Sikora94}
{Sikora}, M., {Begelman}, M.~C., \& {Rees}, M.~J. 1994, \apj, 421, 153

\bibitem[{{Smith} {et~al.}(1976){Smith}, {Smith}, \& {Spinrad}}]{Smith76}
{Smith}, H.~E., {Smith}, E.~O., \& {Spinrad}, H. 1976, \pasp, 88, 621

\bibitem[{{Sol} {et~al.}(2013){Sol}, {Zech}, {Boisson}, {Barres de Almeida},
  {Biteau}, {Contreras}, {Giebels}, {Hassan}, {Inoue}, {Katarzy{\'n}ski},
  {Krawczynski}, {Mirabal}, {Poutanen}, {Rieger}, {Totani}, {Benbow},
  {Cerruti}, {Errando}, {Fallon}, {de Gouveia Dal Pino}, {Hinton}, {Inoue},
  {Lenain}, {Neronov}, {Takahashi}, {Takami}, {White}, \& {CTA
  Consortium}}]{Sol13}
{Sol}, H., {Zech}, A., {Boisson}, C., {et~al.} 2013, Astroparticle Physics, 43,
  215

\bibitem[{{Sowards-Emmerd} {et~al.}(2005){Sowards-Emmerd}, {Romani},
  {Michelson}, {Healey}, \& {Nolan}}]{Sowards05}
{Sowards-Emmerd}, D., {Romani}, R.~W., {Michelson}, P.~F., {Healey}, S.~E., \&
  {Nolan}, P.~L. 2005, \apj, 626, 95

\bibitem[{{Stecker} {et~al.}(1996){Stecker}, {de Jager}, \& {Salamon}}]{SDS}
{Stecker}, F.~W., {de Jager}, O.~C., \& {Salamon}, M.~H. 1996, \apjl, 473, L75

\bibitem[{{Stickel} {et~al.}(1988){Stickel}, {Fried}, \& {Kuehr}}]{Stickel88}
{Stickel}, M., {Fried}, J.~W., \& {Kuehr}, H. 1988, \aap, 191, L16

\bibitem[{{Stickel} {et~al.}(1989){Stickel}, {Fried}, \& {Kuehr}}]{Stickel89}
---. 1989, \aaps, 80, 103

\bibitem[{{Stickel} {et~al.}(1991){Stickel}, {Padovani}, {Urry}, {Fried}, \&
  {Kuehr}}]{Stickel91}
{Stickel}, M., {Padovani}, P., {Urry}, C.~M., {Fried}, J.~W., \& {Kuehr}, H.
  1991, \apj, 374, 431

\bibitem[{{Stocke} {et~al.}(1991){Stocke}, {Morris}, {Gioia}, {Maccacaro},
  {Schild}, {Wolter}, {Fleming}, \& {Henry}}]{Stocke91}
{Stocke}, J.~T., {Morris}, S.~L., {Gioia}, I.~M., {et~al.} 1991, \apjs, 76, 813

\bibitem[{{Taylor} {et~al.}(2011){Taylor}, {Vovk}, \& {Neronov}}]{Taylor11}
{Taylor}, A.~M., {Vovk}, I., \& {Neronov}, A. 2011, \aap, 529, A144

\bibitem[{{Thompson} {et~al.}(1990){Thompson}, {Djorgovski}, \& {de
  Carvalho}}]{Thompson90}
{Thompson}, D.~J., {Djorgovski}, S., \& {de Carvalho}, R. 1990, \pasp, 102,
  1235

\bibitem[{{Turriziani} {et~al.}(2007){Turriziani}, {Cavazzuti}, \&
  {Giommi}}]{ROXA}
{Turriziani}, S., {Cavazzuti}, E., \& {Giommi}, P. 2007, \aap, 472, 699

\bibitem[{{Urry} \& {Padovani}(1995)}]{Urry95}
{Urry}, C.~M., \& {Padovani}, P. 1995, \pasp, 107, 803

\bibitem[{{White} {et~al.}(1988){White}, {Jauncey}, {Wright}, {Batty},
  {Savage}, {Peterson}, \& {Gulkis}}]{White88}
{White}, G.~L., {Jauncey}, D.~L., {Wright}, A.~E., {et~al.} 1988, \apj, 327,
  561

\bibitem[{{White} {et~al.}(2000){White}, {Becker}, {Gregg},
  {Laurent-Muehleisen}, {Brotherton}, {Impey}, {Petry}, {Foltz}, {Chaffee},
  {Richards}, {Oegerle}, {Helfand}, {McMahon}, \& {Cabanela}}]{White00}
{White}, R.~L., {Becker}, R.~H., {Gregg}, M.~D., {et~al.} 2000, \apjs, 126, 133

\bibitem[{{Williams}(2005)}]{milagroul}
{Williams}, D.~A. 2005, in American Institute of Physics Conference Series,
  Vol. 745, High Energy Gamma-Ray Astronomy, ed. F.~A. {Aharonian}, H.~J.
  {V{\"o}lk}, \& D.~{Horns}, 499--504

\end{thebibliography}

\newpage
\begin{appendix}
\section{New redshift estimates}
\label{appendixredshift}

\begin{figure}[ht!]
\centering
\makebox[\textwidth][c]{\includegraphics[width=1.1\textwidth]{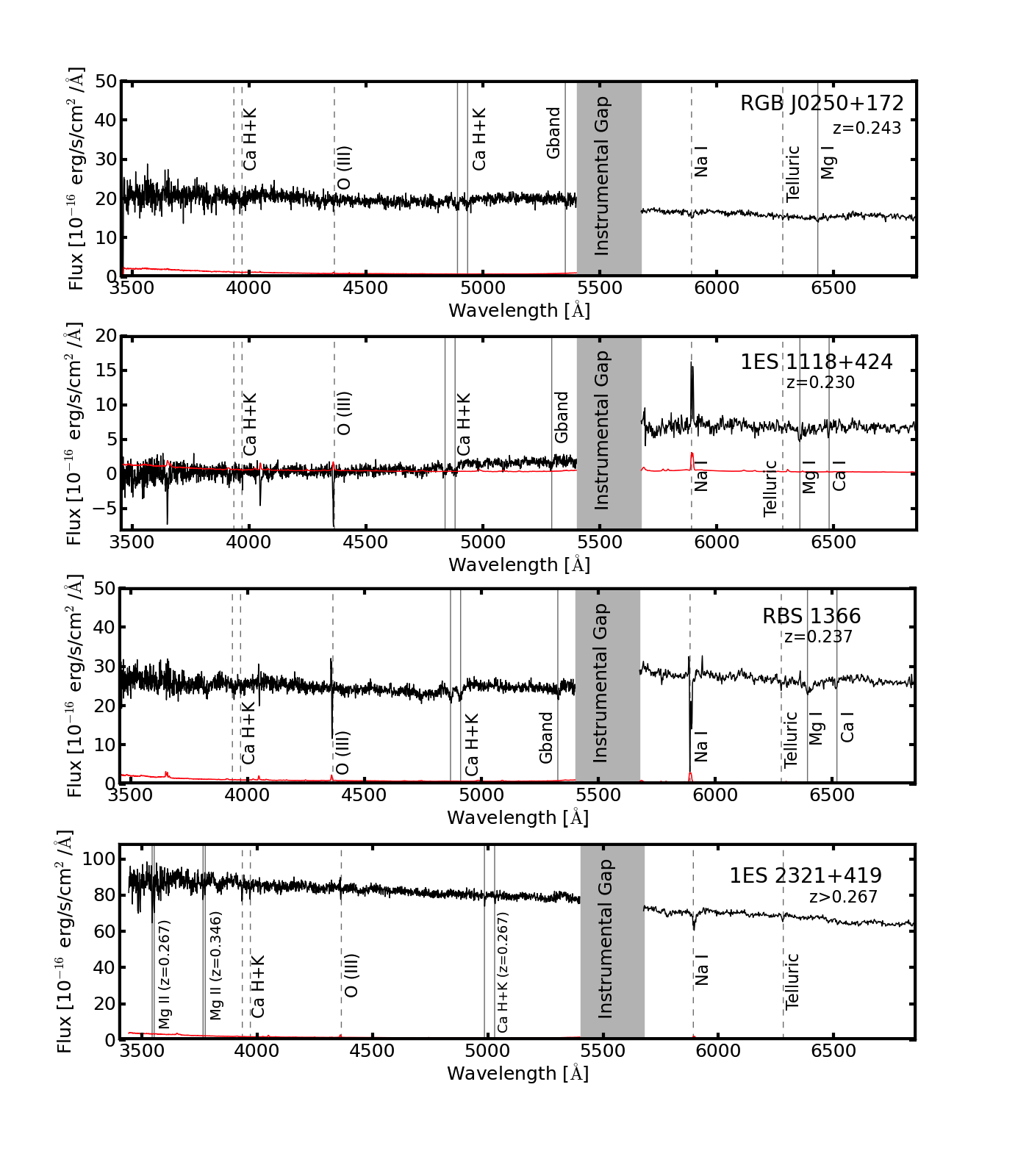}}
\caption{Spectra shown from top to bottom: RGB J0250+172 (August 15, 2010), 1ES 1118+424 (February 14, 2013), RBS 1366 (May 30, 2014), 1ES 2321+419 (October 28, 2014). Dashed lines indicate telluric and Galactic features. Red lines indicate the error array for each observation; some are not visible due to high S/N. Solid gray lines indicate features at non-zero redshift.}
\label{fig:one}
\end{figure}

In this appendix, we present optical spectra of 18 blazars taken at Lick Observatory in an attempt to spectroscopically measure their redshift. The sources selected for spectroscopy were selected independently from the sources selected in the main text, so the overlap is not complete. The spectra we show were taken between August, 2010 and October, 2014. During the observations at Lick we often observed the same source more than once. Duplicate observations are noted in Table \ref{newredshift}, but we only show one spectrum per source in the figures. Four of these spectra (shown in Figure \ref{fig:one}) have host galaxy features which allow an accurate redshift determination, and are discussed in the main text.

\begin{deluxetable*}{c c c c c c c c }
\tablecolumns{8} 
\tablewidth{0pc} 
\tablecaption{Blazars observed at Lick Observatory using the Shane 3m Kast Spectrograph \label{newredshift}}\\
\tablehead{
Target Name$^{a}$&Obs. Date& Obs. Date & Exposure & Signal to noise$^b$&Standard Star&z&Fig.\ $^c$\\
& [UT] & [MJD - 50000] & [s] & & & & 
}
\hline
\startdata
RBS 0082 & August 13, 2010  & 5421 & 3600 & 50, 77& BD+28 4211& &\ref{fig:two} \\
1ES 0033+595 & August 22, 2012  & 6161 & 3600 &20, 88&BD+28 4211& & \\
1ES 0033+595 & December 4, 2013 & 6630 & 3600 &2.8, 34&G191B2B& & \ref{fig:two}\\
1RXS J0045.3+2127 &August 22, 2012  & 6161 &1800&56, 106 &BD+28 4211& & \\ 
1RXS J0045.3+2127 &October 28, 2014 & 6958 &1800&57, 121&Feige 110& &\ref{fig:two}\\
RGB J0250+172  & August 15, 2010 & 5423 &5400&21, 44&BD+28 4211&0.243 &\ref{fig:one}\\
1ES 0446+449& February 14, 2013 & 6337 &3600 &n/a, 65&  HZ2 &  & \ref{fig:three}\\ 
RGB J0505+612 & February 14, 2013 & 6337 &\phn 900 &0.4, 4.3&  HZ2 &  &\ref{fig:three}\\
2FGL J0540.4+5822&October 28, 2014 & 6958 &3600 &14, 46&G19B2B& &\ref{fig:three} \\
B2 0912+29 & April 7, 2013& 6389 & 3600 &57, 179& Feige 34 &  & \\
B2 0912+29 & January 4, 2013& 6299 & 3600 &81, 144& Feige 34 &  &\ref{fig:four} \\
B2 0912+29 & December 4, 2013& 6630 & 3600 &48, 102 & G191B2B &  &\\
RBS 0929 & April 7, 2013 & 6389 & 3600 &13, 33& Feige 34 &  &\ref{fig:four}\\
RGB J1037+571 & February 14, 2013 & 6337  & 3600 &54, 128& Feige 34 &  &\ref{fig:four}\\
1ES 1118+424 & February 14, 2013   & 6337 & 3600 &1.5, 22& Feige 34& 0.230 &\ref{fig:one}\\
PG 1246+586 & April 7, 2013 & 6389 & 3600 &80, 229& GD 153 &  &\\
PG 1246+586 & May 29, 2014 & 6806 & 3600 &86, 175& HZ44 & &\\
PG 1246+586 & May 30, 2014 & 6807 & 3600 &80, 203& HZ44 & &\ref{fig:five}\\
RBS 1366 & April 7, 2013 & 6389  &\phn 900 &7, 34& BD+33 2642 & & \\
RBS 1366 & May 30, 2014 & 6807 & 3600 &26, 72& BD+33 2642 & 0.237 &\ref{fig:one}\\
1RXS J144053.2+061013 &January 4, 2013 & 6299 &3800&13, 53& BD+33 2642& & \ref{fig:five}\\ 
RGB J1725+118 & June 12, 2013  & 6455 & 3600 &48, 129& BD+33 2642 & &\\
RGB J1725+118 & May 29, 2014 & 6806  & 3600 &81, 189& BD+33 2642 & &\\
RGB J1725+118 & May 30, 2014 & 6807 & 3600 &111, 205& BD+33 2642 & &\ref{fig:five}\\
RGB J1903+556 & June 13, 2013 & 6456 & 1800 &18, 46& BD+28 4211 & &\ref{fig:six}\\
BZB J2243+2021 &August 13, 2010 & 5421 &3600&92, 167& BD+28 4211&&\ref{fig:six}\\
1ES 2321+419 &August 22, 2012 & 6161 &3000&20, 33&BD+28 4211&&\\
1ES 2321+419 &October 28, 2014& 6958 &3600&52, 121& Feige 110&$>0.267$&\ref{fig:one}\\
\enddata
\tablecomments{$^a$ See Table \ref{Sourcelist} in main text for coordinates.\\
$^b$ This is the average signal to noise per pixel. The first number is for the blue side CCD, between 3500 and 5400 A. The second number is for the red side CCD, between 5700 and 6800 Angstroms\\
$^c$Several targets have spectra from multiple nights. Only one spectrum per target is shown; this column indicates the corresponding figure, if applicable.
}
\end{deluxetable*}

\begin{figure}[H]
\centering
\makebox[\textwidth][c]{\includegraphics[width=1.1\textwidth]{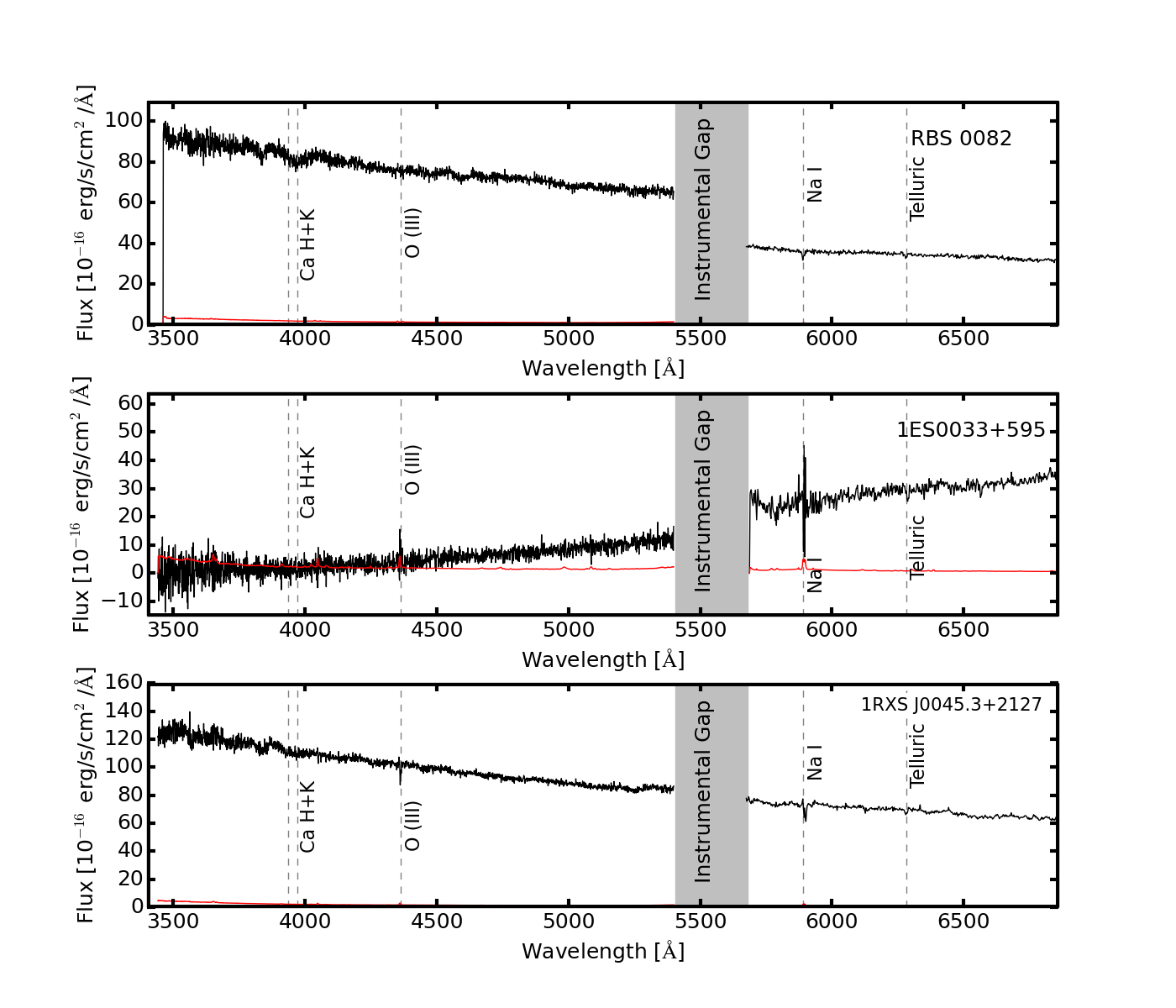}}
\caption{Spectra shown from top to bottom: RBS 0082 (August 13, 2010), 1ES 0033+595 (December 4, 2013), 1RXS J0045.3+2127 (October 28, 2014). Dashed lines indicate telluric and Galactic features. Red lines indicate the error array for each observation; some are not visible due to high S/N.}
\label{fig:two}
\end{figure}

\begin{figure}[H]
\centering
\makebox[\textwidth][c]{\includegraphics[width=1.1\textwidth]{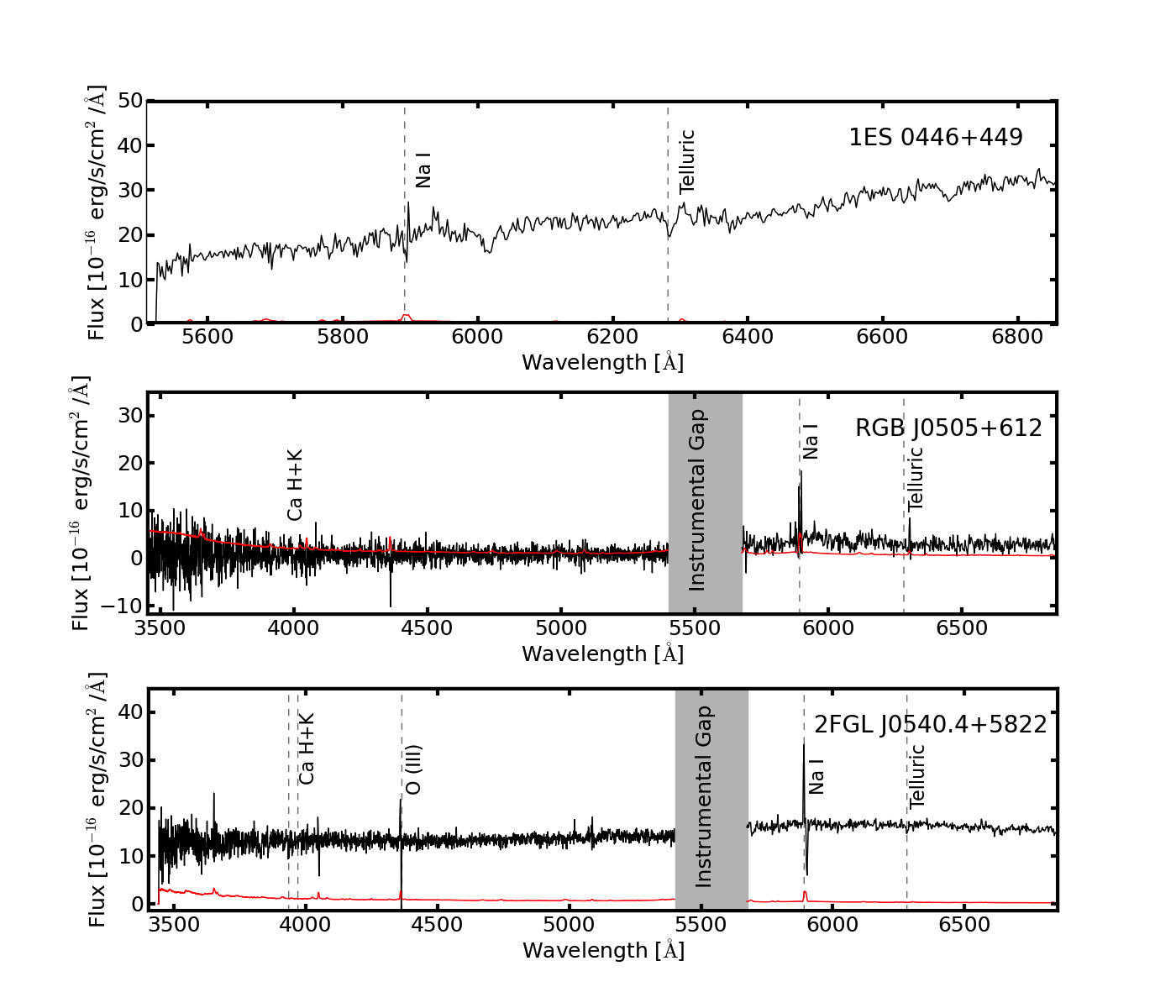}}
\caption{Spectra shown from top to bottom:  1ES 0446+449 (February 14, 2013), RGB J0505+612 (February 14, 2013), 2FGL J0540.4+5822 (October 28, 2014). The spectrum of 1ES 0446+449 (February 14, 2013) only includes the red arm data (5500-8000 Angstroms) because there were complications in reducing the blue arm data. Dashed lines indicate telluric and Galactic features. Red lines indicate the error array for each observation; some are not visible due to high S/N.}
\label{fig:three}
\end{figure}

\begin{figure}[H]
\centering
\makebox[\textwidth][c]{\includegraphics[width=1.1\textwidth]{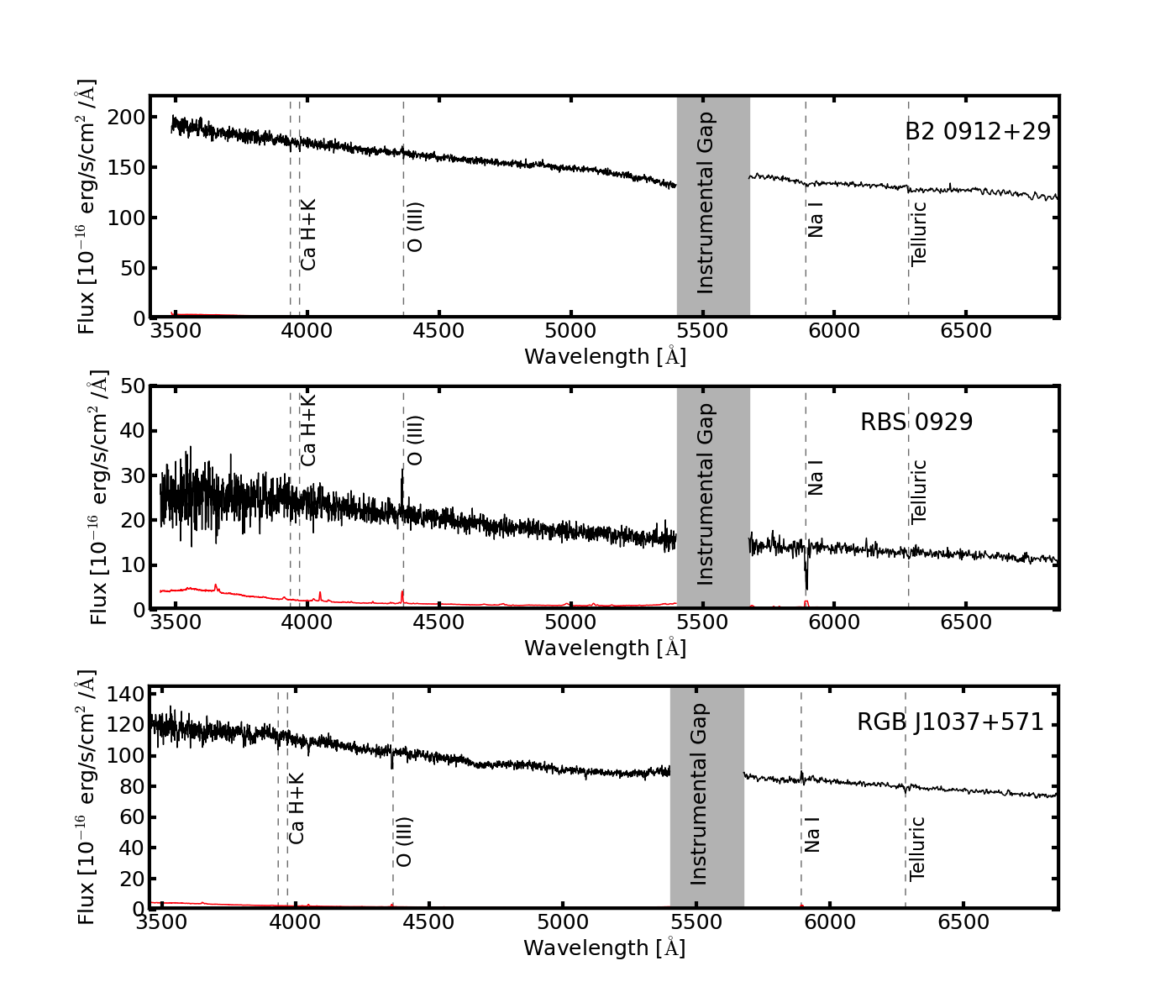}}
\caption{Spectra shown from top to bottom: B2 0912+29 (January 4, 2013), RBS 0929 (April 7, 2013), RGB J1037+571 (February 14, 2013). Dashed lines indicate telluric and Galactic features. Red lines indicate the error array for each observation; some are not visible due to high S/N.}
\label{fig:four}
\end{figure}

\begin{figure}[H]
\centering
\makebox[\textwidth][c]{\includegraphics[width=1.1\textwidth]{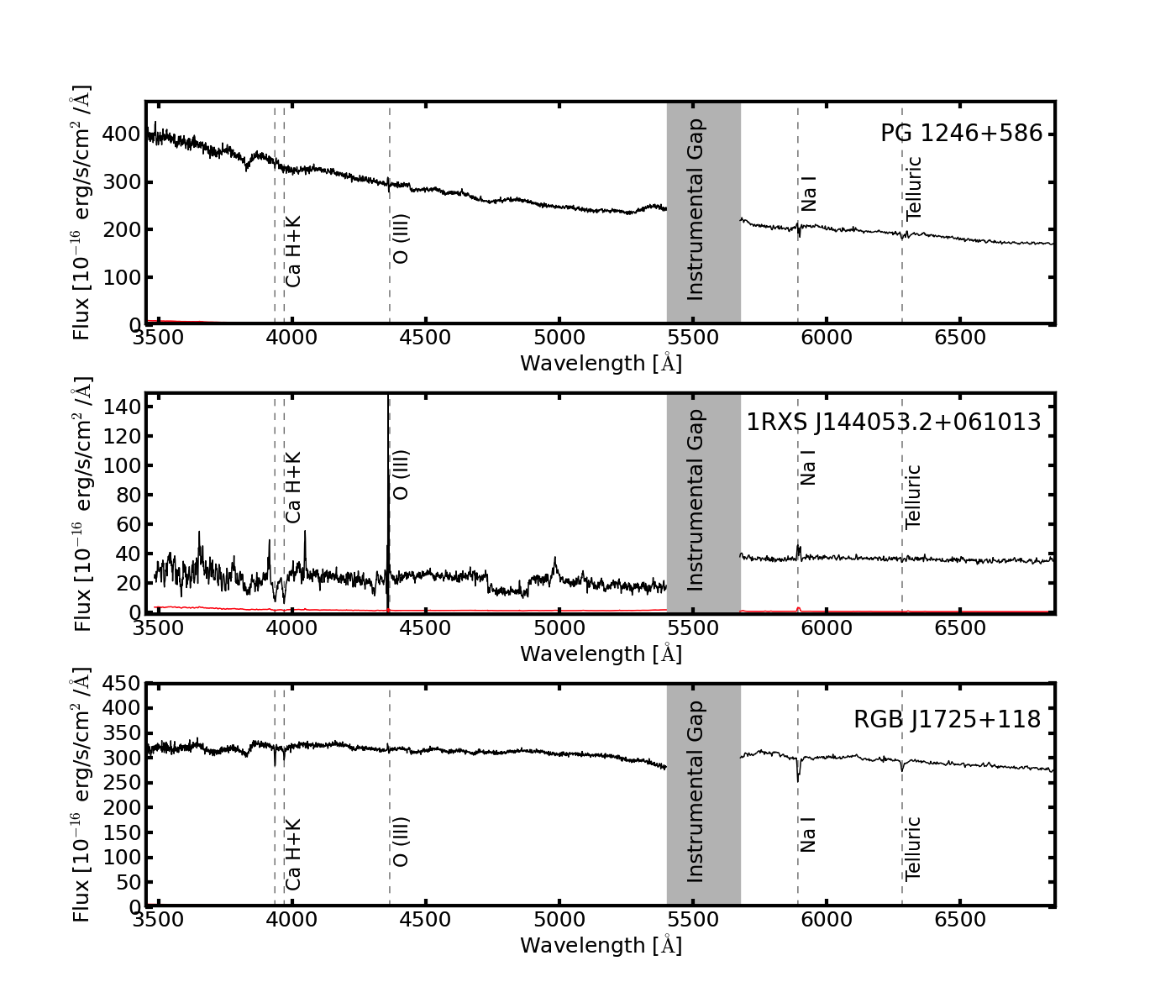}}
\caption{Spectra shown from top to bottom: PG 1246+586 (May 30, 2014), 1RXS J144053.2+061013 (January 4, 2013), RGB J1725+118 (May 30, 2014). Dashed lines indicate telluric and Galactic features. Red lines indicate the error array for each observation; some are not visible due to high S/N.}
\label{fig:five}
\end{figure}

\begin{figure}[H]
\centering
\makebox[\textwidth][c]{\includegraphics[width=1.1\textwidth]{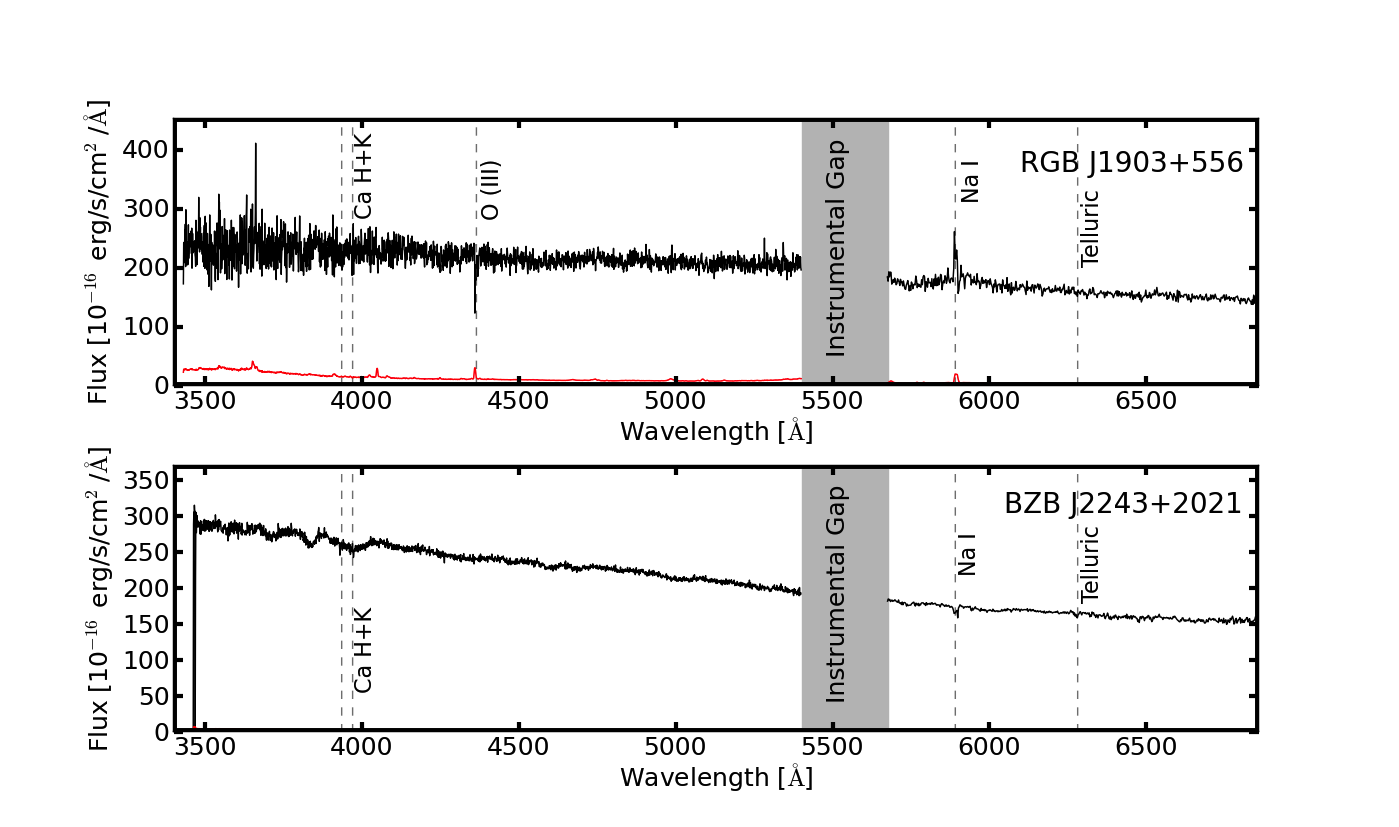}}
\caption{Spectra shown from top to bottom:  RGB J1903+556 (June 13, 2013), BZB J2243+2021 (August 13, 2010). Dashed lines indicate telluric and Galactic features. Red lines indicate the error array for each observation; some are not visible due to high S/N.}
\label{fig:six}
\end{figure}

\end{appendix}

\end{document}